%% file: main.tex
\documentclass{article}

\usepackage[protrusion=true]{microtype}
\usepackage{jzw_macros}
\usepackage[version = 4]{mhchem}
\usepackage{subcaption}
\usepackage{booktabs}
\usepackage[backend=biber,
            style=phys,
            biblabel=brackets,
            maxnames=3,
           ]{biblatex}
\usepackage{xurl}
\usepackage{hyperref}
\hypersetup{colorlinks = true,
            allcolors = blue}

\usepackage{authblk}

\newcommand{\showmatmethods}[1]{}
\title{olLOSC: Unified and efficient density functional approximation to correct
       delocalization error in molecules and periodic materials}

\author[1,$\dagger$]{Yichen Fan}
\author[1,$\dagger$,$\ddag$]{Jacob Z. Williams}
\author[1,2,*]{Weitao Yang}
\affil[1]{Duke University Department of Chemistry, Durham, NC 27708}
\affil[2]{Duke University Department of Physics, Durham, NC 27708}
\affil[$\dagger$]{These authors contributed equally to this work.}
\affil[*]{Correspondence: \href{mailto:weitao.yang@duke.edu}
                               {\texttt{weitao.yang@duke.edu}}}
\affil[$\ddag$]{Current address: 
                Theoretical Division, Los Alamos National Laboratory,
                Los Alamos, NM, 87544}

\date{\today}

\begin{document}
\maketitle
\begin{abstract}
Density functional theory (DFT) is the most promising method for calculating
quantum properties of molecules and materials at moderate and large scales.
However, commonly used density functional approximations (DFAs) have systematic
delocalization error, as demonstrated by underestimated band gaps,
over-delocalized charges, and energy level misalignment at interfaces, which
limits its quantitative prediction.  Extensive efforts, such as the $GW$
approximation to many-body perturbation theory, system-specific tuning of DFA
parameters, and correction functionals have been developed to address
delocalization error. However, an accurate, efficient, and unified solution to
describe total energy, charge density and band structure for both finite systems
and materials is still not available. Building on the linear-response localized
orbital scaling correction  (lrLOSC), we introduce olLOSC: a localized orbital
scaling correction with curvature calculated by orbital-free electronic linear
response. olLOSC has comparable accuracy to lrLOSC, but is much more
computationally efficient. olLOSC corrects delocalization error---especially 
underestimated gaps, but also the total energy---both in molecules and in
materials with small and  moderate band gaps, within the same orbital-free
approximation. Critically, with a a unified  approximation, olLOSC opens the
path for robust and efficient DFT applications across  molecules, materials,
and interfaces.
\end{abstract}

\subsection*{Significance statement}
Key properties of metals in biology and for developing advanced technologies like
catalysts and solar energy converters depend on the quantum-mechanical properties,
especially electronic energy levels and charge distribution, of their component 
molecules and materials. Density functional theory is perhaps the most popular
method for calculating these properties: it is typically much more efficient 
than its competitors. However, it suffers from the systematic delocalization
error, which limits the accuracy of its band gaps and energy level alignment.
There are many corrections for delocalization error; they struggle to apply
universally or are computationally expensive. We present the orbital-free
linear response localized orbital scaling correction (olLOSC). olLOSC corrects
delocalization error in a wide variety of molecules and in materials with small
and moderate band gaps, using the same theory and at a moderate computational
cost comparable to typical DFT calculations.

\input{content}

\subsection*{Data Availability}
The Python (molecules) and R (materials) analysis and figure generation scripts,
as well as the underlying data, are included in the Supporting Information.
olLOSC for materials is implemented as a fork of \texttt{Quantum ESPRESSO} and
is available at \cite{williams_yang_2025}.

\section*{Acknowledgements}
We are grateful to Yang Shen for discussion about the positive semidefiniteness
of $\ket{\delta\rho_{i\nu}}$ under the partial RPA. We acknowledge financial
support from the National Science Foundation (Grant No.~CHE-2154831) and the
National Institutes of Health (Grant No.~R35-GM158181). 

\printbibliography

\end{document}


\maketitle

\section{Additional notation for materials} \label{sec:bz}
Computing LOSC in periodic boundary conditions requires additional
considerations (and more notation). We sample the complex Kohn--Sham orbitals at
$N_k$ different points $\bk$ in the irreducible Brillouin zone. This yields an
additional index $\bk$ to the canonical orbitals. They also obey Bloch's theorem
\cite{bloch1929},
\begin{equation} \label{eq:bloch}
    \psi_{\bk n \sigma}(\br) = e^{i\bk\cdot\br} u_{\bk n\sigma}(\br),
\end{equation}
where $u_{\bk n \sigma}(\br)$ shares the translational symmetry of the
underlying lattice.

The bulk orbitalets must also obey lattice's translational symmetry. This is
shown mathematically by localizing with a different unitary matrix 
$U^{\bk\sigma}$ at each $\bk$-point; the orbitalets are given by
\begin{equation} \label{eq:loc-unitary}
    \ket{\phi_{\bk i \sigma}} = 
    \sum_n U_{in}^{\bk\sigma} \ket{\psi_{\bk n \sigma}},
\end{equation}
the orbitals
and do not mix at different $\bk$. To obtain a more natural representation in
real space, we transform to the (generalized) Wannier representation
\cite{wannier1937}, yielding \emph{dually localized Wannier functions} (DLWFs)
\cite{mahler2022}
\begin{equation} \label{eq:loc-wannier}
    \ket{w_{\bR i\sigma}} = \frac{1}{N_k} \sum_{\bk} e^{-i \bk \cdot \bR}
        \ket{\phi_{i\sigma}}.
\end{equation}
By the same token, maximally localized Wannier functions \cite{marzari1997} are
not an exact analogue of the molecular Foster--Boys orbitals \cite{foster1960};
loosening the translational invariance constraint could, in principle, yield
localized orbitals with smaller spatial variance. (Of course, the periodic
structure of Wannier functions is a major part of their usefulness and appeal.)
Note that, in molecules and materials sampled only at $\Gamma$, $N_k = 1$ and
$\bk = \bZ$. This renders the Fourier transformation of \eqref{eq:loc-wannier}
trivial, and $\ket{\phi_{\bZ i \sigma}} = \ket{\phi_{i\sigma}}$.

\subsection{The monochromatic decomposition}
It is well known \cite{baroni2001, timrov2018} that linear operators $O$ that
are periodic on a supercell consisting of $N_k$ uniformly sampled primitive
unit cells---or, equivalently, on $N_k$ $\bk$-points sampled in the Brillouin
zone---can be decomposed into monochromatic components, each periodic on the
primitive cell:
\begin{equation}
    O(\br, \br') = 
    \sum_{\bq} e^{i\bq\cdot\br} O^{\bq}(\br, \br') e^{-i\bq\cdot\br'}.
\end{equation}
(For the weights of each $O^{\bq}$ to be equal, the Brillouin zone must be
sampled uniformly \cite{monkhorst1976}, and the origin $\Gamma = \bZ$ of
reciprocal space must be one of the $\bk$-points.)

This decomposition applies to one-index quantities such as Bloch orbitals
(including their perturbations, useful for density functional perturbation
theory); indeed, it is nothing more than a statement of Bloch's theorem
\eqref{eq:bloch}. It also applies to Wannier functions, with
\begin{equation}
    w_{\bR i \sigma}(\br) = 
    \frac{1}{N_k} \sum_{\bq} e^{i\bq\cdot\br} w_{\bR i \sigma}^{\bq}(\br);
\end{equation}
the factor of $1/N_k$ is a normalization convention.

More recently, Colonna et al. found in \cite{colonna2022} that the
\emph{density} of a generalized Wannier function (such as a LOSC DLWF)
also has this property:
\begin{equation} \label{eq:rhowann_decomp}
    \rho_{\bR i \sigma}(\br) = \abs{w_{\bR i \sigma}(\br)}^2 =
    \frac{1}{N_k} \sum_{\bq} e^{i\bq\cdot\br} \rho_{\bR i \sigma}^{\bq}(\br),
\end{equation}
where
\begin{equation} \label{eq:rhowann_decomp_2}
    w_{\bR i \sigma}^{\bq}(\br) = 
    \frac{1}{N_k} e^{-i\bq\cdot\bR} 
        \sum_{\bk} \conj{\phi_{\bk i \sigma}}(\br) 
                   \phi_{(\bk+\bq) i \sigma}(\br).
\end{equation}
Observe also that
\begin{equation}
    \ket{\rho_{\bR i \sigma}} = 
    \frac{1}{N_k} \sum_{\bq} e^{-i\bq\cdot\bR} \ket{\rho_ {\bZ i \sigma}^{\bq}}.
\end{equation}
Thus, we only compute quantities for the home unit cell $\bR = \bZ$ explicitly;
for the rest, we require only the phase factors $e^{-i\bq\cdot\bR}$. Because the
LOSC curvature operator is linear in the orbitalet densities,
\begin{equation} \label{eq:losc-kappa}
    \kappa_{\bR ij \sigma} = 
    \mel{\rho_{\bZ i \sigma}}
        {f_{\Hxc}^{\sigma\sigma} + 
         \sum_{\nu\tau} 
            f_{\Hxc}^{\sigma\nu} \chi^{\nu\tau} f_{\Hxc}^{\tau\sigma}
        }
        {\rho_{\bR j \sigma}},
\end{equation}
we can apply the monochromatic decomposition wholesale, obtaining
\begin{equation} \label{eq:kappa-monochromatic}
\begin{split}
    \kappa_{\bR ij \sigma} &=
    \sum_{\bq} e^{-i\bq\cdot \bR} 
        \mel{\rho_{\bZ i \sigma}^{\bq}}
            {f_{\Hxc}^{\bq;\sigma\sigma} + 
             \sum_{\nu\tau}
                f_{\Hxc}^{\bq;\sigma\nu} 
                \chi^{\bq;\nu\tau} 
                f_{\Hxc}^{\bq;\tau\sigma}
            }
            {\rho_{\bZ j \sigma}^{\bq}} \\ &=
    \sum_{\bq} e^{-i\bq\cdot\bR} \kappa_{\bZ ij \sigma}^{\bq}.
\end{split}
\end{equation}

The monochromatic decomposition
$\kappa_{\bR ij\sigma}=\sum_{\bq} e^{-i\bq\cdot\bR}\kappa_{\bZ ij\sigma}^{\bq}$
provides a substantial improvement to computational scaling. Na\"ively, quantities
like $\kappa$ and $\ket{\rho_{\bR i \sigma}}$ are periodic on the supercell
real- and reciprocal-space grids, which have $N_k N_G$ elements (where $N_G$ is
the size of the primitive cell grid). Matrix operations on this space scale as
$\Ord(N_k^3 N_G^3)$. The monochromatic decomposition allows us to work with
quantities periodic on the primitive cell, with matrix operations scaling as
$\Ord(N_G^3)$. Each supercell quantity decomposes into $N_k = N_q$
primitive-cell quantities, which means that computing each curvature element
$\kappa_{\bZ ij \sigma}^{\bq}$ scales as $\Ord(N_k N_G^3)$ overall (up to
logarithmic factors due to the fast Fourier transform), saving a factor of
$N_k^2$. In lrLOSC, the density functional perturbation theory approach
\cite{baroni2001} uses the Sternheimer equation \cite{sternheimer1954} to compute
$\chi^{\nu\tau} f_{\Hxc}^{\tau\sigma} \ket{\rho_{\bZ j \sigma}^{\bq}}$ via the
first-order variations $\ket{\delta\psi_{(\bk+\bq)n\sigma}}$ of the Kohn--Sham
orbitals $\ket{\psi_{\bk n \sigma}}$; the reciprocal-space points $\bk$ and
$\bk + \bq$ are paired, so computing $\kappa_{\bZ ij \sigma}^{\bq}$ in lrLOSC
requires $\Ord(N_k^2 N_G^3)$ effort, saving only one factor of $N_k$. For more
details on the monochromatic decomposition, see \cite{colonna2022} and the
Supplemental Material of \cite{williams2024}.

\section{Implementation of the olLOSC curvature}

\subsection{The Thomas--Fermi--von Weizs\"acker kernel}
The Thomas--Fermi kinetic energy is
\begin{equation}
    T_{\TF}[\rho] = c_F \int d\br\, \rho(\br)^{5/3},
\end{equation}
where $c_F = (3/10) (3\pi^2)^{2/3}$ is the Fermi constant. The variation in
$T_{\TF}$ with respect to $\rho^\sigma$ is
\begin{equation}
    \frac{\delta T_{\TF}}{\delta \rho^\sigma(\br)} = 
    \frac53 c_F \rho^\sigma(\br)^{2/3},
\end{equation}
and the kernel is
\begin{equation}
    \frac{\delta^2 T_{\TF}}{\delta \rho^\sigma(\br) \delta\rho^\tau(\br')} =
    2^{2/3}\, \frac{10 c_F}{9} \rho^\sigma(\br)^{-1/3} \delta_{\sigma\tau}
        \delta(\br - \br');
\end{equation}
the factor of $2^{2/3}$ is omitted for a spinless calculation \cite{oliver1979}.

The expression for the von Weizs\"acker kernel is known, e.g. Eq.\ (A10) of
\cite{chattaraj2007} and Eq.\ (97) of \cite{dellasala2022}, although the
derivation is very long. In Rydberg units, ($\hbar = 1$, but the electron mass
$m = 1/2$), it is
\begin{equation}
\label{eq:vw}
\begin{split}
    \frac{\delta^2 T_\vW}{\delta \rho_\sigma(\br) \delta\rho_\sigma(\br')} &=
    \frac14 \frac{\delta v_\sigma(\br)}{\delta \rho_\sigma(\br')} \\ &=
    \frac12 \Big[
        \frac{\nabla\rho_\sigma(\br') \cdot \nabla\delta(\br - \br')}
             {\rho_\sigma(\br')^2} +
        \frac{\nabla^2\rho_\sigma(\br')}{\rho_\sigma(\br')^2} 
            \delta(\br - \br') - \\ &\qquad\qquad
        \frac{\nabla\rho_\sigma(\br')\cdot\nabla\rho_\sigma(\br')}
             {\rho_\sigma(\br')^3} \delta(\br - \br') -
        \frac{\nabla^2 \delta(\br - \br')}{\rho_\sigma(\br')}
    \Big].
\end{split}
\end{equation}

Together, we have
\begin{equation}
    f_{\TF\vW}^{\sigma\tau}(\br, \br') = 
    f_{\TF\vW}^\sigma(\br, \br') \delta_{\sigma\tau} =
    \left[
        f_{\TF}^\sigma(\br) \delta(\br - \br') + 
        \lambda f_{\vW}^{\sigma}(\br, \br')
    \right] \delta_{\sigma\tau},
\end{equation}
where $\lambda \geq 0$ is the amount of von Weizs\"acker correction. Note that
$f_{\vW}$ is \emph{semilocal}; its action on $\ket{\delta\rho}$ at $\br$
depends on $\ket{\delta\rho}$, $\nabla \ket{\delta\rho}$, and
$\nabla^2 \ket{\delta\rho}$, but still only requires them to be evaluated at
$\br$.

\subsection{olLOSC in molecules}
Molecules have few enough basis functions that we can almost invert $\chi$
in \eqref{eq:losc-kappa} directly. The last step is provided by the
resolution-of-the-identity (RI) approximation
\cite{vahtras_etal_1993, ren2012a} with auxiliary basis functions. Within the
RI approximation, the four-center Coulomb integral
\begin{equation}
    (ij|kl) = 
    \iint d\br\, d\br'\, 
        \frac{\conj{\psi}_{i}(\br) \psi_j(\br) \conj{\psi}_k(\br') \psi_l(\br')}
             {\abs{\br - \br'}}
\end{equation}
can be evaluated in terms of the products of three-center and two-center
integrals:
\begin{equation} \label{eq:ri}
    (ij|kl) \approx \sum_{\mu\nu} C_{ij}^{\mu} V_{\mu\nu}^{-1} C_{kl}^{\nu}.
\end{equation}
Here, we define
\begin{equation}
\label{eq:riv}
    V_{\mu\nu} = \iint d\br d\br' \cfrac{P_\mu(\br)P_\nu(\br')}{|\br - \br'|}
\end{equation}
and
\begin{equation}
    C_{ij}^\mu = \sum_{\nu} (ij|\nu) V_{\nu\mu}^{-1};
\end{equation}
here,
\begin{equation}
    (ij|\nu) = 
    \iint d\br\, d\br'\,
        \frac{\conj{\psi}_i(\br) \psi_j(\br) P_\nu(\br')}
             {\abs{\br - \br'}}.
\end{equation}
$\mu, \nu = 1, 2, .. N_{aux}$ label the auxiliary basis functions $P_{\mu}$.

We also use the same set of auxiliary basis functions to discretize the
Thomas--Fermi--von Weizs\"acker (TFvW) kinetic energy kernel introduced in
Equation \ref{eq:vw}, following the discussion in York and Yang
\cite{york1996}. In our molecular implementation, we evaluate the TFvW
kernel in the auxiliary basis and store the resulting coefficients, which are
\begin{equation}
\label{eq:vwMatrix}
    f^{\TF\vW}_{\mu\nu,\sigma} = \iint d\br d\br' P_{\mu}(\br) 
    \frac{\delta^2 (T_{\TF} + \lambda T_\vW)}{\delta \rho_\sigma(\br) \delta\rho_\sigma(\br')} P_{\nu}(\br')
\end{equation}

In the random phase approximation, the inverse of the many-body static linear
response function $\chi$ is
\begin{equation}
    \label{eq:lr}
    \chi^{-1}_{\sigma\tau}(\br, \br') = \cfrac{\delta V^\sigma(\br)}{\delta \rho^\tau(\br')} = \cfrac{\delta (T^{\sigma} \delta_{\sigma\tau} + V_{\Har}^{\sigma})}{\delta \rho^{\tau}(\br')}.
\end{equation}
We can evaluate the matrix elements of $\chi^{-1}$ as the sum of the kinetic
kernel \eqref{eq:vwMatrix} and Hartree kernel \eqref{eq:riv}. We also implement
a beyond-RPA olLOSC scheme, which adds $V_{\xc}^\sigma$ to the right-hand side
of \eqref{eq:lr}. For results on this, see \ref{sub:beyond-rpa}.

\subsection{olLOSC in materials}
To compute the screened density response $\ket{\delta\rho_{\bZ i \tau}}$ in
materials, we solve the coupled linear equations
\begin{equation} \label{eq:drho-of-lin}
    \begin{pmatrix}
        f_{\kin}^{\alpha \alpha} + f_{\Har}^{\alpha \alpha} &
        f_{\kin}^{\alpha \beta} + f_{\Har}^{\alpha \beta} \\[0.5em]
        f_{\kin}^{\beta \alpha} + f_{\Har}^{\beta \alpha} &
        f_{\kin}^{\beta \beta} + f_{\Har}^{\beta \beta}
    \end{pmatrix} 
    \begin{pmatrix}
        \ket{\delta\rho_{\bZ i \alpha}^{\vphantom{\beta}}} \\[0.5em]
        \ket{\delta\rho_{\bZ i \beta}^{\vphantom{\beta}}}
    \end{pmatrix}
    =
    \begin{pmatrix}
        \delta\mu^\alpha - \ket{\delta V_{\bR i \alpha}} \\[0.5em]
        \delta\mu^\beta - \ket{\delta V_{\bR i \beta}}
    \end{pmatrix}.
\end{equation}
Because the Thomas--Fermi--von Weisz\"acker kernel is diagonal in spin, this
reduces to
\begin{equation} \label{eq:drho-of-tfvw}
    \begin{pmatrix}
        f_{\TF\vW}^{\alpha} + f_{\Har}^{\alpha\alpha} &
        f_{\Har}^{\alpha\beta} \\
        f_{\Har}^{\beta\alpha} &
        f_{\TF\vW}^{\beta} + f_{\Har}^{\beta\beta}
    \end{pmatrix}
    \begin{pmatrix}
        \ket{\delta\rho_{\bZ i \alpha}} \\ \ket{\delta\rho_{\bZ i \beta}}
    \end{pmatrix} =
    \begin{pmatrix}
        \delta\mu^\alpha - \ket{\delta V_{\bZ i \alpha}} \\
        \delta\mu^\beta - \ket{\delta V_{\bZ i \beta}}
    \end{pmatrix}.
\end{equation}
Formally, we solve \eqref{eq:drho-of-tfvw} by inverting the left-hand matrix
$M_{\ol}$. However, it is nonlocal on either the real-space or reciprocal-space
grid, and the number of grid points (plane waves) can exceed $10^6$. We
therefore use a Krylov subsapce algorithm to invert it iteratively. Under the
partial RPA, $M_{\ol}$ is Hermitian and positive semidefinite; as we show in
\ref{sec:psd-ol-rpa}, $\ev{M_{\ol}}{x} \geq 0$ for any vector
$\ket{x} = \begin{pmatrix} \ket{x}^\alpha & \ket{x}^\beta \end{pmatrix}^\top$,
so we solve for $\ket{\delta\rho_{\bZ i \sigma}}$ with the conjugate gradient
algorithm.

Instead of having to recalculate the charge conservation terms
$\delta\mu^\sigma$ every iteration, we account for them implicitly by
constraining the optimization. Including $\delta\mu^\sigma$ on the right side
of \eqref{eq:drho-of-lin} is equivalent to requiring
\begin{equation}
    \int d\br\, \delta\rho_{\bZ i \alpha}(\br) =
    \int d\br\, \delta\rho_{\bZ i \beta}(\br) = 0.
\end{equation}
We implement the projected preconditioning method of \cite{gould2001}, using a
preconditioner $P$ that preserves the constraint:
\begin{equation}
    P = 
    \begin{pmatrix} 
        -\chi_{\TF\vW; \HEG}^{\alpha}(\bG) & \\ 
        & -\chi_{\TF\vW; \HEG}^{\beta}(\bG),
    \end{pmatrix},
\end{equation}
where 
\begin{equation} \label{eq:ker-tfvw-heg}
    \chi_{\TF\vW; \HEG}^{\sigma}(\bG) =
    -\left[f_{\TF\vW; \HEG}^{\sigma}\right]^{-1}(\bG) = 
    -\frac{k_F}{m_e \pi^2 (1 + 3 \lambda \eta^2)}
\end{equation}
is the Thomas--Fermi--von Weisz\"acker kernel for the homogeneous electron gas.
$k_F = (3\pi^2 \bar{\rho}^\sigma)^{1/3}$ is the Fermi wavevector given the
average density of spin $\sigma$ in one unit cell,
\begin{equation*}
    \bar{\rho}^\sigma = \int_{\text{UC}} d\br\, \rho^\sigma(\br);
\end{equation*}
$\lambda$ is the fraction of von Weisz\"acker exchange in
$f_{\TF\vW} = f_{\TF} + \lambda f_{\vW}$; $m_e$ is the mass of the electron,
unity in atomic units but $1/2$ in Rydberg units, favored by
\texttt{Quantum ESPRESSO}; and $\eta = \abs{\bG}/k_F$ is the reduced wavevector
magnitude. Note that $f_{\TF\vW}$ (and $\chi_{\TF\vW}$ are local in both spin and
reciprocal space, with $f_{\TF\vW}^{\sigma\nu}(\bG, \bG') =
f_{\TF\vW}^{\sigma}(\bG) \delta_{\sigma\nu} \delta_{\bG\bG'}$.

\subsection{\texorpdfstring{$M_{\ol}$}{M} is positive semidefinite under
            the partial RPA}
\label{sec:psd-ol-rpa}
We decompose $M_{\ol}$ into three components:
\begin{equation}
    M_{\ol} =
    \begin{pmatrix}
    f_{\TF}^\alpha & \\ & f_{\TF}^\beta
\end{pmatrix} +
\begin{pmatrix}
    f_{\vW}^\alpha & \\ & f_{\vW}^\beta
\end{pmatrix} +
\begin{pmatrix}
    f_{\Har}^{\bq} & f_{\Har}^{\bq} \\ f_{\Har}^{\bq} & f_{\Har}^{\bq}
\end{pmatrix} \coloneqq
    M_1 + M_2 + M_3.
\end{equation}
By linearity, if $M_1$, $M_2$, and $M_3$ are each positive semidefinite, then
\begin{equation}
    \ev{M_{\ol}}{x} = \ev{M_1}{x} + \ev{M_2}{x} + \ev{M_3}{x} \geq 0,
\end{equation}
and so is $M_{\ol}$.

\subsubsection{The Thomas--Fermi kernel}
This is the simplest to show. The Thomas--Fermi kernel is
\begin{equation}
    f_{\TF}^\sigma(\br) = 
    \frac{\delta^2 T_{\TF}}{\delta\rho_\sigma(\br) \delta\rho_\tau(\br')} =
    \frac{10 c_F}{9} \left[ \rho_\sigma(\br) \right]^{-1/3}\, 
        \delta_{\sigma\tau} \delta(\br - \br'),
\end{equation}
where the constant $c_F = (3/10) \left(3 n_s \pi^2\right)^{2/3} > 0$. (The
number of spin components is $n_s$, here $n_s = 2$; and in Rydberg units, where
$\hbar = 1$ but the electron mass $m_e = 1/2$, $c_F$ must be multiplied by 2.)

Given any nonzero two-spin vector $x$,
\begin{equation}
\begin{split}
x^\dagger M_1 x &=
\begin{pmatrix} \bra{x_\alpha} & \bra{x_\beta} \end{pmatrix}
\begin{pmatrix} f_{\TF}^{\alpha} & 0 \\ 0 & f_{\TF}^\beta \end{pmatrix}
\begin{pmatrix} \ket{x_\alpha} \\ \ket{x_\beta} \end{pmatrix} \\ &=
\ev{f_{\TF}^\alpha}{x_\alpha} + \ev{f_{\TF}^\beta}{x_\beta} \\ &=
\iint d\br\, d\br'\, 
    \conj{x_\alpha}(\br) f_{\TF}^\alpha(\br, \br') x_\alpha(\br') +
\iint d\br\, d\br'\, 
    \conj{x_\beta}(\br) f_{\TF}^\beta(\br, \br') x_\beta(\br') \\ &=
\frac{10 c_F}{9} \int d\br\,
    \left(
    \abs{x_\alpha(\br)}^2 \left[ \rho_\alpha(\br) \right]^{-1/3} + 
    \abs{x_\beta(\br)}^2 \left[ \rho_\beta(\br) \right]^{-1/3} 
    \right) \\ &\geq 0,
\end{split}
\end{equation}
since $\rho^\sigma(\br) \geq 0$.

\subsubsection{The von Weizs\"acker kernel}
Like the Thomas--Fermi kernel, the vW kernel is diagonal in spin, so
$x^\dagger M_2 x = \ev{f_{\vW}^\alpha}{x_\alpha} + \ev{f_{\vW}^\beta}{x_\beta}$.
It suffices to consider $\ev{f_{\vW}^\alpha}{x_\alpha}$; exactly the same
argument will apply to the spin-$\beta$ term.

Observe that
\begin{equation}
\begin{split}
    T_\vW^\alpha[\rho] &= 
    \frac\lambda8 \int d\br\,
        \frac{\abs{\nabla \rho_\alpha(\br)}^2}{\rho_\alpha(\br)} =
    \frac\lambda8 \int d\br\,
        \frac{\nabla\rho_\alpha(\br) \cdot \nabla\rho_\alpha(\br)}
             {\rho_\alpha(\br)} =
    \frac\lambda8 \int d\br\, 
        \frac{\gamma_{\alpha\alpha}(\br)}{\rho_\alpha(\br)} \\ &=
    \frac\lambda8 \int d\br\, f(\br, \rho_\alpha, \gamma_{\alpha\alpha}),
\end{split}
\end{equation}
where the reduced gradient
$\gamma_{\alpha\beta} = \nabla\rho_\alpha \cdot \nabla\rho_\beta$.
Without loss of generality, we ignore the constant $\lambda/8$.

The derivatives of $f$ with respect to $\gamma_{\alpha\alpha}$ and
$\rho_\alpha$ are \cite{hirata1999}
\begin{equation}
\begin{array}{*5{>{\displaystyle}c}}
    \frac{\partial f}{\partial \rho_\alpha} = 
        -\frac{\gamma_{\alpha\alpha}}{\rho_\alpha^2}; &
    \frac{\partial^2 f}{\partial \rho_\alpha^2} =
        +\frac{2\gamma_{\alpha\alpha}}{\rho_\alpha^3}; &
    \frac{\partial f}{\partial \gamma_{\alpha\alpha}} =
        \frac{1}{\rho_\alpha}; &
    \frac{\partial^2 f}{\partial \gamma_{\alpha\alpha}^2} = 0; \\[1em]
    \frac{\partial^2 f}{\partial \rho_\alpha \partial \gamma_{\alpha\alpha}} =
        -\frac{1}{\rho_\alpha^2}.
\end{array}
\end{equation}
Omitting the dependence on $\br$,
\begin{equation}
\begin{split}
    \ev{f_\vW^\alpha}{x_\alpha} &=
        2 \int d\br\, \left( \nabla\conj{x}_\alpha\cdot\nabla x_\alpha\right)
            \frac{\partial f}{\partial \gamma_{\alpha\alpha}} \\ &\qquad+
        \int d\br\, \conj{x_\alpha} 
            \frac{\partial^2 f}{\partial \rho_\alpha^2} x_\alpha \\ &\qquad+
        2 \int d\br\, 
            \left[ 
                \left(\nabla \rho_\alpha \cdot \nabla \conj{x}_\alpha\right) 
                    x_\alpha +
                \conj{x}_\alpha \left(\nabla \rho \cdot \nabla x_\alpha \right)
            \right] 
            \frac{\partial^2 f}
                 {\partial \rho_\alpha \partial \gamma_{\alpha\alpha}} \\ 
            &\qquad+
        4 \int d\br\, 
            \left( \nabla\rho_\alpha \cdot \nabla \conj{x}_\alpha \right)
            \frac{\partial^2 f}{\partial \gamma_{\alpha\alpha}^2}
            \left( \nabla\rho_\alpha \cdot \nabla x_\alpha \right) \\ &=
    2 \int d\br\, \frac{\abs{\nabla x_\alpha}^2}{\rho_\alpha} \\ &\qquad-
    2 \int d\br\, 
        \frac{\conj{x}_\alpha 
              \left(\nabla\rho_\alpha \cdot \nabla x_\alpha\right) + 
              \left(\nabla\rho_\alpha \cdot \nabla \conj{x}_\alpha \right) 
              x_\alpha}
             {\rho_\alpha^2} \\ &\qquad+
    2 \int d\br\, \frac{\gamma_{\alpha\alpha} \abs{x_\alpha}^2}
                       {\rho_\alpha^3}.
\end{split}
\end{equation}
Multiplying by powers of $\rho_\alpha$ to obtain a common denominator, we have
\begin{multline}
2 \int d\br\, \frac{1}{\rho_\alpha^3} 
    \Big[ 
        \rho_\alpha^2 (\nabla \conj{x}_\alpha \cdot \nabla x_\alpha) - 
        \rho_\alpha \conj{x}_\alpha (\nabla\rho_\alpha \cdot \nabla x_\alpha) -
        \\
        \rho_\alpha x_\alpha (\nabla\rho_\alpha \cdot \nabla \conj{x}_\alpha) +
        (\nabla \rho_\alpha \cdot \nabla\rho_\alpha) \abs{x_\alpha}^2
    \Big] \\ =
2 \int d\br\, 
    \frac{\left(\rho_\alpha \nabla x_\alpha - 
                x_\alpha \nabla \rho_\alpha \right)^* \times
          \left(\rho_\alpha \nabla x_\alpha -
                x_\alpha \nabla \rho_\alpha \right)}
         {\rho_\alpha^3} \\ =
2 \int d\br\, 
    \frac{\abs*{\rho_\alpha \nabla x_\alpha - x_\alpha \nabla \rho_\alpha}^2}
         {\rho_\alpha^3};
\end{multline}
since $\rho_\alpha \geq 0$, $\ev{f_{\vW}^\alpha}{x_\alpha} \geq 0$ (as long as
the denominator does not diverge).

Replacing the spin index by means that $\ev{f_{\vW}^\beta}{x_\beta} \geq 0$,
whence $x^\dagger M_2 x \geq 0$.

\subsubsection{The Hartree kernel}
We work in reciprocal space, where $f_{\Har}^{\bq}$ is local, and note that
positive semidefiniteness is invariant under the choice of basis. By the shift
theorem of Fourier transforms,
\begin{equation}
    f_{\Har}^{\bq}(\bG) = 
    \FT{\left[f_{\Har}^{\bq}(\br)\right]} =
    \FT{\left[ \frac{e^{i\bq\cdot\abs{\br'-\br}}}{\abs{\br'-\br}}\right]} =
    \frac{4\pi}{\abs{\bG + \bq}^2} > 0.
\end{equation}
(The divergence at $\abs{\bG + \bq} = 0$ can be handled in various ways; it's
enough to know that $f_{\Har}(\abs{\bG + \bq} = 0) \geq 0$.)

The expectation value of any two-spin vector $x$ is
\begin{multline}
\begin{pmatrix}
    \bra{x_\alpha} & \bra{x_\beta}
\end{pmatrix}
\begin{pmatrix}
    f_{\Har}^{\bq} & f_{\Har}^{\bq} \\ f_{\Har}^{\bq} & f_{\Har}^{\bq}
\end{pmatrix}
\begin{pmatrix}
    \ket{x_\alpha} \\ \ket{x_\beta}
\end{pmatrix} \\ =
\ev{f_{\Har}^{\bq}}{x_\alpha} + \mel{x_\alpha}{f_{\Har}^{\bq}}{x_\beta} +
    \mel{x_\beta}{f_{\Har}^{\bq}}{x_\alpha} + \ev{f_{\Har}^{\bq}}{x_\beta} \\=
    \sum_{\bG} \frac{4\pi}{\abs{\bq + \bG}^2} \times
        \Big( 
            \conj{x_\alpha}(\bG) x_\alpha(\bG) + 
            \conj{x_\alpha}(\bG) x_\beta(\bG) \\ \qquad +
            \conj{x_\beta}(\bG) x_\alpha(\bG) +
            \conj{x_\beta}(\bG) x_\beta(\bG)
        \Big).
\end{multline}
We simplify the parts depending on $x_\sigma$:
\begin{multline}
\conj{x_\alpha}(\bG) x_\alpha(\bG) + \conj{x_\alpha}(\bG) x_\beta(\bG) +
    \conj{x_\beta}(\bG) x_\alpha(\bG) + \conj{x_\beta}(\bG) x_\beta(\bG) \\ =
\conj{\left[x_\alpha(\bG) + x_\beta(\bG)\right]}
    \left[x_\alpha(\bG) + x_\beta(\bG)\right],
\end{multline}
so that the total expectation value is the Hartree self-energy of the vector
$\ket{x_\alpha + x_\beta}$. Since the Hartree kernel is nonnegative (manifest
in reciprocal space),
\begin{equation}
    \ev{f_{\Har}^{\bq}}{x_\alpha + x_\beta} \geq 0,
\end{equation}
with equality only if $\ket{x_\beta} = -\ket{x_\alpha}$.

\section{The LOSC Hamiltonian}
Using the chain rule from the generalized complex (or Wirtinger) derivative
\cite{wirtinger1927}, the LOSC Hamiltonian is given by
\begin{equation}
  \Delta h = \frac{\delta \Delta E}{\delta \rho}
    = \sum_\sigma \Delta h^\sigma,
\end{equation}
with
\begin{equation}
    \Delta h^\sigma = 
    \sum_{\bR ij} \left[ 
        \frac{\partial \Delta E}{\partial \lambda_{\bR ij \sigma}}
        \frac{\delta \lambda_{\bR ij \sigma}}{\delta \rho}
      +
        \frac{\partial \Delta E}{\partial \lambda_{\bR ij \sigma}^*}
        \frac{\delta \lambda_{\bR ij \sigma}^*}{\delta \rho}
      \right].
\end{equation}
We can split the correction by spins:
\begin{equation}
    \Delta E = 
    \sum_\sigma \sum_{\bR ij} \conj{\lambda_{\bR ij \sigma}}
        \left( \delta_{\bR ij} - \lambda_{\bR ij \sigma} \right)
        \kappa_{\bR ij \sigma}.
\end{equation}
Recall that $\lambda_{\bR ij \sigma} = \mel{w_{\bZ i}}{\rho}{w_{\bR j}}$. We
are also assuming that $\kappa_{\bR ij \sigma}$ is fixed; this is the
\emph{frozen-orbitalet approximation} for SCF calculation.
The orbitalets can be relaxed in the so-called macro-SCF loop, although this
has only minor effects \cite{mei2020a}.

An elementary result from the calculus of variations gives us
\begin{equation}
  \frac{\delta \lambda_{\bR ij \sigma}}{\delta \rho}
    = \frac{\delta \mel{w_{\bZ i}}{\rho}{w_{\bR j}}}{\delta \rho}
    = \ketbra{w_{\bR j}}{w_{\bZ i}};
\end{equation}
similarly, by the Hermiticity of $\lambda$,
\begin{equation} 
  \frac{\delta \lambda_{\bR ij \sigma}^*}{\delta \rho}
    = \frac{\delta \mel{w_{\bR j}}{\rho}{w_{\bZ i}}}{\delta \rho}
    = \ketbra{w_{\bZ i}}{w_{\bR j}}.
\end{equation}
Next, we observe that
\begin{equation}
  \frac{\partial \Delta E}{\partial \lambda_{\bR ij \sigma}}
    = \frac{\partial \left[ \frac12 \lambda_{\bR ij \sigma}^* 
              \left(\delta_{\bR ij \sigma} - \lambda_{\bR ij \sigma}\right) 
              \kappa_{\bR ij \sigma} \right]}
            {\partial \lambda_{\bR ij \sigma}}
    = -\frac12 \lambda_{\bR ij \sigma}^* \kappa_{\bR ij \sigma},
\end{equation}
and similarly
\begin{equation}
  \frac{\partial \Delta E}{\partial \lambda_{\bR ij \sigma}^*}
    = \frac{\partial \left[ \frac12 \lambda_{\bR ij \sigma}^* 
              \left(\delta_{\bR ij \sigma} - \lambda_{\bR ij \sigma}\right) 
              \kappa_{\bR ij \sigma} \right]}
            {\partial \lambda_{\bR ij \sigma}^*}
    = -\frac12 \left( \delta_{\bR ij \sigma} - \lambda_{\bR ij \sigma} \right) \kappa_{\bR ij \sigma}.
\end{equation}
Thus
\begin{multline}
  \Delta h^\sigma 
  = \sum_{\bR ij} \Big[
    -\frac12 \lambda_{\bR ij \sigma}^* \kappa_{\bR ij \sigma} 
        \ketbra{w_{\bR j}}{w_{\bZ i}} \\ +
     \frac12 \left(\delta_{\bR ij \sigma} - \lambda_{\bR ij \sigma}\right)\kappa_{\bR ij \sigma}
        \ketbra{w_{\bZ i}}{w_{\bR j}} \Big].  
\end{multline}

To simplify this, we separate diagonal and off-diagonal terms, with
\begin{multline}
  \Delta h^\sigma = \sum_i \frac12 \kappa_{ii\bZ} \Big[ 
    \left(1 - \lambda_{ii\bZ}\right) \ketbra{w_{\bZ i}}{w_{\bZ i}} -
    \lambda_{ii\bZ} \ketbra{w_{\bZ i}}{w_{\bZ i}} \Big]  \\
    - \sum_{\bR j \neq \bZ i} \frac12 \kappa_{\bR ij \sigma} \left[
      \lambda_{\bR ij \sigma} \ketbra{w_{\bZ i}}{w_{\bR j}} + 
      \lambda_{\bR ij \sigma}^* \ketbra{w_{\bR j}}{w_{\bZ i}} \right].
\end{multline}

Since the diagonal elements of $\lambda$ and $\kappa$ are real, the first
summand simplifies to
\begin{multline}
  \frac12 \kappa_{ii\bZ} \left[
      \left(1 - \lambda_{ii\bZ}\right) \ketbra{w_{\bZ i}}{w_{\bZ i}} -
      \lambda_{ii\bZ} \ketbra{w_{\bZ i}}{w_{\bZ i}} \right] \\
    = \kappa_{ii\bZ} \left( \frac12 - \lambda_{ii\bZ} \right) \op{w_{\bZ i}},
\end{multline}
and the second to
\begin{multline}
  \sum_{\bR j \neq \bZ i} \frac12 \kappa_{\bR ij \sigma} \left[
      \lambda_{\bR ij \sigma} \ketbra{w_{\bZ i}}{w_{\bR j}} + 
      \lambda_{\bR ij \sigma}^* \ketbra{w_{\bR j}}{w_{\bZ i}} \right] \\
    = \sum_{\bR j \neq \bZ i} \kappa_{\bR ij \sigma}
      \frac{\lambda_{\bR ij \sigma} + \lambda_{\bR ij \sigma}^*}{2}
      \ketbra{w_{\bZ i}}{w_{\bR j}} \\
    = \sum_{\bR j \neq \bZ i} \kappa_{\bR ij \sigma} \Real{\lambda_{\bR ij \sigma}}
      \ketbra{w_{\bZ i}}{w_{\bR j}}.
\end{multline}
(In the above, we used the fact that
$\left(\ketbra{w_{\bZ i}}{w_{\bR j}}\right)^\dagger = 
\ketbra{w_{\bR j}}{w_{\bZ i}}$, with $(\lambda^\dagger)_{\bR ij \sigma} =
(\lambda^\top)_{\bR ij \sigma}^* = \lambda_{ji(-\bR)}^* = \lambda_{\bR ij \sigma}$.
Observe that the coefficients for each outer product of DLWFs come in complex
conjugate pairs.)

Combining, we have
\begin{equation}
\begin{split}
  \Delta h 
    = \sum_\sigma \Delta h^\sigma
    &= \sum_\sigma \sum_{\bR ij} \kappa_{\bR ij \sigma} 
       \left( 
        \frac12 \delta_{\bR ij \sigma} - \Real{\lambda_{\bR ij \sigma}} 
       \right)
       \ketbra{w_{\bZ i}}{w_{\bR j}} \\
    &= \sum_\sigma \sum_{\bR ij \sigma} \frac12 \kappa_{\bR ij \sigma} 
       \left( \frac12 \delta_{\bR ij \sigma} - \lambda_{\bR ij \sigma} \right)
       \ketbra{w_{\bZ i}}{w_{\bR j}} + h.c.\\
\end{split}
\end{equation}

In finite molecules, the Hamiltonian is
\begin{align}
    \Delta h = \sum_\sigma \Delta h^\sigma = \sum_\sigma\left(\sum_{i} (\frac{1}{2} - \lambda_{ii}^\sigma) \kappa_{ii}^\sigma|\phi_i^\sigma\rangle \langle \phi_i^\sigma| -\sum_{i\neq j} \kappa^\sigma_{ij} |\phi_i^\sigma\rangle \langle \phi_j^\sigma|\right),
 \end{align}
 with $|\phi_i\rangle$ as the localized molecular orbitals. 

\section{Choice of parameters} \label{sec:param}
There are two parameters that can be tuned in olLOSC: the space/energy mixing
$0 \leq \gamma \leq 1$ in the orbitalets, and the fraction $\lambda \geq 0$
of von Weizs\"acker kinetic energy added to the Thomas--Fermi kernel.

\subsection{In materials}
\begin{figure}
\centering
\begin{subfigure}{0.45\textwidth}
    \includegraphics[width=\linewidth]{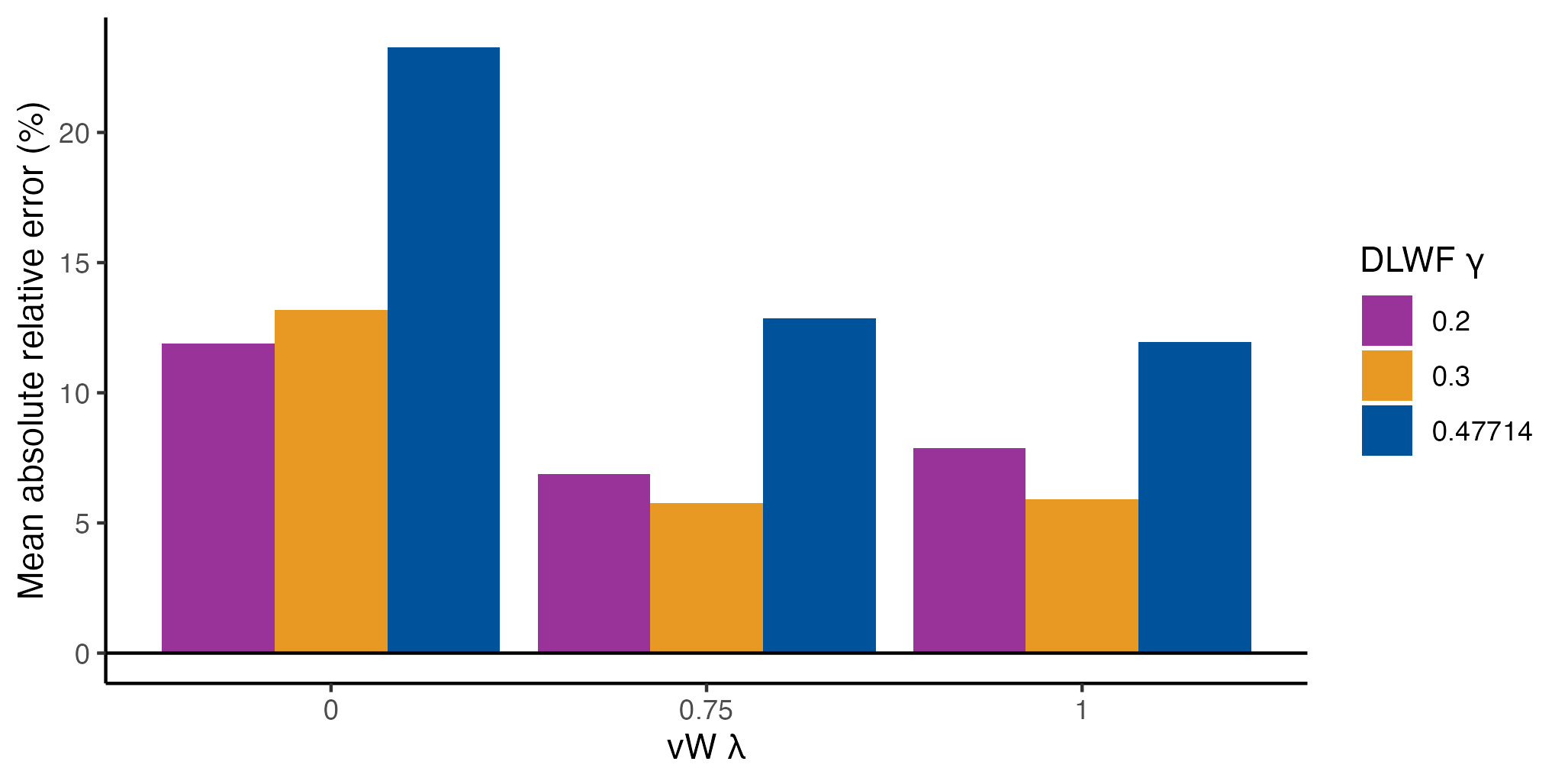}
    \subcaption{Mean absolute error (\%) for small-gap
                ($\leq \SI{8}{\electronvolt}$) systems.}
\end{subfigure} \hfill
\begin{subfigure}{0.45\textwidth}
    \includegraphics[width=\linewidth]{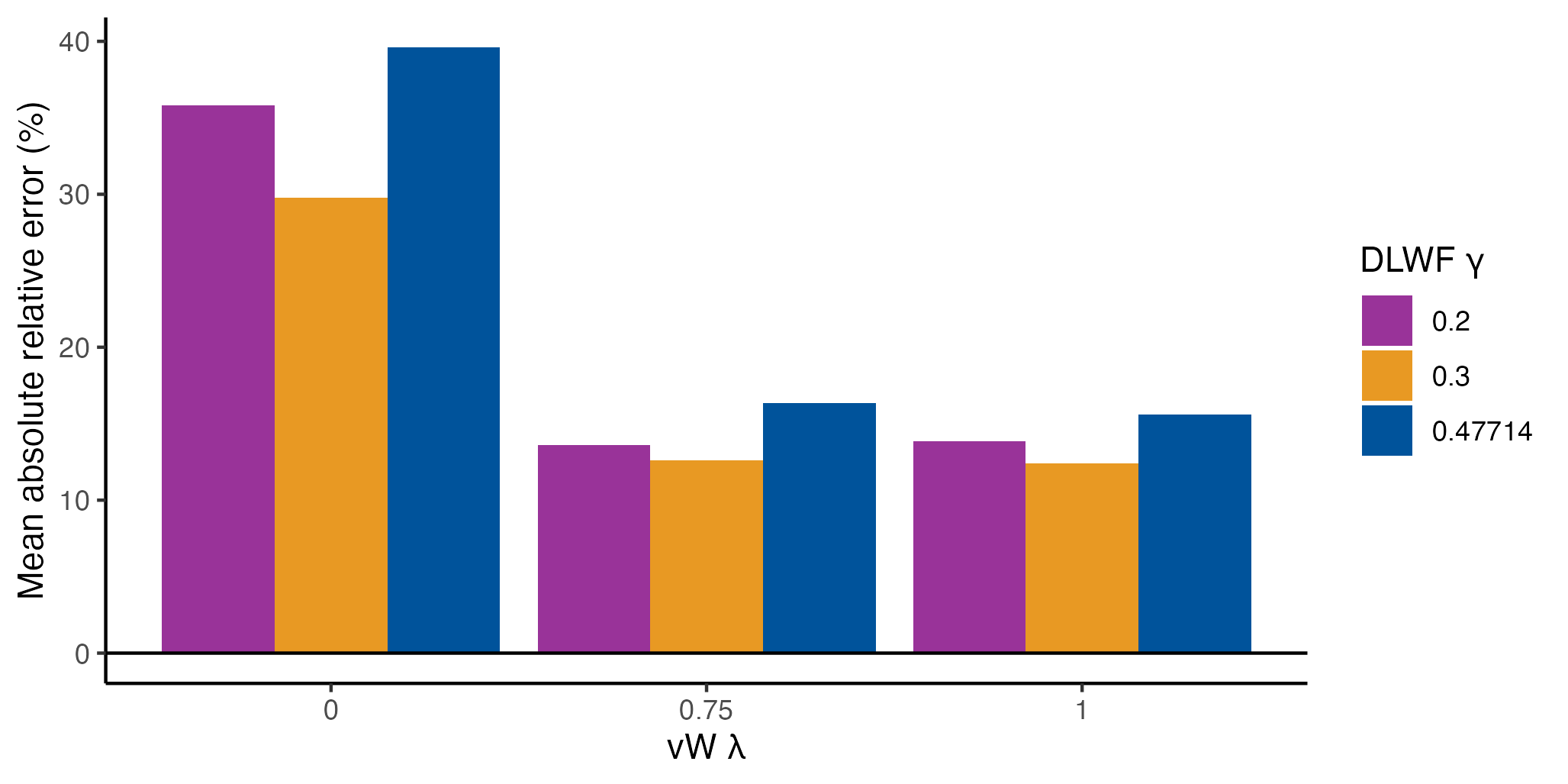}
    \subcaption{Mean absolute error (\%) for all systems.}
\end{subfigure}

\vspace{\baselineskip}

\begin{subfigure}{0.45\textwidth}
    \includegraphics[width=\linewidth]{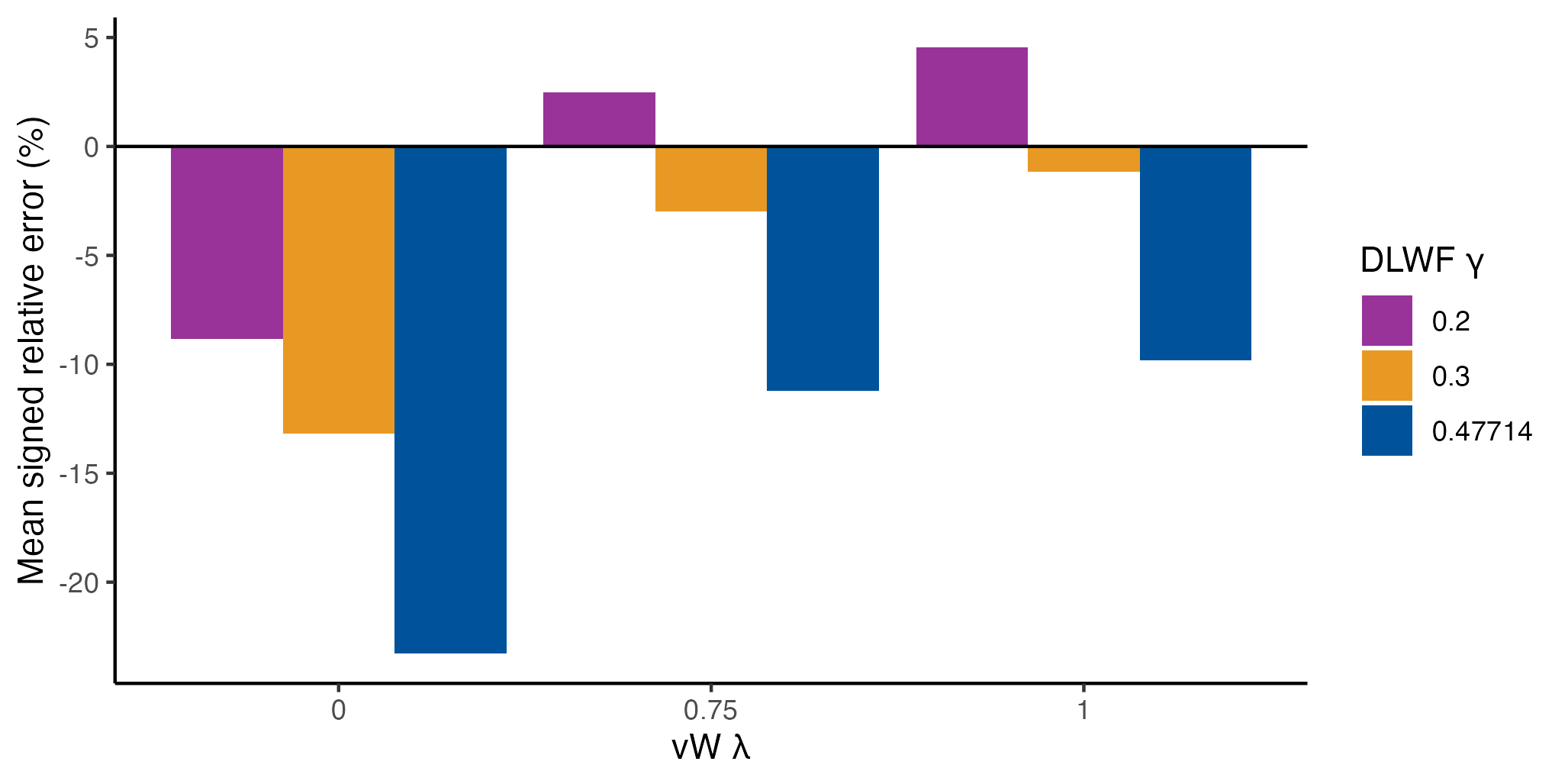}
    \subcaption{Mean signed error (\%) for small-gap
                ($\leq \SI{8}{\electronvolt}$) systems.}
\end{subfigure} \hfill
\begin{subfigure}{0.45\textwidth}
    \includegraphics[width=\linewidth]{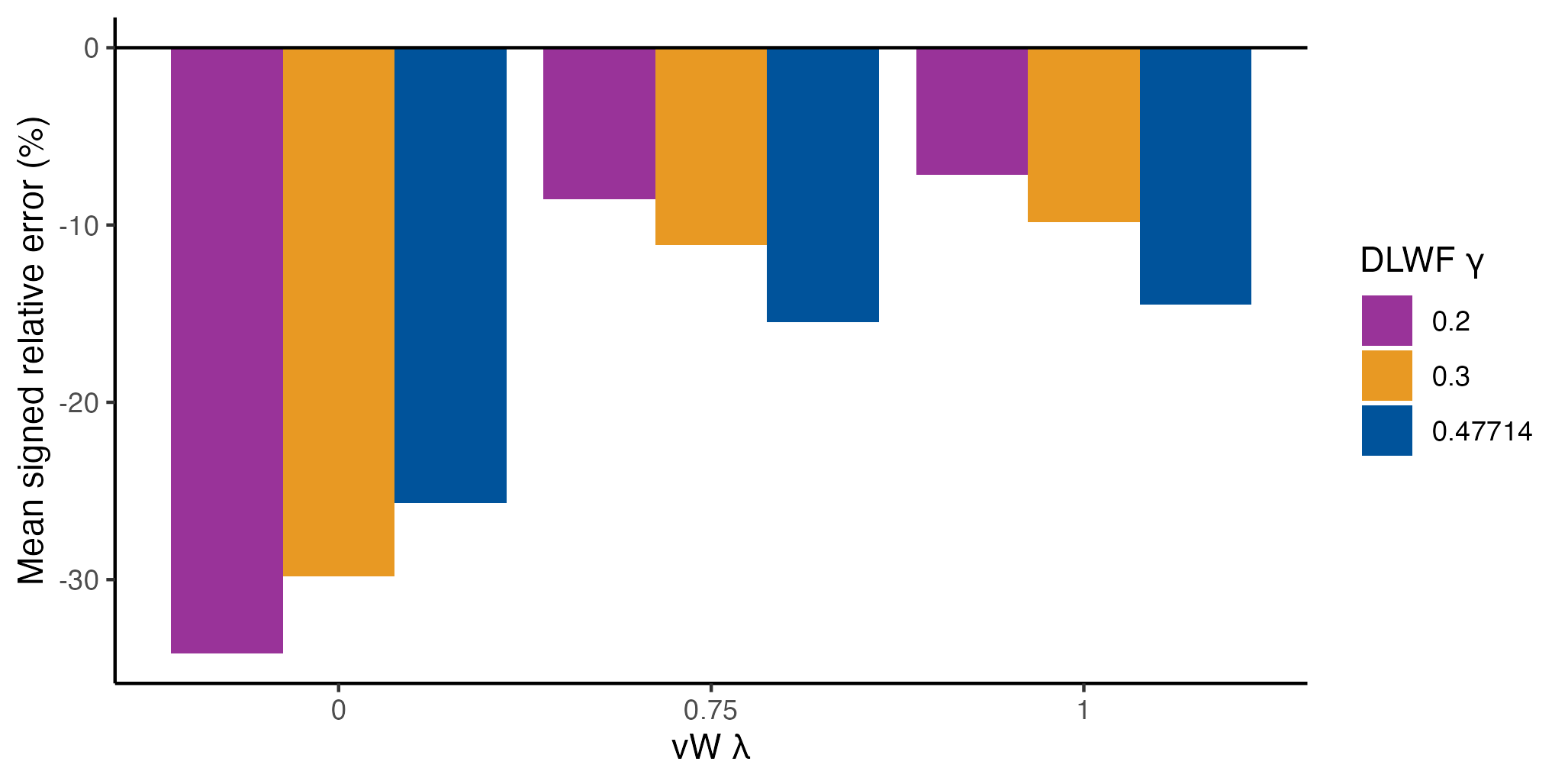}
    \subcaption{Mean signed error (\%) for all systems.}
\end{subfigure}

\vspace{\baselineskip}

\begin{subfigure}{0.45\textwidth}
    \includegraphics[width=\linewidth]{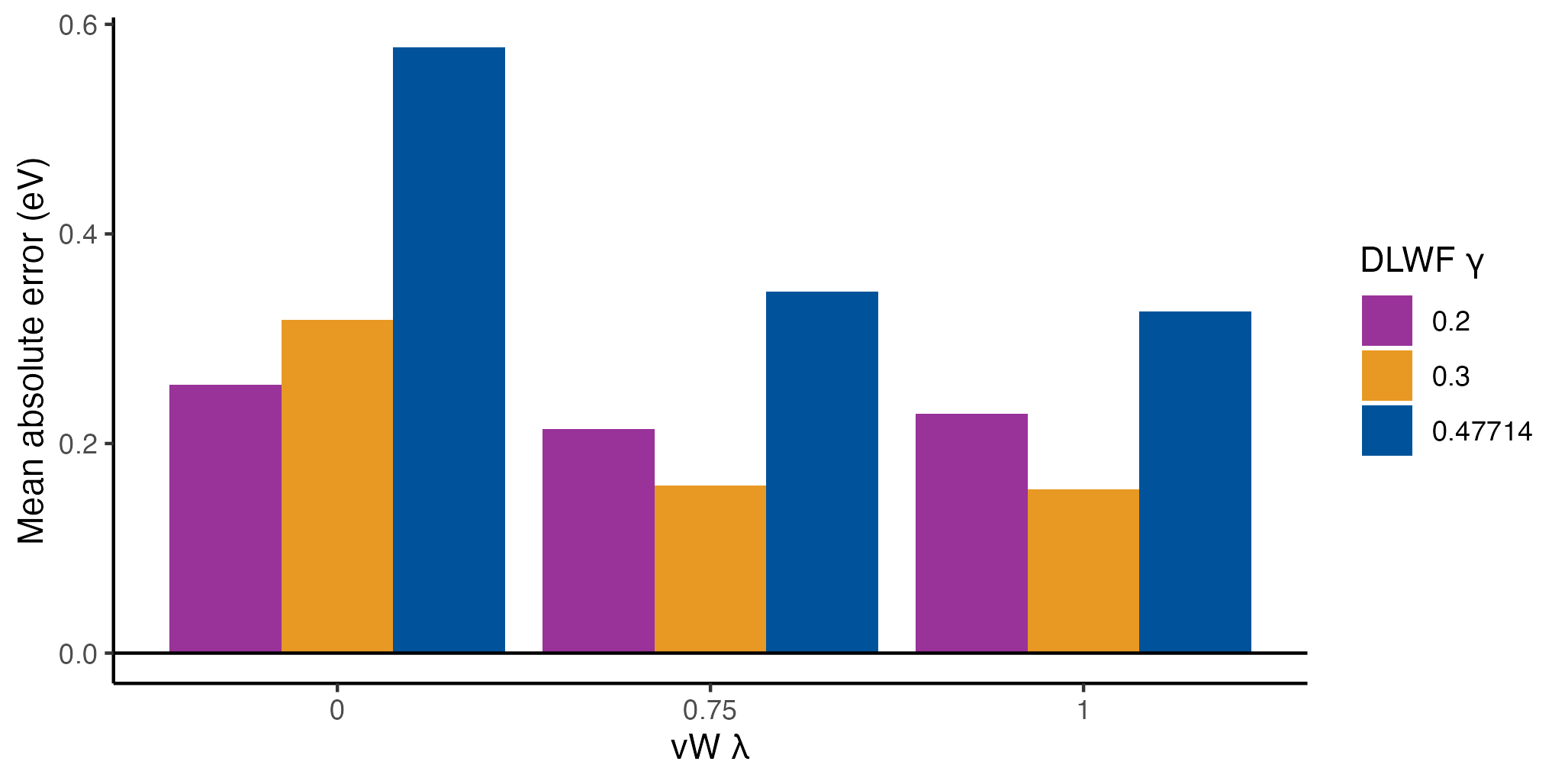}
    \subcaption{Mean absolute error (\si{\electronvolt}) for small-gap
                ($\leq \SI{8}{\electronvolt}$) systems.}
\end{subfigure} \hfill
\begin{subfigure}{0.45\textwidth}
    \includegraphics[width=\linewidth]{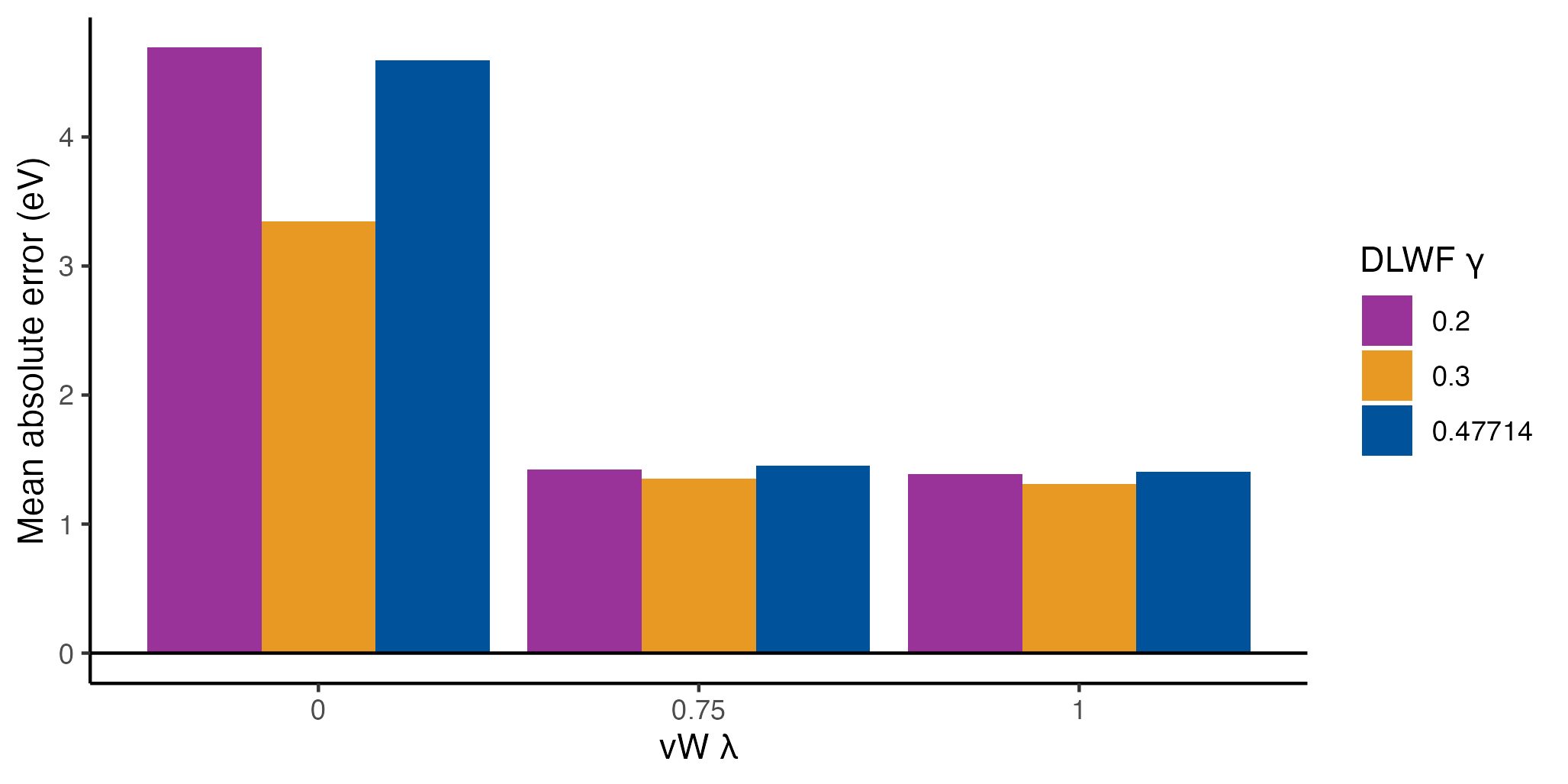}
    \subcaption{Mean absolute error (\si{\electronvolt}) for all systems.}
\end{subfigure}

\vspace{\baselineskip}

\begin{subfigure}{0.45\textwidth}
    \includegraphics[width=\linewidth]{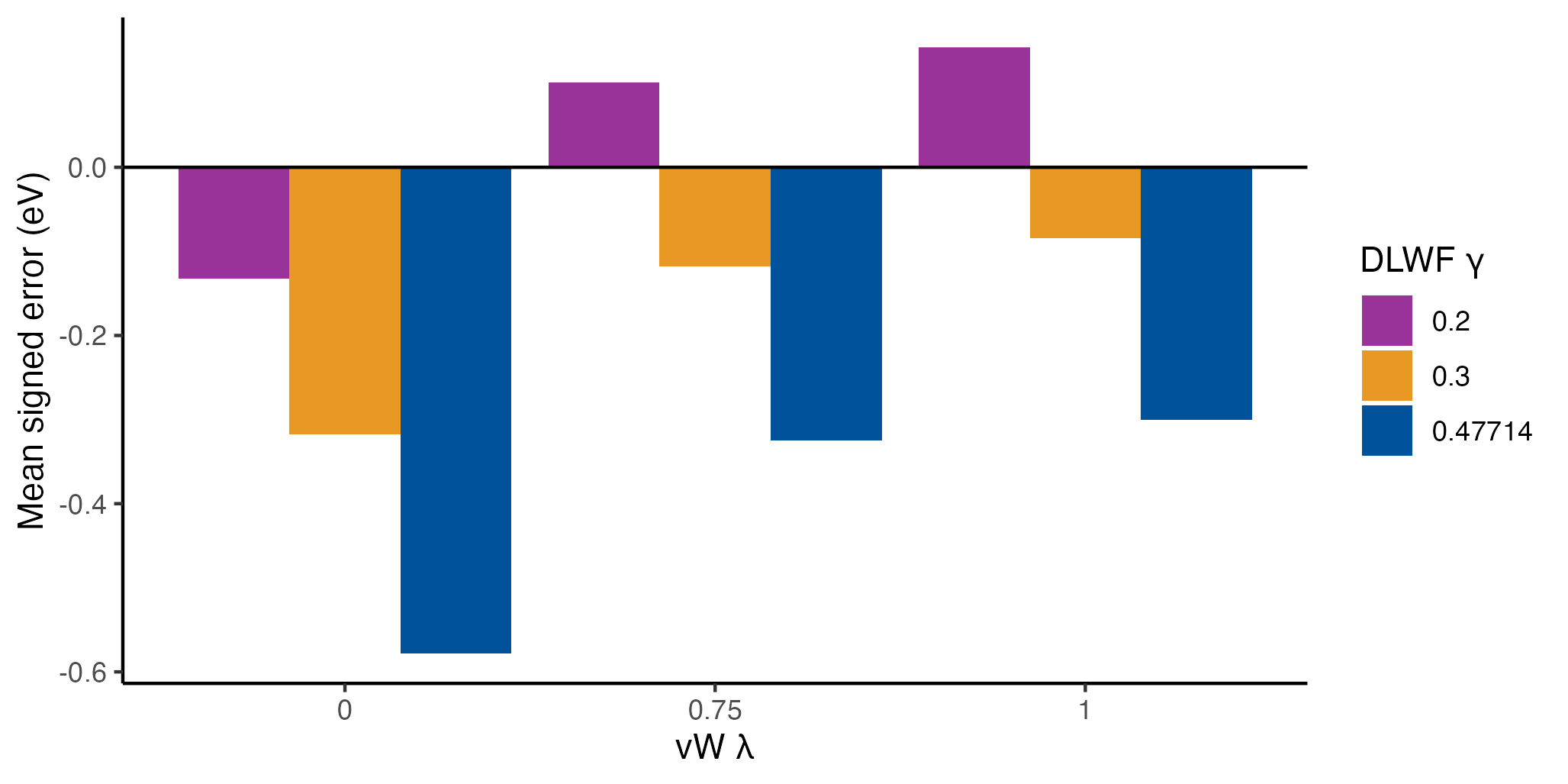}
    \subcaption{Mean signed error (\si{\electronvolt}) for small-gap
                ($\leq \SI{8}{\electronvolt}$) systems.}
\end{subfigure} \hfill
\begin{subfigure}{0.45\textwidth}
    \includegraphics[width=\linewidth]{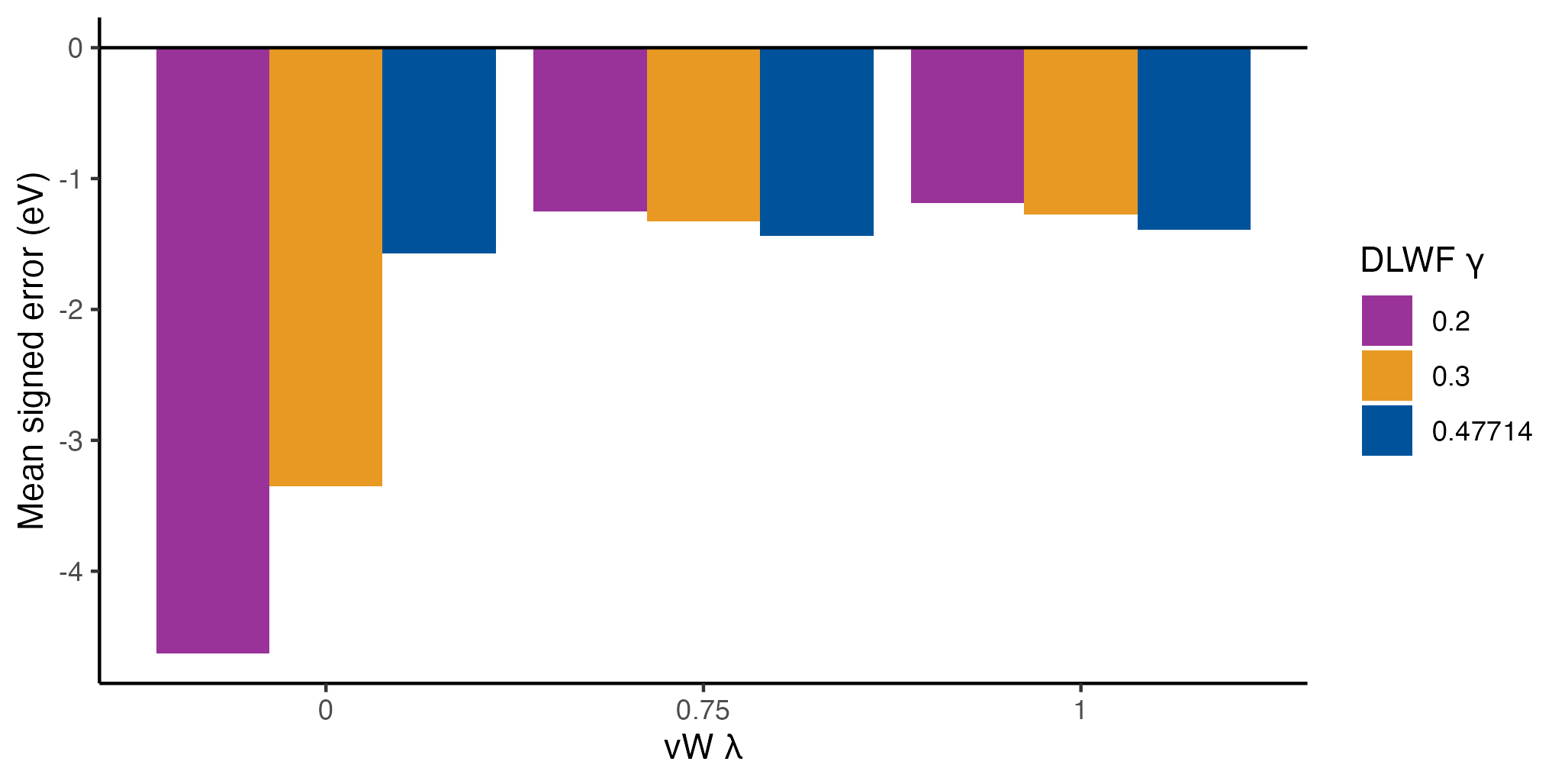}
    \subcaption{Mean signed error (\si{\electronvolt}) for all systems.}
\end{subfigure}

\caption{Errors by DLWF mixing $\gamma$ and the vW $\lambda$ in the materials
         tested. (a--d): Relative errors (\%); (e--h): unscaled errors
         (\si{\electronvolt}). (a, c, e, g): Small-gapped systems (gaps
         $\leq \SI{8}{\electronvolt}$), for which olLOSC is most accurate;
         (b, d, f, h): All systems. (a, b, e, f): Absolute values;
         (c, d, g, h): Signed errors.}
\label{fig:gl_bulk}
\end{figure}

The results of our parameter sweep are shown in Fig.~\ref{fig:gl_bulk}. Since
olLOSC is most accurate for systems with smaller gaps
($\leq \SI{8}{\electronvolt}$), we chose parameters based on their performance
in the left column of the figure. It was clear that some nonzero fraction of
von Weizs\"acker kinetic energy is required: for all values of the DLWF
localization parameter $\gamma$, there was a large negative mean signed error
(c, g), indicating systematic underestimation of the band gap. 

If $\gamma = 0.47714$, the value used in previous works on LOSC
\cite{li2018, su2020, mei2020a, mahler2022b}, then 100\% von Weizs\"acker
correction is preferred ($\lambda = 1$). However, $\gamma = 0.30$ yields
improved performance in some metrics: $\gamma = 0.30, \lambda = 1$ has almost
zero mean signed percent error (c), while $\gamma = 0.30, \lambda = 0.75$ has
minimal mean signed error (eV; g). 

We chose $\gamma = 0.30$ for this work to reduce the appearance of local minima
in the DLWF minimization (see below), and $\lambda = 0.75$ to minimize the
systematic error.

\subsubsection{Convergence in Brillouin zone sampling}
We investigate the convergence with respect to the $\bk$-mesh in
Fig.~\ref{fig:k_bulk}. (Recall that olLOSC requires a uniform $\bk$-sampling
centered at $\Gamma$ for its monochromatic decomposition.) In general, the
change in olLOSC gap is fairly small as the $\bk$-sampling is increased from
$6 \times 6 \times 6$ to $8 \times 8 \times 8$. Noticeable differences appear
in Si, C, BP, and SiC because their conduction band minimum is not at a
high-symmetry $\bk$-point; thus, the olLOSC gap change is matched by a similar
one in the PBE calculation. 

Large-gapped systems tend to have a slight reduction in the band gap as the
$\bk$-sampling is increased separate from the DFA. This behavior generally
coincides with differences in the `degeneracy pattern' of the DLWFs, as noted
in \cite{mahler2022}; their convergence is often more difficult in larger-gapped
systems. As olLOSC is most applicable to materials with smaller gaps, we did
not investigate further. The large deviation in AlP when $\gamma = 0.47714$ is
also due to a qualitative difference in the DLWFs (see Section \ref{sec:dlwf}
below); nevertheless, it is partially to avoid this possibility that we choose
$\gamma = 0.30$ in the main text.

\subsection{In molecules}
The results of our parameter sweep in molecular systems are shown in
Fig.~\ref{fig:gl_rel_mol}  and Fig.~\ref{fig:gl_unscale_mol}; $\lambda$ is one
of 50\%, 75\%, and 100\%, and $\gamma \in \{0.2, 0.3, 0.47714\}$.
Size-consistency is of particular importance in molecular calculations.
Small molecules are generally spatially localized but delocalized in energy,
have the opposite problem. We therefore partitioned the molecular test based on
system size. The \emph{small-molecule} subset contains the molecules with no
more than five atoms; the \emph{large-molecule} subset contains the rest. 

From the standpoint of mean signed error, the molecular datasets allow greater 
flexibility with respect to $\lambda$. For all three choices of  $\lambda$, the
fundamental gaps predicted by olLOSC with tf+$\lambda$vW screening exhibit a
mean absolute error of roughly \qty{0.4}{\electronvolt}. Calculations on
materials favor larger $\lambda$ values, as can be seen in Tables
\ref{tab:molg3l75} $(\gamma = 0.3, \lambda = 0.75)$ and \ref{tab:smolg48l1}
$(\gamma = 0.47714, \lambda = 1.0)$. Data for the rest of the parameter space
can be found in the SI Dataset \cite{fan2026}.

\subsection{olLOSC beyond the random phase approximation} \label{sub:beyond-rpa}
As we mentioned earlier, the RPA is an approximation; at first glance, it
should appear that we should avoid it if possible. 
In the subsequent discourse, a comparative analysis will be conducted between the beyond RPA olLOSC scheme and the RPA-based olLOSC scheme. Owing to certain numerical instability, wherein the preconditioner employed in bulk calculations may lose its positive-definiteness when the potential term includes exchange and correlation effects, the beyond RPA olLOSC protocol demonstrates robustness exclusively in molecular systems. In Fig. \ref{fig:gl_ofxc_rel_mol} and Fig. \ref{fig:gl_ofxc_unscale_mol}, parameter scanning is performed analogously to the conventional RPA-based olLOSC. For two parameter sets with particular interest, \{$\gamma = 0.3, \lambda = 0.75$\} and \{$\gamma = 0.47714, \lambda = 1$\}, we also provide their numerical results in Tab.~\ref{tab:rpa}. The beyond-RPA olLOSC routine yields superior results for small-sized molecules, yet underperforms in large-sized molecules, with comparable outcomes observed for both sets of parameters. Based on this comparison and on the basis of self-consistence and robustness, it is preferred to adapt the RPA-based olLOSC as a conventional olLOSC routine.

\subsection{Total energy}
\subsubsection{Materials}
The materials we tested are in their equilibrium geometry, so we expect only a 
small correction from olLOSC to the total energy; and that is what we see.
For the production dataset seen in the main text with
$(\gamma = 0.30, \lambda = 0.75)$ (Fig.~1; Table \ref{tab:bulk}), the
maximum energy correction is $\SI{3.60e-3}{\rydberg}$, for silicon; it also has
the largest relative energy change, $-0.0213\%$. 

\subsubsection{Molecules}
Total energies in the underlying DFA calculations for molecules are not
size-consistent \cite{li2018}, so a large $\Delta E_{\LOSC}$ is expected for
large molecules. Tables \ref{tab:molg3l75} and \ref{tab:molg48l1} include
$E_{\DFA}$ and $\Delta E_{\LOSC}$ for the large molecules we tested. We observed
olLOSC energy corrections up to a few \si{\milli\hartree}. The largest is for
TCNQ (tetracyanoquinodimethane), which is a highly conjugated molecule (thus,
significant delocalization error is expected). It has
$\Delta E_{\LOSC} = \SI{16.349}{\milli\hartree}$ when
$\gamma = 0.30$ and $\lambda = 0.75$. Note that increased spatial localization
(a smaller $\gamma$) yields a larger $\Delta E$ because non-integer local
occupations $\lambda_{\bR ij \sigma}$ are penalized less. Thus, when
$\gamma = 0.47714$ and $\lambda = 1$, $\Delta E_{\LOSC}$ for TCNQ is
$\SI{11.383}{\milli\hartree}$.
The reaction barrier height is heavily based on total energy calculations. Tab. \ref{tab:rb} includes mean signed error (MSE) and mean absolute error (MAE) for reaction barrier heights using the HTBH38 and NHTBH38 datasets\cite{ZhengRB2009, Peverati2014}, where reference reaction barrier heights are considered to be the 'REF1' (exclude relativistic effect) within these datasets. A $\gamma$ value of 0.47714 generally replicates the results of PBE reaction barrier heights, whereas a $\gamma$ value of 0.3 yields the lowest mean absolute error.

\section{Local minima in dually localized Wannier functions} \label{sec:dlwf}
DLWFs \cite{mahler2022} are computed variationally, minimizing for each spin
component the cost function
\begin{equation}
\begin{split}
    F^\sigma[\gamma] &=
    (1 - \gamma) \sum_i 
        \left[
            \ev{\br}{w_{\bZ i \sigma}}^2 - \ev{r^2}{w_{\bZ i \sigma}}
        \right] \\ &\qquad +
    C \gamma \sum_i
        \left[ 
            \ev{h}{w_{\bZ i \sigma}}^2 - \ev{h^2}{w_{\bZ i \sigma}}
        \right] \\ &=
    (1 - \gamma) \sum_i \Delta r_{\bZ i \sigma}^2 + 
        C \gamma \sum_i \Delta h_{\bZ i \sigma}^2
\end{split}
\end{equation}
where $0 \leq \gamma \leq 1$ and $C = 1$. (Note that
$w_{\bR i \sigma}(\br) = w_{\bZ i \sigma}(\br - \bR)$, so we do not need to sum
over the unit cell index $\bR$; that would only multiply $F^\sigma[\gamma]$ by
a constant, which does not affect the optimization.)

We minimize $F^\sigma$ by modifying the method used for maximally localized
Wannier functions (MLWFs) \cite{marzari1997}. The difference is in the cost
function, which includes the Hamiltonian, as first suggested by 
\cite{gygi2003}. The key difference is that DLWFs are constructed from
\emph{both} valence and conduction bands. We first disentangle the conduction'
manifold \cite{souza2001}, obtaining optimally smooth pseudo-bands that span the
virtual subspace. (If the  disentangled bands are analytic, the MLWFs they
specify decay exponentially \cite{monaco2018}.) Once the bands are chosen, we
formulate $F^\sigma[\gamma]$ as a function of the elements of the localization
unitaries $U^{\bk \sigma}$, which localize the disentangled bands:
\begin{equation}
    \phi_{\bk i \sigma}(\br) = 
    \sum_n U^{\bk\sigma}_{in} \psi_{\bk n \sigma}(\br).
\end{equation}
Then we minimize $F^\sigma$ with the conjugate gradient method.

It was noted as early as \cite{marzari1997} that, depending on the quality of
the initial guess, the MLWF localization scheme can become trapped in local
minima of the cost function. More than two decades of work has rendered MLWFs
extremely robust to local minima, to the point that there are automated and
high-throughput workflows for Wannierization \cite{vitale2020}. This scheme
scheme relies on the selected columns of the density matrix (SCDM)
algorithm \cite{damle2015, damle2017a} for its initial guess. SCDM produces
orthonormal, localized Wannier-like functions \emph{deterministically} by
exploiting the exponential decay of the one-electron reduced density matrix
\cite{prodan2005},
\begin{equation}
    \gamma_\sigma(\br, \br') \sim e^{-\abs{\br - \br'}}.
\end{equation}

We, too, use SCDM to generate initial guesses for DLWFs, but our minimization is
still sensitive to local minima. This is due largely to the disentangled
conduction bands, which have large energy variance $\Delta h_{\bZ i \sigma}^2$.
We avoid this as best we can by minimizing $F^\sigma$ several times, over a
variety of conjugate gradient step sizes, and choosing the smallest cost
function value overall. 

Mahler et al. also observed that the qualitative nature of the DLWFs changes as
a function of $\gamma$ (see Fig.~1 of \cite{mahler2022}). At $\gamma = 0$ (the
MLWF limit), all the WFs are degenerate; they are also typically equivalent up
to symmetry operations in the primitive cell. At small positive $\gamma$, the
valence and conduction DLWFs split off from one another. They have different
shapes and different average energies $\evo{h}_{\bZ i \sigma}$. Splitting of
degenerate subsets continues as $\gamma$ is increased further.

\begin{figure}
\centering
\begin{subfigure}{0.45\textwidth}
    \includegraphics[width=\linewidth]{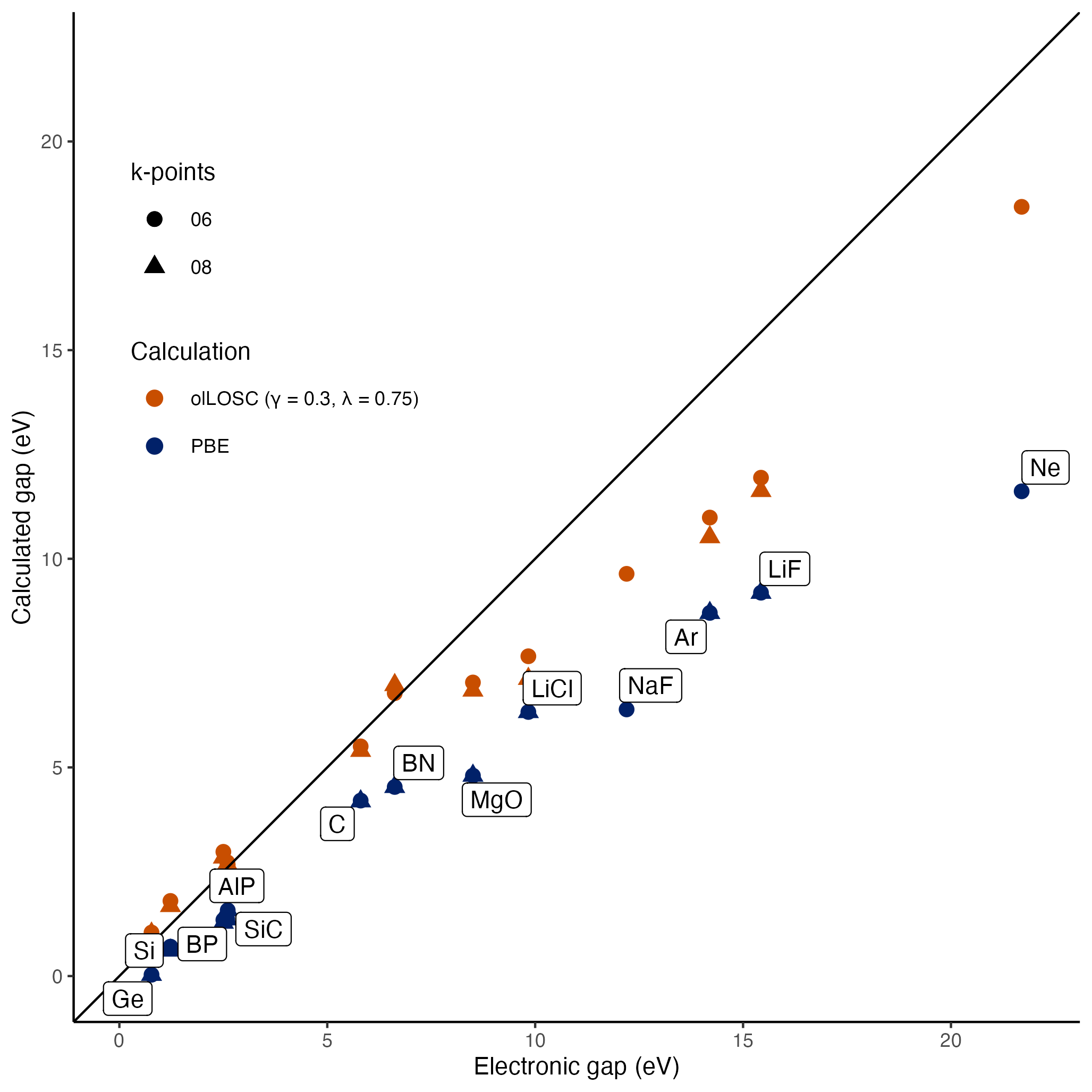}
    \subcaption{$\gamma = 0.3, \lambda = 0.75$ (all systems).}
    \label{fig:k_bulk_g030l075}
\end{subfigure} \hfill
\begin{subfigure}{0.45\textwidth}
    \includegraphics[width=\linewidth]{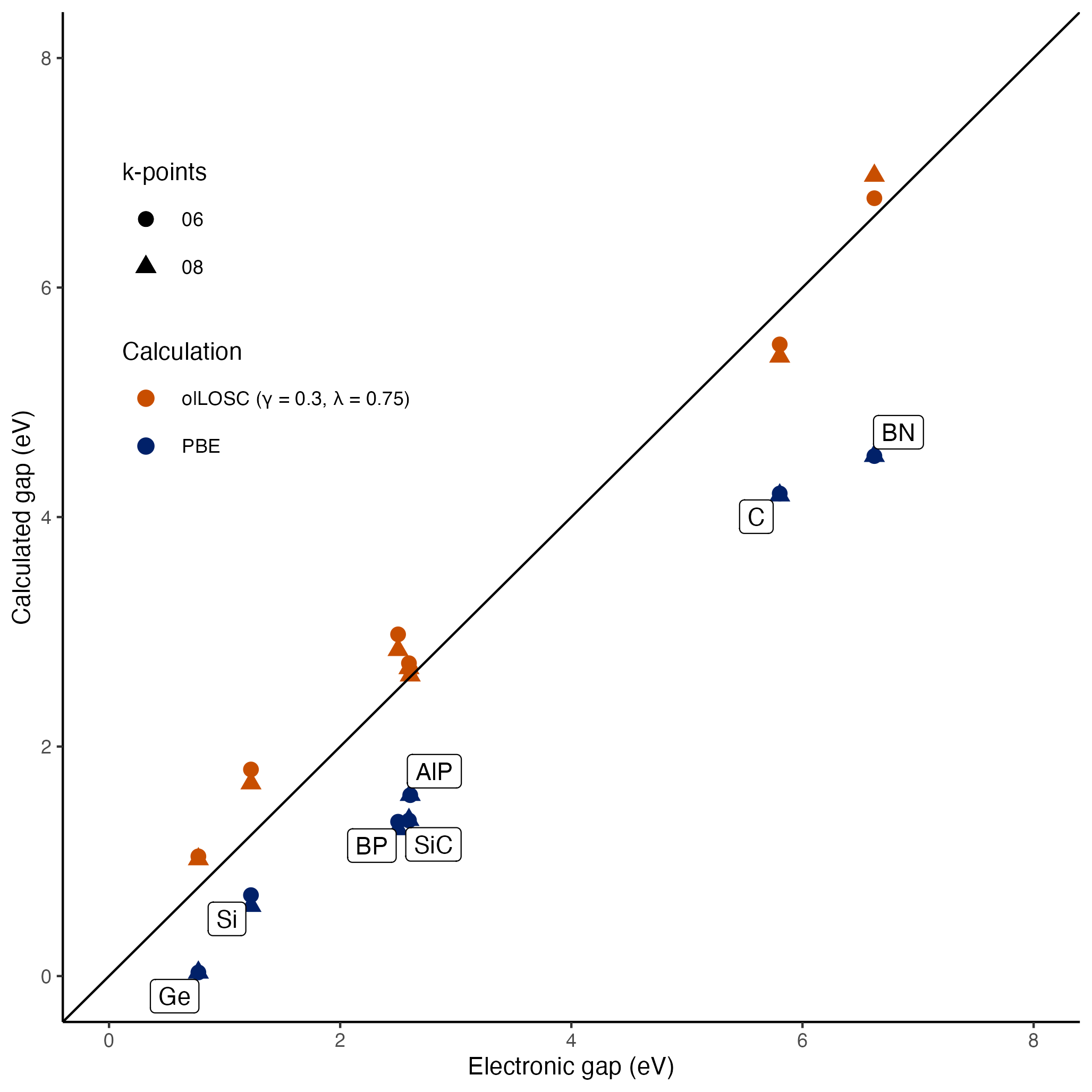}
    \subcaption{$\gamma = 0.3, \lambda = 0.75$ (gap
                $\leq \SI{8}{\electronvolt}$)).}
    \label{fig:k_bulk_g030l075_small}
\end{subfigure}
\par\bigskip
\begin{subfigure}{0.45\textwidth}
    \includegraphics[width=\linewidth]{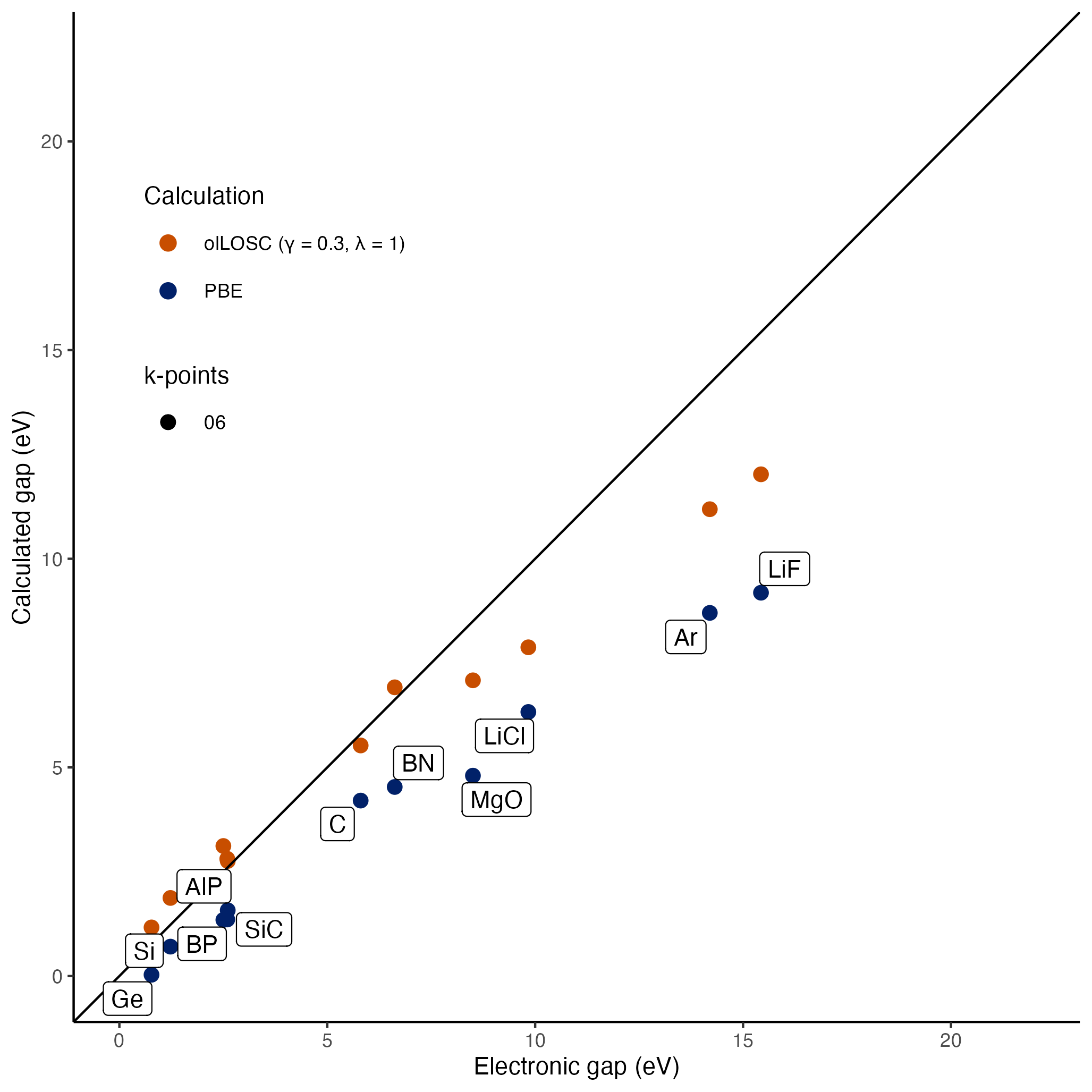}
    \subcaption{$\gamma = 0.30, \lambda = 1$ (all systems).}
    \label{fig:k_bulk_g030l100}
\end{subfigure} \hfill
\begin{subfigure}{0.45\textwidth}
    \includegraphics[width=\linewidth]{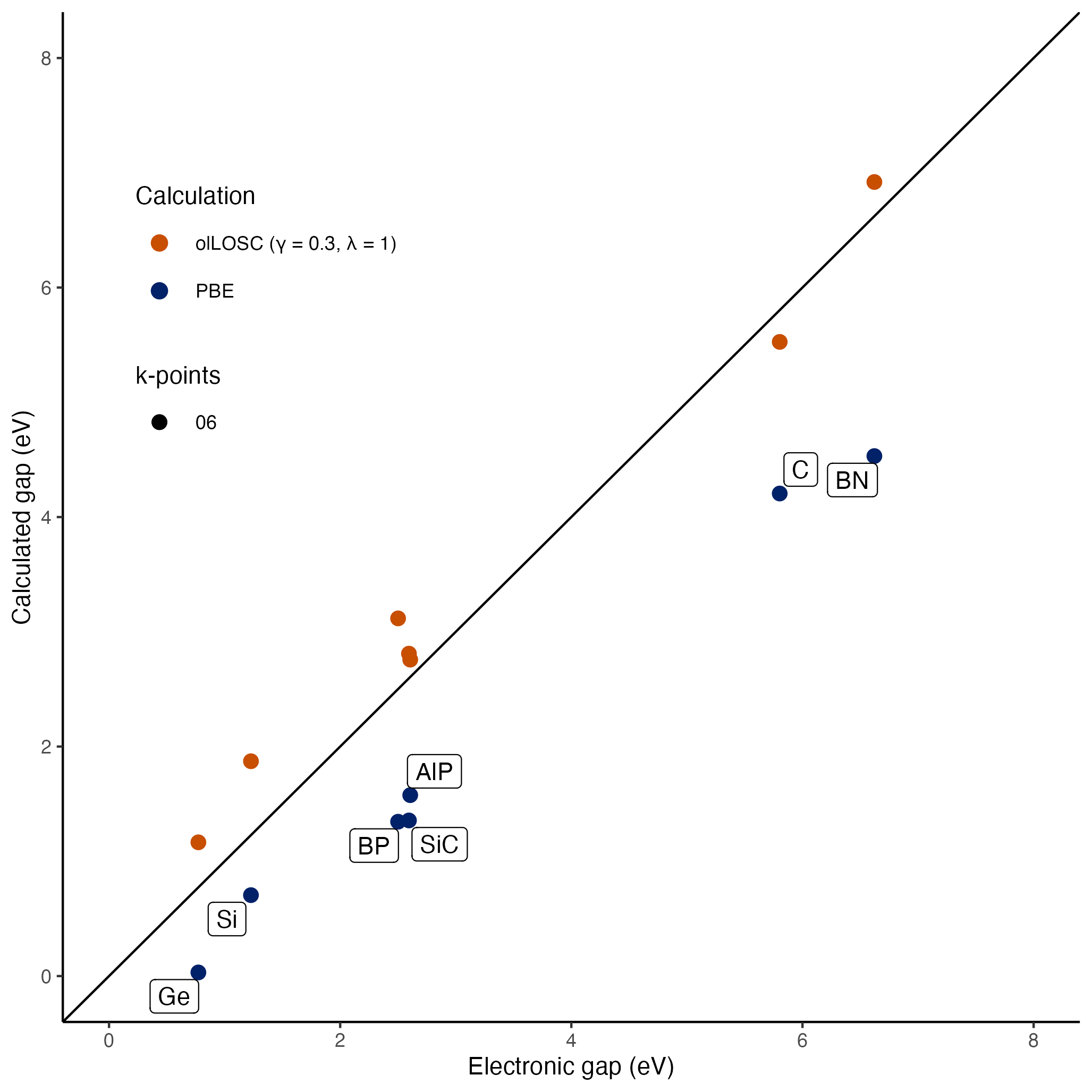}
        \subcaption{$\gamma = 0.30, \lambda = 1$ (gap
                $\leq \SI{8}{\electronvolt}$)).}
        \label{fig:k_bulk_g030l100_small}
\end{subfigure}
\caption{olLOSC convergence with respect to $\bk$-points for materials at the
         two best parameter choices. Left column: All systems. Right column:
         Small-gapped systems ($\leq \SI{8}{\electronvolt}$). Note that NaF and
         Ne were only computed with a $6 \times 6 \times 6$ $\bk$-mesh.}
\label{fig:k_bulk}
\end{figure}

\begin{landscape}
\begin{figure}
\centering
\includegraphics[width=0.95 \linewidth]{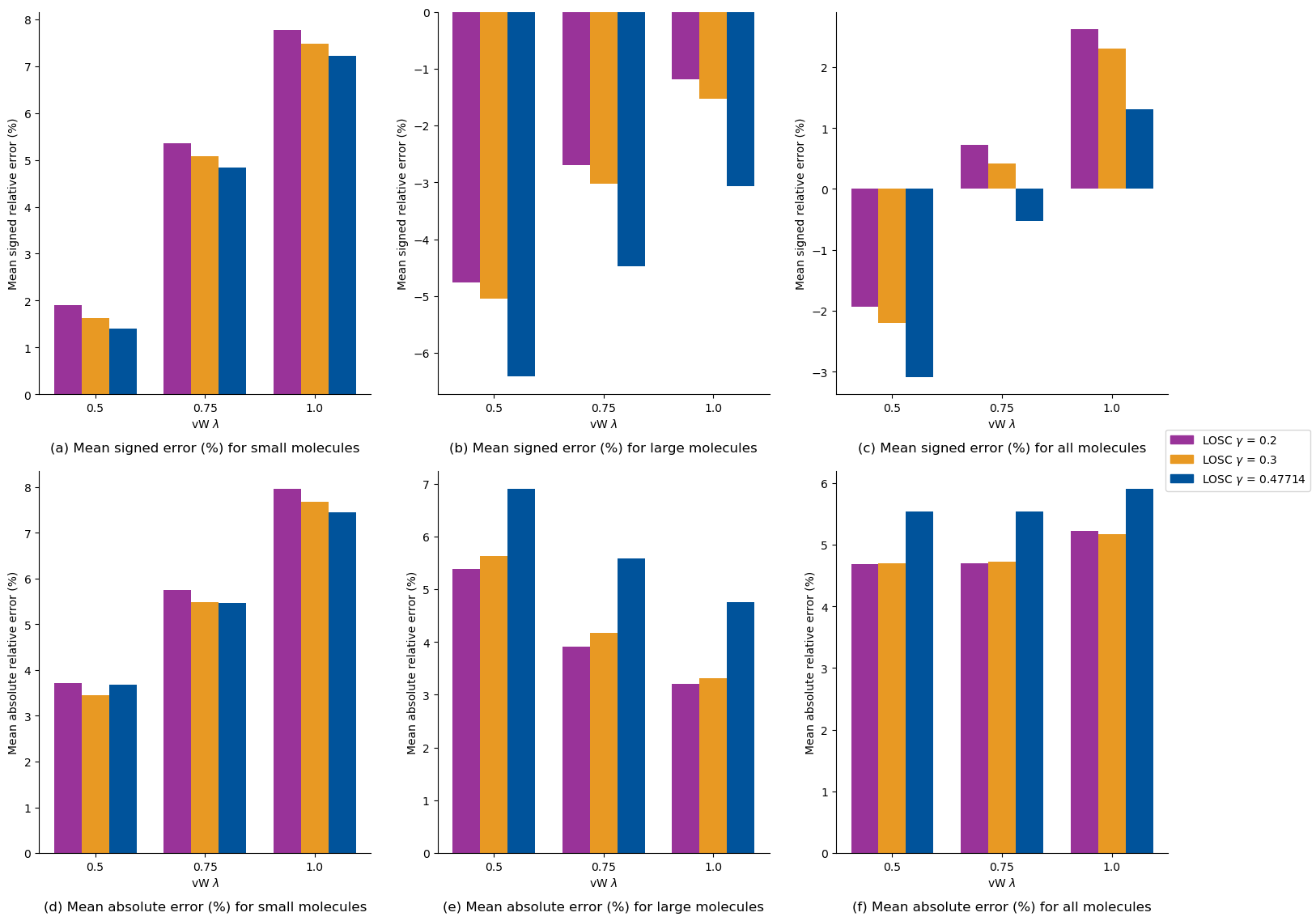}
\caption{Relative errors (\%) by LOSC $\gamma$ and the vW $\lambda$ in the molecules
         tested. (a, d): Small molecules. 
         (b, e): Large molecules.
         (c, f): All molecular systems.}
\label{fig:gl_rel_mol}
\end{figure}
\end{landscape}

\begin{landscape}
\begin{figure}
\centering
\includegraphics[width=0.95 \linewidth]{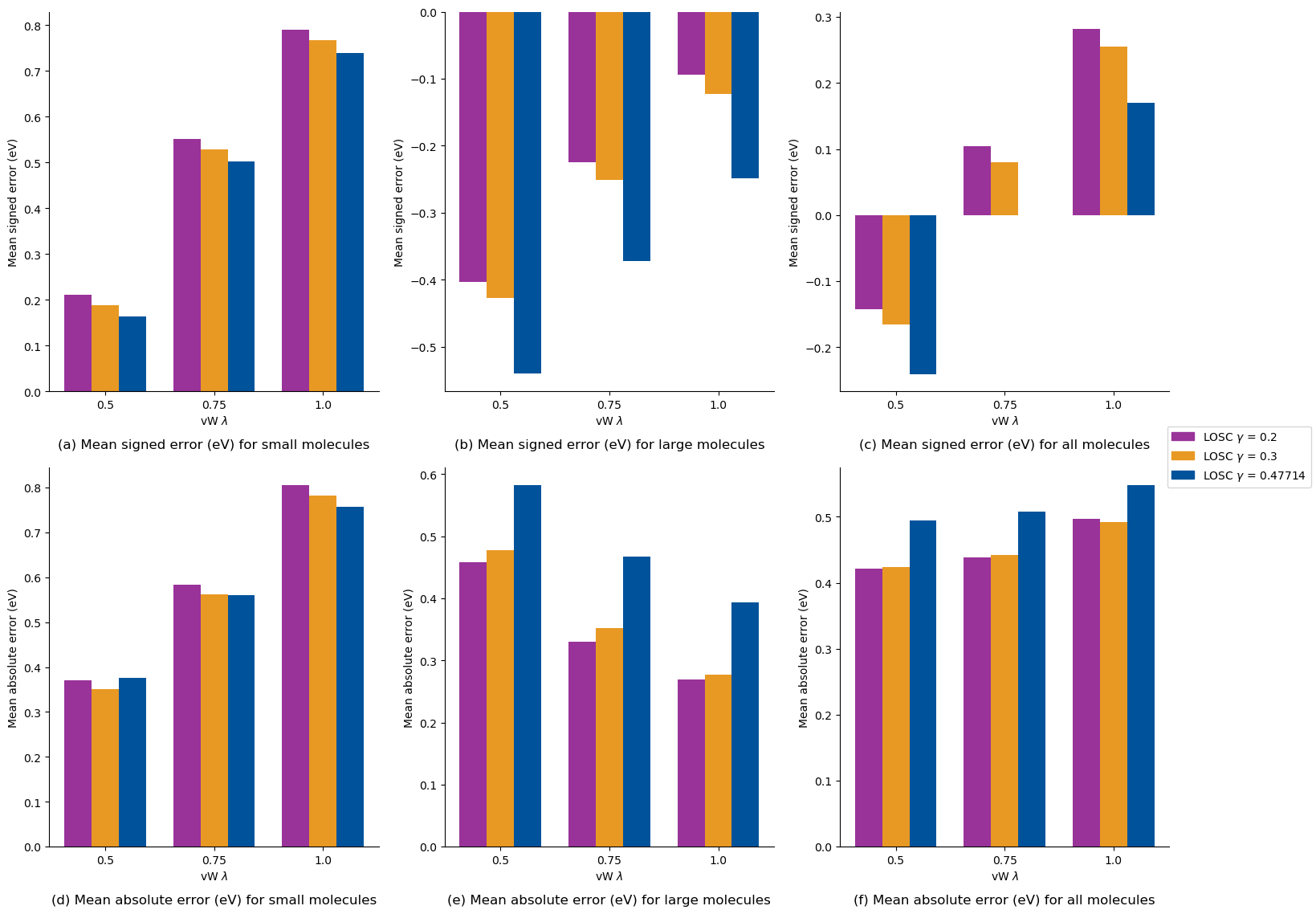}
\caption{Unscaled errors (\si{\electronvolt}) by LOSC $\gamma$ and the vW $\lambda$ 
        in the molecules tested. (a, d): Small molecules. 
         (b, e): Large molecules.
         (c, f): All molecular systems.}
\label{fig:gl_unscale_mol}
\end{figure}
\end{landscape}

\begin{landscape}
\begin{figure}
\centering
\includegraphics[width=0.95 \linewidth]{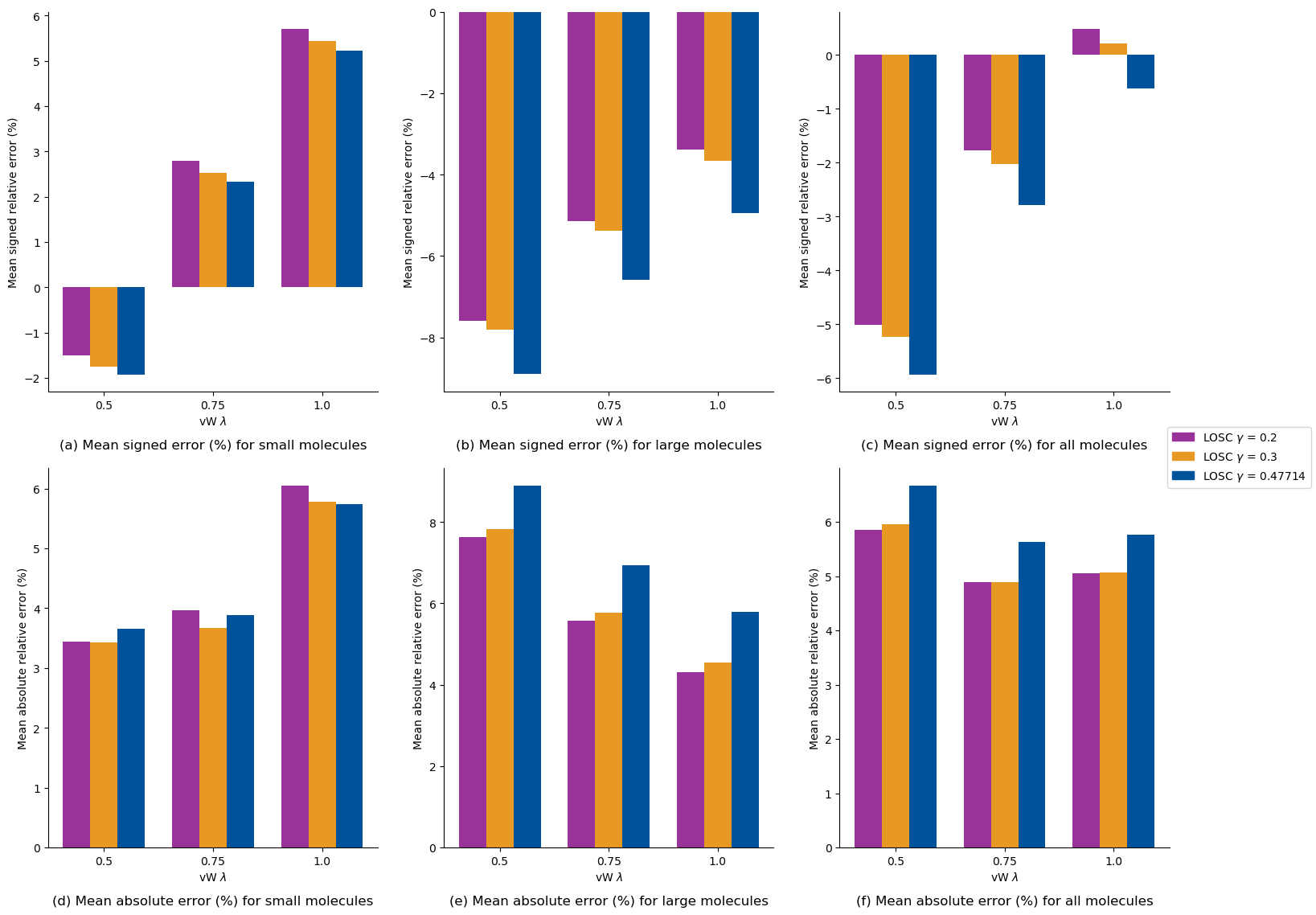}
\caption{of-LOSC beyond RPA Relative errors (\%) by LOSC $\gamma$ and the vW $\lambda$ in the molecules
         tested. (a, d): Small molecules. 
         (b, e): Large molecules.
         (c, f): All molecular systems.}
\label{fig:gl_ofxc_rel_mol}
\end{figure}
\end{landscape}

\begin{landscape}
\begin{figure}
\centering
\includegraphics[width=0.95 \linewidth]{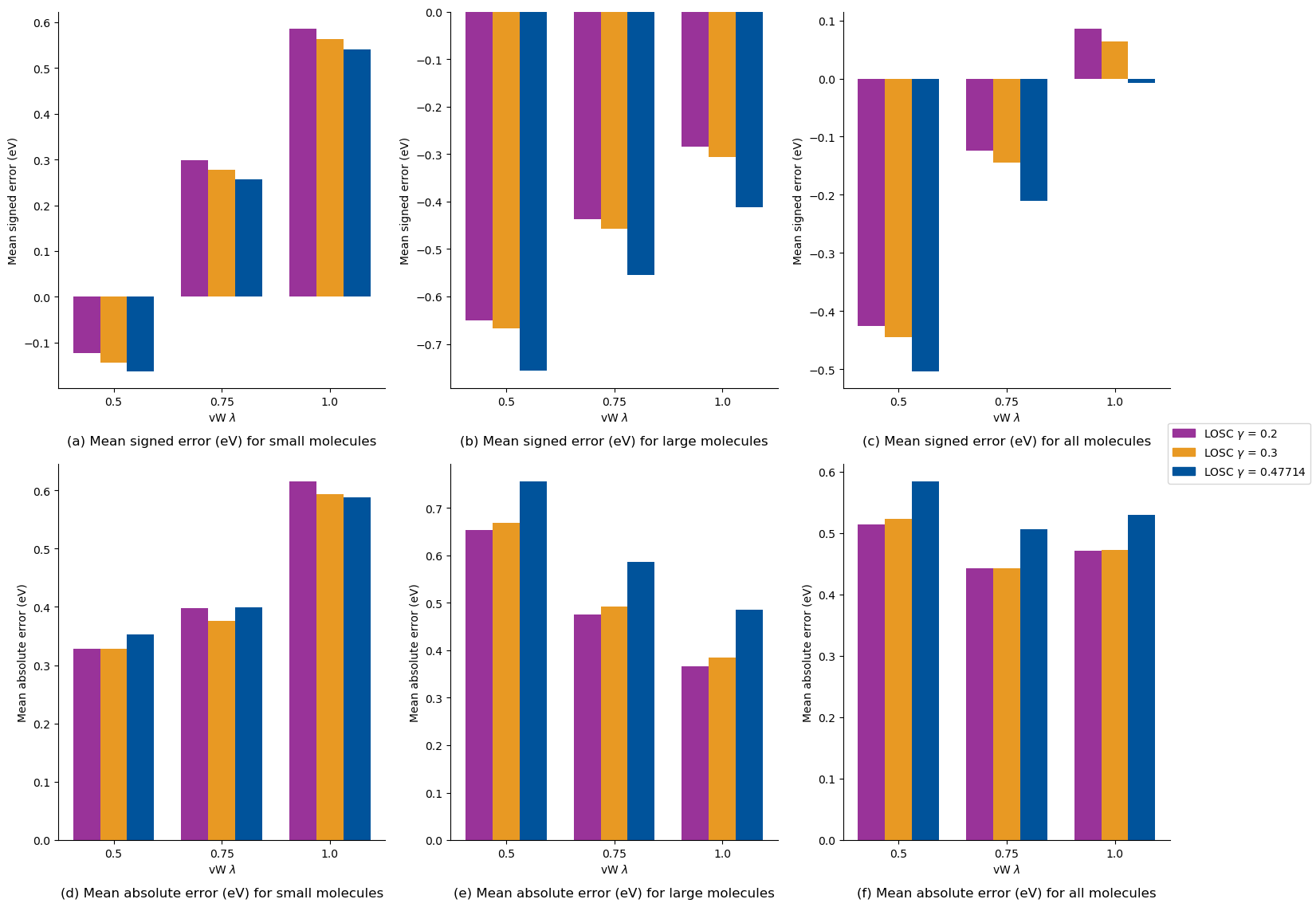}
\caption{of-LOSC beyond RPA unscaled errors (\si{\electronvolt}) by LOSC $\gamma$ and the vW $\lambda$ 
        in the molecules tested. (a, d): Small molecules. 
         (b, e): Large molecules.
         (c, f): All molecular systems.}

\label{fig:gl_ofxc_unscale_mol}
\end{figure}
\end{landscape}

We see similar patterns in the DLWFs for these systems, but the degeneracy
pattern varies somewhat with the $\bk$-sampling---especially for our largest
$\gamma$, $0.47714$. As a particularly noticeable example, observe the large
difference (\SI{0.442}{\electronvolt}) in the olLOSC gap
($\gamma = 0.47714, \lambda = 1$) of BP between $\bk = 6$ and $\bk = 8$
(Fig. \ref{fig:k_bulk} \textbf{(d)} and Table \ref{tab:bulk_gl}). We attribute
this difference to the DLWFs. Table \ref{tab:bulk_dlwf_bp} demonstrates how BP
has two `classes' of occupied DLWFs (with local occupations
$\lambda_{\bZ ii \sigma} > 0.99$) when computed with the $6 \times 6 \times 6$
$\bk$-mesh: one has $\evo{h}_{\bZ i \sigma} \approx \SI{-3.1}{\electronvolt}$,
and three are nearly degenerate, with
$\evo{h}_{\bZ i \sigma} \approx \SI{4.7}{\electronvolt}$. In the
$8 \times 8 \times 8$ $\bk$-sampled calculation, the lowest-energy DLWF is
qualitatively very similar, but the other three occupied DLWFs have split
into two sub-bands. One has
$\evo{h}_{\bZ i \sigma} \approx \SI{3.0}{\electronvolt}$, and two are nearly
degenerate, with $\evo{h}_{\bZ i \sigma} \approx \SI{5.5}{\electronvolt}$.
They also have qualitatively different spatial spreads, as seen by the values of
$\evo{r^2}_{\bZ i \sigma}$ in Table \ref{tab:bulk_dlwf_bp}.

Differences in spatial localization change, especially, the Coulomb interaction
$\mel{\rho_{\bZ i \sigma}}{f_{\Har}^{\sigma\sigma}}{\rho_{\bR j \sigma}}$ in the
curvature, leading to a different correction to the band gap. In particular,
observe from Fig. \ref{fig:k_bulk} \textbf{(d)} that the $6 \times 6 \times 6$
calculation has a substantially higher gap than its $8 \times 8 \times 8$
$\bk$-sampled counterpart, corresponding to the increased spatial localization
(smaller $\evo{r^2}$) of its occupied DLWFs.

We are currently seeking an alternative to the DLWF minimization that avoids
this weakness. Since there is no physical principle that specifies what $\gamma$
should be, we believe that a deterministic algorithm---yielding Wannier
functions \emph{approximately} localized in both space and energy---will
improve the robustness of olLOSC.

\clearpage


\begin{landscape}
\begin{figure}
\centering
\includegraphics[width=0.95 \linewidth]{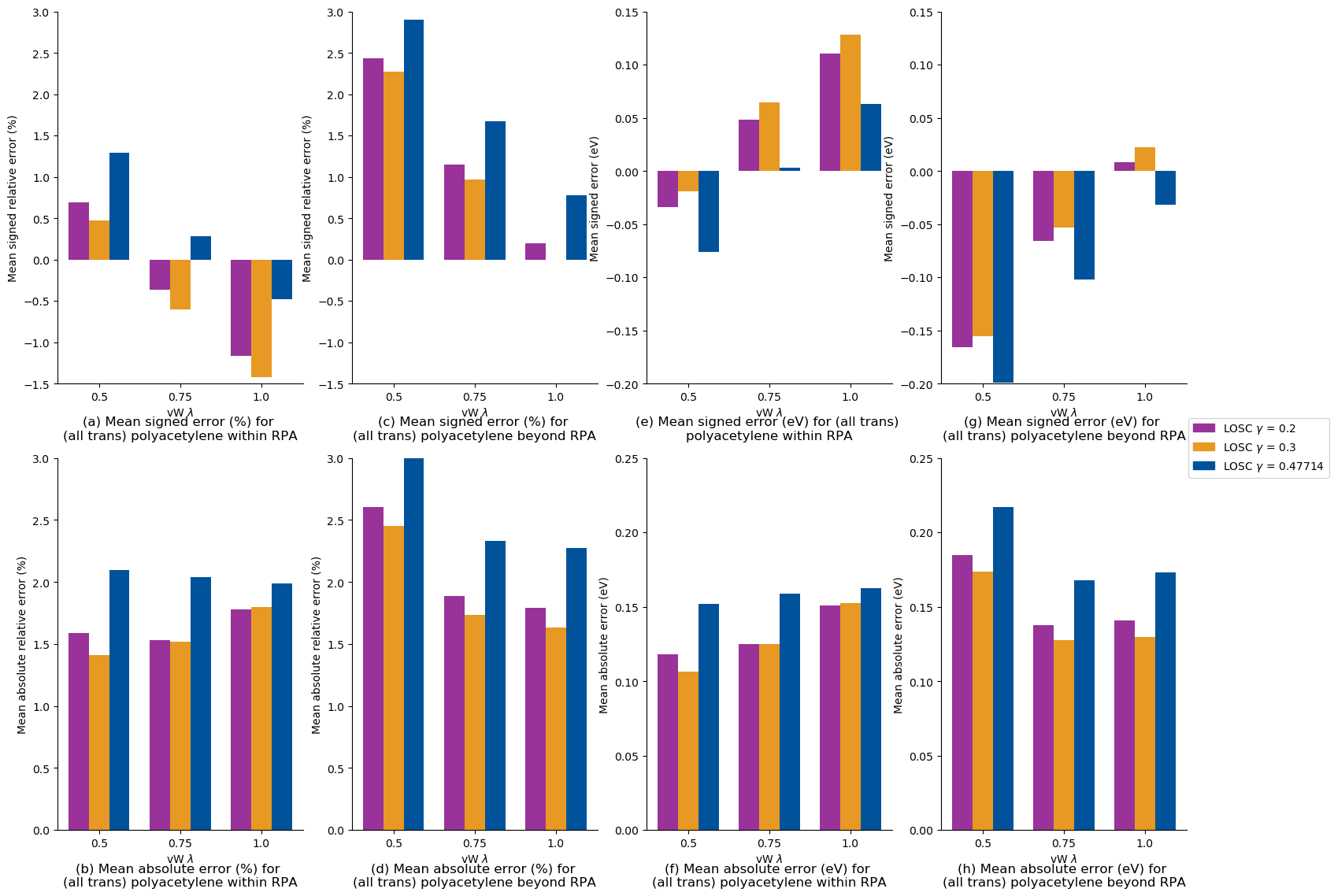}
\caption{olLOSC errors in ionization energy for all-trans polyacetylene with different LOSC $\gamma$ and vW $\lambda$. (a,b): Percentage error for olLOSC within RPA; (c,d): Percentage error for olLOSC beyond RPA; (e,f): Unscaled error for olLOSC within RPA; (g,h): Unscaled error for olLOSC beyond RPA.}
\label{fig:gl_polymer}
\end{figure}
\end{landscape}

\begin{table}\centering
\caption{The bulk systems studied. The lattice is defined by the Strukturbericht
symbol (A4: diamond; B3: zincblende; B1: halite; B3: wurtzite; A1: simple
cubic). $a$ is the experimental lattice parameter, in angstrom; these are taken
from \cite{heyd2005} when available, \cite{wyckoff1973} otherwise. The gap is
experimental, taken from \cite{tran2009} (LiF, LiCl, Ne, Ar); \cite{poole1975}
(NaF); \cite[p.~184]{madelung2004} (MgO); and \cite{heyd2005} and containing
references (the rest). ZPR is zero-point renormalization, computed in the
indicated reference for each system.
}
\label{tab:bulk_ref}
\begin{tabular*}{\hsize}{@{\extracolsep{\fill}}rrrrrr}
\toprule
System & Lattice & $a$ (\si{\angstrom}) & 
Gap (\si{\electronvolt}) & 
ZPR (\si{\electronvolt}) & 
Gap $-$ ZPR (\si{\electronvolt}) \\ 
\midrule
  C & A4 & $3.57$ & $5.48$ & $-0.32$ \cite{engel2022}     & 5.80 \\ 
  Si & A4 & $5.43$ & $1.17$ & $-0.06$ \cite{engel2022}    & 1.23 \\ 
  SiC & B3 & $4.36$ & $2.42$ & $-0.17$ \cite{engel2022}   & 2.59 \\ 
  Ge & A4 & $5.66$ & $0.74$ & $-0.03$ \cite{miglio2020}   & 0.77 \\ 
  MgO & B1 & $4.21$ & $7.97$ & $-0.53$ \cite{engel2022}   & 7.75 \\ 
  LiF & B1 & $4.02$ & $14.20$ & $-1.23$ \cite{engel2022}  & 15.43 \\ 
  LiCl & B1 & $5.13$ & $9.40$ & $-0.44$ \cite{shang2021}  & 9.84 \\ 
  NaF & B1 & $4.62$ & $11.50$ & $-0.70$ \cite{shang2021}  & 12.20 \\ 
  BN & B3 & $3.62$ & $6.22$ & $-0.40$ \cite{engel2022}    & 6.62 \\ 
  BP & B3 & $4.54$ & $2.40$ & $-0.10$ \cite{shang2021}    & 2.50 \\ 
  AlP & B3 & $5.46$ & $2.51$ & $-0.10$ \cite{engel2022}   & 2.61 \\ 
  Ne & A1 & $4.43$ & $21.70$ & n/a                        & 21.70 \\ 
  Ar & A1 & $5.26$ & $14.20$ & n/a                        & 14.20 \\ 
\bottomrule
\end{tabular*}
\end{table}

\begin{landscape}
\begin{table}\centering
\caption{The dataset seen in Fig.~1 (b) of the main text: the olLOSC parameters
are $\gamma = 0.30$, $\lambda = 0.75$ We use a $6 \times 6 \times 6$
Monkhorst--Pack $\bk$-mesh, except in cases where the material's conduction band
minimum does not lie on a high-symmetry $\bk$-point (as evidenced by the PBE CBM
changing with $\bk$). Thus, we use $8 \times 8 \times 8$ $\bk$-points for C and
SiC, and $10 \times 10 \times 10$ $\bk$-points for BP and Si.
}
\label{tab:bulk}
\begin{tabular*}{\hsize}{@{\extracolsep{\fill}}rrrrrrrrrrrrr}
  \toprule
& & \multicolumn{2}{c}{Total energy (Ry)} &
\multicolumn{2}{c}{VBM (eV)} & \multicolumn{2}{c}{CBM (eV)} & 
\multicolumn{3}{c}{Gap (eV)} & \multicolumn{2}{c}{olLOSC gap error} \\
\cmidrule{3-4} \cmidrule{5-6} \cmidrule{7-8} \cmidrule{9-11} \cmidrule{12-13}
System & $\bk$-mesh & $E_{\text{PBE}}$ & $\Delta E_{\LOSC}$ & PBE & olLOSC & PBE & olLOSC & PBE & olLOSC & Ref. &
eV & \% \\
  \midrule
  AlP & 06 & $-$18.58 & $9.75 \times 10^{-4}$ & 5.021 & 4.359 & 6.597 & 6.709 & 1.576 & 2.350 & 2.606 & $-$0.256 & $-$9.808 \\ 
  Ar & 06 & $-$45.18 & $4.48 \times 10^{-5}$ & $-$3.948 & $-$5.756 & 4.755 & 5.066 & 8.703 & 10.822 & 14.200 & $-$3.378 & $-$23.788 \\ 
  BN & 06 & $-$26.81 & $2.88 \times 10^{-5}$ & 11.146 & 9.772 & 15.677 & 16.427 & 4.531 & 6.655 & 6.622 & 0.033 & 0.500 \\ 
  BP & 10 & $-$19.79 & $8.28 \times 10^{-4}$ & 9.240 & 8.570 & 10.492 & 10.932 & 1.252 & 2.362 & 2.501 & $-$0.139 & $-$5.570 \\ 
  C & 08 & $-$24.08 & $6.85 \times 10^{-5}$ & 13.355 & 12.820 & 17.544 & 18.235 & 4.189 & 5.415 & 5.803 & $-$0.388 & $-$6.693 \\ 
  Ge & 06 & $-$357.28 & $2.13 \times 10^{-3}$ & 9.561 & 9.094 & 9.593 & 9.855 & 0.032 & 0.761 & 0.773 & $-$0.012 & $-$1.604 \\ 
  LiCl & 06 & $-$47.47 & $5.99 \times 10^{-5}$ & 1.667 & 0.422 & 7.996 & 7.842 & 6.329 & 7.420 & 9.836 & $-$2.416 & $-$24.564 \\ 
  LiF & 06 & $-$63.98 & $2.11 \times 10^{-5}$ & 0.970 & $-$1.702 & 10.156 & 10.241 & 9.187 & 11.943 & 15.431 & $-$3.488 & $-$22.605 \\ 
  MgO & 06 & $-$152.10 & $8.69 \times 10^{-5}$ & 8.099 & 6.326 & 12.901 & 13.366 & 4.803 & 7.041 & 8.503 & $-$1.462 & $-$17.200 \\ 
  NaF & 06 & $-$140.97 & $7.98 \times 10^{-5}$ & 0.999 & $-$1.639 & 7.389 & 8.002 & 6.390 & 9.641 & 12.199 & $-$2.558 & $-$20.971 \\ 
  Ne & 06 & $-$66.71 & $2.25 \times 10^{-5}$ & $-$9.071 & $-$12.716 & 2.545 & 5.833 & 11.616 & 18.549 & 21.700 & $-$3.151 & $-$14.521 \\ 
  Si & 10 & $-$16.92 & $2.52 \times 10^{-3}$ & 6.234 & 5.692 & 6.812 & 7.034 & 0.578 & 1.342 & 1.228 & 0.114 & 9.267 \\ 
  SiC & 08 & $-$20.54 & $4.84 \times 10^{-4}$ & 9.496 & 8.635 & 10.857 & 11.051 & 1.361 & 2.416 & 2.595 & $-$0.179 & $-$6.913 \\ 
   \bottomrule
\end{tabular*}
\end{table}
\end{landscape}

\begin{table}\centering
\caption{This table (continued in Table \ref{tab:bulk_gl_2}): Choosing the
$\gamma$ and $\lambda$ parameters in bulk systems (see Fig.~\ref{fig:gl_bulk}).
The $\bk$-mesh is as in Table \ref{tab:bulk}. Reference gap (Ref.) is
experimental $-$ ZPR from Table \ref{tab:bulk_ref}.}
\label{tab:bulk_gl}
\begin{footnotesize}
\begin{tabular}{@{}rrrrrrrrrrrr@{}}
\toprule
& & &
\multicolumn{2}{c}{VBM (eV)} & \multicolumn{2}{c}{CBM (eV)} & 
\multicolumn{3}{c}{Gap (eV)} & \multicolumn{2}{c}{olLOSC gap error} \\
\cmidrule{4-5} \cmidrule{6-7} \cmidrule{8-10} \cmidrule{11-12}
System & DLWF $\gamma$ & vW $\lambda$ & PBE & olLOSC & PBE & olLOSC & PBE & olLOSC & Ref. &
eV & \% \\
  \midrule
  AlP & 0.20 & 0.00 & 5.021 & 4.416 & 6.597 & 6.482 & 1.576 & 2.066 & 2.606 & $-$0.540 & $-$20.733 \\ 
  AlP & 0.20 & 0.75 & 5.021 & 4.349 & 6.597 & 6.710 & 1.576 & 2.361 & 2.606 & $-$0.245 & $-$9.405 \\ 
  AlP & 0.20 & 1.00 & 5.021 & 4.335 & 6.597 & 6.732 & 1.576 & 2.397 & 2.606 & $-$0.209 & $-$8.035 \\ 
  AlP & 0.30 & 0.00 & 5.021 & 4.425 & 6.597 & 6.489 & 1.576 & 2.064 & 2.606 & $-$0.542 & $-$20.814 \\ 
  AlP & 0.30 & 0.75 & 5.021 & 4.359 & 6.597 & 6.709 & 1.576 & 2.350 & 2.606 & $-$0.256 & $-$9.808 \\ 
  AlP & 0.30 & 1.00 & 5.021 & 4.345 & 6.597 & 6.730 & 1.576 & 2.385 & 2.606 & $-$0.221 & $-$8.480 \\ 
  AlP & 0.48 & 0.00 & 5.021 & 4.438 & 6.597 & 6.501 & 1.576 & 2.063 & 2.606 & $-$0.543 & $-$20.837 \\ 
  AlP & 0.48 & 0.75 & 5.021 & 4.373 & 6.597 & 6.713 & 1.576 & 2.340 & 2.606 & $-$0.266 & $-$10.207 \\ 
  AlP & 0.48 & 1.00 & 5.021 & 4.360 & 6.597 & 6.733 & 1.576 & 2.373 & 2.606 & $-$0.232 & $-$8.922 \\ 
  Ar & 0.20 & 0.00 & $-$3.948 & $-$9.236 & 4.755 & $-$9.282 & 8.703 & $-$0.047 & 14.200 & $-$14.247 & $-$100.329 \\ 
  Ar & 0.20 & 0.75 & $-$3.948 & $-$5.761 & 4.755 & 4.582 & 8.703 & 10.343 & 14.200 & $-$3.857 & $-$27.162 \\ 
  Ar & 0.20 & 1.00 & $-$3.948 & $-$5.778 & 4.755 & 4.645 & 8.703 & 10.423 & 14.200 & $-$3.777 & $-$26.599 \\ 
  Ar & 0.30 & 0.00 & $-$3.948 & $-$6.830 & 4.755 & $-$7.730 & 8.703 & $-$0.900 & 14.200 & $-$15.100 & $-$106.339 \\ 
  Ar & 0.30 & 0.75 & $-$3.948 & $-$5.756 & 4.755 & 5.066 & 8.703 & 10.822 & 14.200 & $-$3.378 & $-$23.788 \\ 
  Ar & 0.30 & 1.00 & $-$3.948 & $-$5.774 & 4.755 & 5.119 & 8.703 & 10.893 & 14.200 & $-$3.307 & $-$23.291 \\ 
  Ar & 0.48 & 0.00 & $-$3.948 & $-$5.601 & 4.755 & $-$6.469 & 8.703 & $-$0.868 & 14.200 & $-$15.068 & $-$106.115 \\ 
  Ar & 0.48 & 0.75 & $-$3.948 & $-$5.752 & 4.755 & 5.109 & 8.703 & 10.861 & 14.200 & $-$3.339 & $-$23.513 \\ 
  Ar & 0.48 & 1.00 & $-$3.948 & $-$5.769 & 4.755 & 5.158 & 8.703 & 10.928 & 14.200 & $-$3.272 & $-$23.044 \\ 
  BN & 0.20 & 0.00 & 11.146 & 9.860 & 15.677 & 16.548 & 4.531 & 6.689 & 6.622 & 0.067 & 1.010 \\ 
  BN & 0.20 & 0.75 & 11.146 & 9.764 & 15.677 & 16.745 & 4.531 & 6.981 & 6.622 & 0.359 & 5.423 \\ 
  BN & 0.20 & 1.00 & 11.146 & 9.743 & 15.677 & 16.781 & 4.531 & 7.037 & 6.622 & 0.415 & 6.275 \\ 
  BN & 0.30 & 0.00 & 11.146 & 9.867 & 15.677 & 16.250 & 4.531 & 6.383 & 6.622 & $-$0.239 & $-$3.617 \\ 
  BN & 0.30 & 0.75 & 11.146 & 9.772 & 15.677 & 16.427 & 4.531 & 6.655 & 6.622 & 0.033 & 0.500 \\ 
  BN & 0.30 & 1.00 & 11.146 & 9.752 & 15.677 & 16.458 & 4.531 & 6.706 & 6.622 & 0.084 & 1.272 \\ 
  BN & 0.48 & 0.00 & 11.146 & 10.372 & 15.677 & 16.311 & 4.531 & 5.939 & 6.622 & $-$0.683 & $-$10.307 \\ 
  BN & 0.48 & 0.75 & 11.146 & 10.333 & 15.677 & 16.400 & 4.531 & 6.067 & 6.622 & $-$0.555 & $-$8.380 \\ 
  BN & 0.48 & 1.00 & 11.146 & 10.325 & 15.677 & 16.418 & 4.531 & 6.093 & 6.622 & $-$0.529 & $-$7.982 \\ 
  BP & 0.20 & 0.00 & 9.240 & 8.629 & 10.492 & 10.850 & 1.252 & 2.221 & 2.501 & $-$0.280 & $-$11.188 \\ 
  BP & 0.20 & 0.75 & 9.240 & 8.556 & 10.492 & 10.935 & 1.252 & 2.378 & 2.501 & $-$0.123 & $-$4.910 \\ 
  BP & 0.20 & 1.00 & 9.240 & 8.540 & 10.492 & 10.949 & 1.252 & 2.409 & 2.501 & $-$0.092 & $-$3.675 \\ 
  BP & 0.30 & 0.00 & 9.240 & 8.641 & 10.492 & 10.848 & 1.252 & 2.208 & 2.501 & $-$0.293 & $-$11.731 \\ 
  BP & 0.30 & 0.75 & 9.240 & 8.570 & 10.492 & 10.932 & 1.252 & 2.362 & 2.501 & $-$0.139 & $-$5.570 \\ 
  BP & 0.30 & 1.00 & 9.240 & 8.554 & 10.492 & 10.946 & 1.252 & 2.392 & 2.501 & $-$0.109 & $-$4.362 \\ 
  BP & 0.48 & 0.75 & 9.240 & 8.591 & 10.492 & 10.819 & 1.252 & 2.227 & 2.501 & $-$0.274 & $-$10.944 \\ 
  BP & 0.48 & 1.00 & 9.240 & 8.576 & 10.492 & 10.827 & 1.252 & 2.251 & 2.501 & $-$0.250 & $-$9.996 \\ 
  C & 0.20 & 0.00 & 13.355 & 12.390 & 17.544 & 18.504 & 4.189 & 6.114 & 5.803 & 0.311 & 5.352 \\ 
  C & 0.20 & 0.75 & 13.355 & 12.303 & 17.544 & 18.638 & 4.189 & 6.335 & 5.803 & 0.532 & 9.164 \\ 
  C & 0.20 & 1.00 & 13.355 & 12.283 & 17.544 & 18.664 & 4.189 & 6.381 & 5.803 & 0.578 & 9.962 \\ 
  C & 0.30 & 0.00 & 13.355 & 12.852 & 17.544 & 18.158 & 4.189 & 5.307 & 5.803 & $-$0.496 & $-$8.554 \\ 
  C & 0.30 & 0.75 & 13.355 & 12.820 & 17.544 & 18.235 & 4.189 & 5.415 & 5.803 & $-$0.388 & $-$6.693 \\ 
  C & 0.30 & 1.00 & 13.355 & 12.814 & 17.544 & 18.249 & 4.189 & 5.435 & 5.803 & $-$0.368 & $-$6.333 \\ 
  C & 0.48 & 0.00 & 13.355 & 12.933 & 17.544 & 18.034 & 4.189 & 5.101 & 5.803 & $-$0.702 & $-$12.104 \\ 
  C & 0.48 & 0.75 & 13.355 & 12.911 & 17.544 & 18.086 & 4.189 & 5.175 & 5.803 & $-$0.628 & $-$10.815 \\ 
  C & 0.48 & 1.00 & 13.355 & 12.906 & 17.544 & 18.096 & 4.189 & 5.190 & 5.803 & $-$0.613 & $-$10.569 \\ 
  Ge & 0.20 & 0.00 & 9.561 & 9.192 & 9.593 & 9.748 & 0.032 & 0.555 & 0.773 & $-$0.218 & $-$28.150 \\ 
  Ge & 0.20 & 0.75 & 9.561 & 9.060 & 9.593 & 9.855 & 0.032 & 0.795 & 0.773 & 0.022 & 2.885 \\ 
  Ge & 0.20 & 1.00 & 9.561 & 9.031 & 9.593 & 9.874 & 0.032 & 0.843 & 0.773 & 0.070 & 9.056 \\ 
  Ge & 0.30 & 0.00 & 9.561 & 9.216 & 9.593 & 9.747 & 0.032 & 0.530 & 0.773 & $-$0.243 & $-$31.384 \\ 
  Ge & 0.30 & 0.75 & 9.561 & 9.094 & 9.593 & 9.855 & 0.032 & 0.761 & 0.773 & $-$0.012 & $-$1.604 \\ 
  Ge & 0.30 & 1.00 & 9.561 & 9.067 & 9.593 & 9.873 & 0.032 & 0.806 & 0.773 & 0.033 & 4.295 \\ 
  Ge & 0.48 & 0.00 & 9.561 & 9.359 & 9.593 & 9.735 & 0.032 & 0.377 & 0.773 & $-$0.396 & $-$51.294 \\ 
  Ge & 0.48 & 0.75 & 9.561 & 9.288 & 9.593 & 9.842 & 0.032 & 0.553 & 0.773 & $-$0.220 & $-$28.396 \\ 
  Ge & 0.48 & 1.00 & 9.561 & 9.273 & 9.593 & 9.860 & 0.032 & 0.587 & 0.773 & $-$0.186 & $-$24.036 \\ 
  \midrule
  \midrule
\end{tabular}
\end{footnotesize}
\end{table}

\begin{table}\centering
\caption{Choosing the $\gamma$ and $\lambda$ parameters in bulk systems
(continued).}
\label{tab:bulk_gl_2}
\begin{footnotesize}
\begin{tabular}{@{}rrrrrrrrrrrr@{}}
\midrule
\midrule
& & &
\multicolumn{2}{c}{VBM (eV)} & \multicolumn{2}{c}{CBM (eV)} & 
\multicolumn{3}{c}{Gap (eV)} & \multicolumn{2}{c}{olLOSC gap error} \\
\cmidrule{4-5} \cmidrule{6-7} \cmidrule{8-10} \cmidrule{11-12}
System & DLWF $\gamma$ & vW $\lambda$ & PBE & olLOSC & PBE & olLOSC & PBE & olLOSC & Ref. &
eV & \% \\
  \midrule
  LiCl & 0.20 & 0.00 & 1.667 & 0.538 & 7.996 & 4.990 & 6.329 & 4.452 & 9.836 & $-$5.384 & $-$54.737 \\ 
  LiCl & 0.20 & 0.75 & 1.667 & 0.420 & 7.996 & 7.749 & 6.329 & 7.329 & 9.836 & $-$2.507 & $-$25.491 \\ 
  LiCl & 0.20 & 1.00 & 1.667 & 0.404 & 7.996 & 7.800 & 6.329 & 7.396 & 9.836 & $-$2.441 & $-$24.812 \\ 
  LiCl & 0.30 & 0.00 & 1.667 & 0.540 & 7.996 & 5.522 & 6.329 & 4.982 & 9.836 & $-$4.854 & $-$49.349 \\ 
  LiCl & 0.30 & 0.75 & 1.667 & 0.422 & 7.996 & 7.842 & 6.329 & 7.420 & 9.836 & $-$2.416 & $-$24.564 \\ 
  LiCl & 0.30 & 1.00 & 1.667 & 0.406 & 7.996 & 7.886 & 6.329 & 7.480 & 9.836 & $-$2.356 & $-$23.952 \\ 
  LiCl & 0.48 & 0.00 & 1.667 & 0.545 & 7.996 & 5.959 & 6.329 & 5.414 & 9.836 & $-$4.422 & $-$44.957 \\ 
  LiCl & 0.48 & 0.75 & 1.667 & 0.427 & 7.996 & 7.940 & 6.329 & 7.513 & 9.836 & $-$2.323 & $-$23.618 \\ 
  LiCl & 0.48 & 1.00 & 1.667 & 0.411 & 7.996 & 7.978 & 6.329 & 7.567 & 9.836 & $-$2.269 & $-$23.072 \\ 
  LiF & 0.20 & 0.00 & 0.970 & $-$1.543 & 10.156 & 7.622 & 9.187 & 9.165 & 15.431 & $-$6.265 & $-$40.603 \\ 
  LiF & 0.20 & 0.75 & 0.970 & $-$1.711 & 10.156 & 10.167 & 9.187 & 11.878 & 15.431 & $-$3.553 & $-$23.024 \\ 
  LiF & 0.20 & 1.00 & 0.970 & $-$1.738 & 10.156 & 10.225 & 9.187 & 11.963 & 15.431 & $-$3.468 & $-$22.472 \\ 
  LiF & 0.30 & 0.00 & 0.970 & $-$1.534 & 10.156 & 7.911 & 9.187 & 9.445 & 15.431 & $-$5.986 & $-$38.795 \\ 
  LiF & 0.30 & 0.75 & 0.970 & $-$1.702 & 10.156 & 10.241 & 9.187 & 11.943 & 15.431 & $-$3.488 & $-$22.605 \\ 
  LiF & 0.30 & 1.00 & 0.970 & $-$1.729 & 10.156 & 10.295 & 9.187 & 12.024 & 15.431 & $-$3.407 & $-$22.079 \\ 
  LiF & 0.48 & 0.00 & 0.970 & $-$1.526 & 10.156 & 8.257 & 9.187 & 9.783 & 15.431 & $-$5.648 & $-$36.605 \\ 
  LiF & 0.48 & 0.75 & 0.970 & $-$1.695 & 10.156 & 10.330 & 9.187 & 12.025 & 15.431 & $-$3.406 & $-$22.076 \\ 
  LiF & 0.48 & 1.00 & 0.970 & $-$1.722 & 10.156 & 10.379 & 9.187 & 12.100 & 15.431 & $-$3.331 & $-$21.584 \\ 
  MgO & 0.20 & 0.00 & 8.099 & 6.446 & 12.901 & 12.508 & 4.803 & 6.061 & 8.503 & $-$2.442 & $-$28.716 \\ 
  MgO & 0.20 & 0.75 & 8.099 & 6.312 & 12.901 & 13.246 & 4.803 & 6.934 & 8.503 & $-$1.569 & $-$18.453 \\ 
  MgO & 0.20 & 1.00 & 8.099 & 6.288 & 12.901 & 13.282 & 4.803 & 6.994 & 8.503 & $-$1.509 & $-$17.748 \\ 
  MgO & 0.30 & 0.00 & 8.099 & 6.461 & 12.901 & 12.905 & 4.803 & 6.444 & 8.503 & $-$2.059 & $-$24.209 \\ 
  MgO & 0.30 & 0.75 & 8.099 & 6.326 & 12.901 & 13.366 & 4.803 & 7.041 & 8.503 & $-$1.462 & $-$17.200 \\ 
  MgO & 0.30 & 1.00 & 8.099 & 6.302 & 12.901 & 13.394 & 4.803 & 7.092 & 8.503 & $-$1.411 & $-$16.592 \\ 
  MgO & 0.48 & 0.00 & 8.099 & 6.470 & 12.901 & 13.090 & 4.803 & 6.620 & 8.503 & $-$1.883 & $-$22.146 \\ 
  MgO & 0.48 & 0.75 & 8.099 & 6.335 & 12.901 & 13.409 & 4.803 & 7.074 & 8.503 & $-$1.429 & $-$16.811 \\ 
  MgO & 0.48 & 1.00 & 8.099 & 6.311 & 12.901 & 13.428 & 4.803 & 7.117 & 8.503 & $-$1.386 & $-$16.304 \\ 
  NaF & 0.20 & 0.00 & 0.999 & $-$1.485 & 7.389 & 6.396 & 6.390 & 7.881 & 12.199 & $-$4.318 & $-$35.396 \\ 
  NaF & 0.20 & 0.75 & 0.999 & $-$1.652 & 7.389 & 7.999 & 6.390 & 9.652 & 12.199 & $-$2.547 & $-$20.883 \\ 
  NaF & 0.20 & 1.00 & 0.999 & $-$1.679 & 7.389 & 8.035 & 6.390 & 9.714 & 12.199 & $-$2.485 & $-$20.370 \\ 
  NaF & 0.30 & 0.00 & 0.999 & $-$1.470 & 7.389 & 6.540 & 6.390 & 8.010 & 12.199 & $-$4.189 & $-$34.337 \\ 
  NaF & 0.30 & 0.75 & 0.999 & $-$1.639 & 7.389 & 8.002 & 6.390 & 9.641 & 12.199 & $-$2.558 & $-$20.971 \\ 
  NaF & 0.30 & 1.00 & 0.999 & $-$1.666 & 7.389 & 8.035 & 6.390 & 9.700 & 12.199 & $-$2.499 & $-$20.482 \\ 
  NaF & 0.48 & 0.00 & 0.999 & $-$1.458 & 7.389 & 6.743 & 6.390 & 8.201 & 12.199 & $-$3.998 & $-$32.772 \\ 
  NaF & 0.48 & 0.75 & 0.999 & $-$1.627 & 7.389 & 8.005 & 6.390 & 9.632 & 12.199 & $-$2.567 & $-$21.039 \\ 
  NaF & 0.48 & 1.00 & 0.999 & $-$1.654 & 7.389 & 8.034 & 6.390 & 9.688 & 12.199 & $-$2.511 & $-$20.586 \\ 
  Ne & 0.20 & 0.00 & $-$9.071 & $-$104.215 & 2.545 & $-$109.062 & 11.616 & $-$4.847 & 21.700 & $-$26.547 & $-$122.336 \\ 
  Ne & 0.20 & 0.75 & $-$9.071 & $-$12.721 & 2.545 & 6.016 & 11.616 & 18.737 & 21.700 & $-$2.963 & $-$13.654 \\ 
  Ne & 0.20 & 1.00 & $-$9.071 & $-$12.756 & 2.545 & 6.177 & 11.616 & 18.933 & 21.700 & $-$2.767 & $-$12.749 \\ 
  Ne & 0.30 & 0.00 & $-$9.071 & $-$109.389 & 2.545 & $-$96.808 & 11.616 & 12.581 & 21.700 & $-$9.119 & $-$42.021 \\ 
  Ne & 0.30 & 0.75 & $-$9.071 & $-$12.716 & 2.545 & 5.833 & 11.616 & 18.549 & 21.700 & $-$3.151 & $-$14.521 \\ 
  Ne & 0.30 & 1.00 & $-$9.071 & $-$12.751 & 2.545 & 5.980 & 11.616 & 18.732 & 21.700 & $-$2.968 & $-$13.680 \\ 
  Ne & 0.48 & 0.00 & $-$9.071 & $-$74.744 & 2.545 & $-$36.434 & 11.616 & 38.310 & 21.700 & 16.610 & 76.544 \\ 
  Ne & 0.48 & 0.75 & $-$9.071 & $-$12.712 & 2.545 & 5.619 & 11.616 & 18.330 & 21.700 & $-$3.370 & $-$15.529 \\ 
  Ne & 0.48 & 1.00 & $-$9.071 & $-$12.747 & 2.545 & 5.751 & 11.616 & 18.499 & 21.700 & $-$3.201 & $-$14.753 \\ 
  Si & 0.20 & 0.00 & 6.234 & 5.691 & 6.812 & 6.973 & 0.578 & 1.282 & 1.228 & 0.054 & 4.406 \\ 
  Si & 0.20 & 0.75 & 6.234 & 5.618 & 6.812 & 7.034 & 0.578 & 1.416 & 1.228 & 0.188 & 15.309 \\ 
  Si & 0.20 & 1.00 & 6.234 & 5.601 & 6.812 & 7.042 & 0.578 & 1.441 & 1.228 & 0.213 & 17.386 \\ 
  Si & 0.30 & 0.00 & 6.234 & 5.755 & 6.812 & 6.973 & 0.578 & 1.218 & 1.228 & $-$0.010 & $-$0.790 \\ 
  Si & 0.30 & 0.75 & 6.234 & 5.692 & 6.812 & 7.034 & 0.578 & 1.342 & 1.228 & 0.114 & 9.267 \\ 
  Si & 0.30 & 1.00 & 6.234 & 5.677 & 6.812 & 7.042 & 0.578 & 1.365 & 1.228 & 0.137 & 11.132 \\ 
  Si & 0.48 & 0.75 & 6.234 & 5.735 & 6.812 & 7.033 & 0.578 & 1.298 & 1.228 & 0.070 & 5.692 \\ 
  Si & 0.48 & 1.00 & 6.234 & 5.722 & 6.812 & 7.041 & 0.578 & 1.319 & 1.228 & 0.091 & 7.427 \\ 
  SiC & 0.20 & 0.00 & 9.496 & 8.690 & 10.857 & 10.961 & 1.361 & 2.271 & 2.595 & $-$0.324 & $-$12.470 \\ 
  SiC & 0.20 & 0.75 & 9.496 & 8.617 & 10.857 & 11.184 & 1.361 & 2.567 & 2.595 & $-$0.028 & $-$1.094 \\ 
  SiC & 0.20 & 1.00 & 9.496 & 8.601 & 10.857 & 11.216 & 1.361 & 2.615 & 2.595 & 0.020 & 0.759 \\ 
  SiC & 0.30 & 0.00 & 9.496 & 8.707 & 10.857 & 10.902 & 1.361 & 2.195 & 2.595 & $-$0.400 & $-$15.399 \\ 
  SiC & 0.30 & 0.75 & 9.496 & 8.635 & 10.857 & 11.051 & 1.361 & 2.416 & 2.595 & $-$0.179 & $-$6.913 \\ 
  SiC & 0.30 & 1.00 & 9.496 & 8.619 & 10.857 & 11.069 & 1.361 & 2.450 & 2.595 & $-$0.145 & $-$5.580 \\ 
  SiC & 0.48 & 0.00 & 9.496 & 8.863 & 10.857 & 10.894 & 1.361 & 2.031 & 2.595 & $-$0.564 & $-$21.738 \\ 
  SiC & 0.48 & 0.75 & 9.496 & 8.817 & 10.857 & 11.008 & 1.361 & 2.191 & 2.595 & $-$0.404 & $-$15.584 \\ 
  SiC & 0.48 & 1.00 & 9.496 & 8.807 & 10.857 & 11.021 & 1.361 & 2.214 & 2.595 & $-$0.381 & $-$14.678 \\ 
  \bottomrule
\end{tabular}
\end{footnotesize}
\end{table}

\begin{landscape}
\begin{table}\centering
\begin{scriptsize}
\caption{$\bk$-point convergence of bulk systems. VBM: valence band maximum;
CBM: conduction band minimum. Reference values are the electronic gaps from 
Table \ref{tab:bulk_ref}. Gap error is $E_g^{\text{olLOSC}} - E_g^{\text{ref}}$.
}
\label{tab:conv-bulk-k}
\begin{tabular*}{\hsize}{@{\extracolsep{\fill}}llrrrrrrrrrrrrr}
  \toprule
 & & & & \multicolumn{2}{c}{Total energy (Ry)} &
\multicolumn{2}{c}{VBM (eV)} & \multicolumn{2}{c}{CBM (eV)} & 
\multicolumn{3}{c}{Gap (eV)} & \multicolumn{2}{c}{olLOSC gap error} \\
\cmidrule{5-6} \cmidrule{7-8} \cmidrule{9-10} \cmidrule{11-13} \cmidrule{14-15}
System & $\bk$-mesh & $\gamma$ & $\lambda$ & $E_{\text{PBE}}$ & $\Delta E_{\LOSC}$ & PBE & olLOSC & PBE & olLOSC & PBE & olLOSC & Ref. &
eV & \% \\
  \midrule
  AlP & 06 & 0.30 & 0.75 & $-$18.58 & $1.64 \times 10^{-3}$ & 5.021 & 4.150 & 6.597 & 6.816 & 1.576 & 2.666 & 2.606 & 0.060 & 2.306 \\ 
  AlP & 08 & 0.30 & 0.75 & $-$18.58 & $2.13 \times 10^{-3}$ & 5.016 & 4.172 & 6.595 & 6.792 & 1.579 & 2.620 & 2.606 & 0.014 & 0.553 \\ 
  AlP & 06 & 0.30 & 1.00 & $-$18.58 & $1.81 \times 10^{-3}$ & 5.021 & 4.099 & 6.597 & 6.856 & 1.576 & 2.757 & 2.606 & 0.151 & 5.810 \\ 
  Ar & 06 & 0.30 & 0.75 & $-$45.18 & $5.44 \times 10^{-5}$ & $-$3.948 & $-$5.909 & 4.755 & 5.079 & 8.703 & 10.988 & 14.200 & $-$3.212 & $-$22.620 \\ 
  Ar & 08 & 0.30 & 0.75 & $-$45.18 & $5.07 \times 10^{-5}$ & $-$3.948 & $-$5.759 & 4.755 & 4.760 & 8.703 & 10.519 & 14.200 & $-$3.681 & $-$25.923 \\ 
  Ar & 06 & 0.30 & 1.00 & $-$45.18 & $6.65 \times 10^{-5}$ & $-$3.948 & $-$6.034 & 4.755 & 5.154 & 8.703 & 11.188 & 14.200 & $-$3.012 & $-$21.213 \\ 
  BN & 06 & 0.30 & 0.75 & $-$26.81 & $2.97 \times 10^{-5}$ & 11.146 & 9.716 & 15.677 & 16.495 & 4.531 & 6.778 & 6.622 & 0.156 & 2.362 \\ 
  BN & 08 & 0.30 & 0.75 & $-$26.81 & $4.12 \times 10^{-5}$ & 11.143 & 9.800 & 15.676 & 16.775 & 4.533 & 6.975 & 6.622 & 0.353 & 5.331 \\ 
  BN & 06 & 0.30 & 1.00 & $-$26.81 & $3.24 \times 10^{-5}$ & 11.146 & 9.641 & 15.677 & 16.560 & 4.531 & 6.920 & 6.622 & 0.298 & 4.494 \\ 
  BP & 06 & 0.30 & 0.75 & $-$19.79 & $7.10 \times 10^{-4}$ & 9.246 & 8.290 & 10.591 & 11.268 & 1.345 & 2.978 & 2.501 & 0.477 & 19.060 \\ 
  BP & 08 & 0.30 & 0.75 & $-$19.79 & $1.03 \times 10^{-3}$ & 9.241 & 8.316 & 10.522 & 11.158 & 1.281 & 2.842 & 2.501 & 0.341 & 13.643 \\ 
  BP & 06 & 0.30 & 1.00 & $-$19.79 & $7.85 \times 10^{-4}$ & 9.246 & 8.215 & 10.591 & 11.332 & 1.345 & 3.117 & 2.501 & 0.616 & 24.630 \\ 
  C & 06 & 0.30 & 0.75 & $-$24.07 & $3.92 \times 10^{-5}$ & 13.358 & 12.785 & 17.563 & 18.289 & 4.205 & 5.505 & 5.803 & $-$0.299 & $-$5.144 \\ 
  C & 08 & 0.30 & 0.75 & $-$24.08 & $6.78 \times 10^{-5}$ & 13.355 & 12.825 & 17.544 & 18.224 & 4.189 & 5.399 & 5.803 & $-$0.404 & $-$6.962 \\ 
  C & 06 & 0.30 & 1.00 & $-$24.07 & $4.04 \times 10^{-5}$ & 13.358 & 12.778 & 17.563 & 18.304 & 4.205 & 5.526 & 5.803 & $-$0.277 & $-$4.770 \\ 
  Ge & 06 & 0.30 & 0.75 & $-$357.28 & $3.11 \times 10^{-3}$ & 9.561 & 8.932 & 9.593 & 9.975 & 0.032 & 1.043 & 0.773 & 0.270 & 34.955 \\ 
  Ge & 08 & 0.30 & 0.75 & $-$357.28 & $4.39 \times 10^{-3}$ & 9.554 & 8.933 & 9.585 & 9.951 & 0.031 & 1.018 & 0.773 & 0.245 & 31.759 \\ 
  Ge & 06 & 0.30 & 1.00 & $-$357.28 & $3.53 \times 10^{-3}$ & 9.561 & 8.866 & 9.593 & 10.032 & 0.032 & 1.165 & 0.773 & 0.392 & 50.776 \\ 
  LiCl & 06 & 0.30 & 0.75 & $-$47.47 & $9.32 \times 10^{-5}$ & 1.667 & 0.228 & 7.996 & 7.892 & 6.329 & 7.664 & 9.836 & $-$2.172 & $-$22.085 \\ 
  LiCl & 08 & 0.30 & 0.75 & $-$47.47 & $2.26 \times 10^{-5}$ & 1.667 & 0.303 & 7.996 & 7.427 & 6.329 & 7.124 & 9.836 & $-$2.712 & $-$27.573 \\ 
  LiCl & 06 & 0.30 & 1.00 & $-$47.47 & $1.23 \times 10^{-4}$ & 1.667 & 0.109 & 7.996 & 7.988 & 6.329 & 7.879 & 9.836 & $-$1.957 & $-$19.899 \\ 
  LiF & 06 & 0.30 & 0.75 & $-$63.98 & $2.10 \times 10^{-5}$ & 0.970 & $-$1.702 & 10.156 & 10.241 & 9.187 & 11.943 & 15.431 & $-$3.488 & $-$22.604 \\ 
  LiF & 08 & 0.30 & 0.75 & $-$63.98 & $2.45 \times 10^{-5}$ & 0.970 & $-$1.539 & 10.156 & 10.088 & 9.187 & 11.626 & 15.431 & $-$3.805 & $-$24.658 \\ 
  LiF & 06 & 0.30 & 1.00 & $-$63.98 & $2.23 \times 10^{-5}$ & 0.970 & $-$1.728 & 10.156 & 10.296 & 9.187 & 12.024 & 15.431 & $-$3.407 & $-$22.077 \\ 
  MgO & 06 & 0.30 & 0.75 & $-$152.10 & $8.57 \times 10^{-5}$ & 8.099 & 6.326 & 12.901 & 13.362 & 4.803 & 7.036 & 8.503 & $-$1.467 & $-$17.257 \\ 
  MgO & 08 & 0.30 & 0.75 & $-$152.10 & $1.02 \times 10^{-4}$ & 8.098 & 6.424 & 12.901 & 13.267 & 4.803 & 6.844 & 8.503 & $-$1.659 & $-$19.515 \\ 
  MgO & 06 & 0.30 & 1.00 & $-$152.10 & $8.94 \times 10^{-5}$ & 8.099 & 6.302 & 12.901 & 13.389 & 4.803 & 7.087 & 8.503 & $-$1.416 & $-$16.649 \\ 
  NaF & 06 & 0.30 & 0.75 & $-$140.97 & $7.97 \times 10^{-5}$ & 0.999 & $-$1.638 & 7.389 & 8.001 & 6.390 & 9.639 & 12.199 & $-$2.560 & $-$20.984 \\ 
  Ne & 06 & 0.30 & 0.75 & $-$66.71 & $2.19 \times 10^{-5}$ & $-$9.071 & $-$12.638 & 2.545 & 5.796 & 11.616 & 18.434 & 21.700 & $-$3.266 & $-$15.050 \\ 
  Si & 06 & 0.30 & 0.75 & $-$16.92 & $2.35 \times 10^{-3}$ & 6.241 & 5.483 & 6.947 & 7.284 & 0.705 & 1.800 & 1.228 & 0.572 & 46.604 \\ 
  Si & 08 & 0.30 & 0.75 & $-$16.92 & $3.20 \times 10^{-3}$ & 6.235 & 5.492 & 6.847 & 7.172 & 0.612 & 1.680 & 1.228 & 0.452 & 36.816 \\ 
  Si & 06 & 0.30 & 1.00 & $-$16.92 & $2.53 \times 10^{-3}$ & 6.241 & 5.441 & 6.947 & 7.313 & 0.705 & 1.872 & 1.228 & 0.644 & 52.443 \\ 
  SiC & 06 & 0.30 & 0.75 & $-$20.54 & $5.10 \times 10^{-4}$ & 9.503 & 8.408 & 10.859 & 11.134 & 1.356 & 2.726 & 2.595 & 0.131 & 5.056 \\ 
  SiC & 08 & 0.30 & 0.75 & $-$20.54 & $6.69 \times 10^{-4}$ & 9.496 & 8.439 & 10.857 & 11.123 & 1.361 & 2.684 & 2.595 & 0.089 & 3.434 \\ 
  SiC & 06 & 0.30 & 1.00 & $-$20.54 & $5.54 \times 10^{-4}$ & 9.503 & 8.356 & 10.859 & 11.166 & 1.356 & 2.810 & 2.595 & 0.215 & 8.270 \\ 
  \bottomrule
\end{tabular*}
\end{scriptsize}
\end{table}
\end{landscape}

\begin{table}\centering
\caption{The DLWFs ($\gamma = 0.47714$) for BP, arranged in order of increasing
energy. The local occupations are the diagonal elements of the density matrix
$\rho$ in the DLWF basis,
$\lambda_{\bZ ii \sigma} = \ev{\rho}{w_{\bZ i \sigma}}$; they are real, with
$0 \leq \lambda_{\bZ ii \sigma} \leq 1$, so they measure DLWF occupation.
Observe the threefold near-degeneracy of the occupied DLWFs in the
$6 \times 6 \times 6$ $\bk$-sampled calculation and how it differs from the
occupied submanifold in the $8 \times 8 \times 8$ calculation.}
\label{tab:bulk_dlwf_bp}
\begin{tabular*}{\hsize}{@{\extracolsep{\fill}}rrrrrr}
    \toprule
    \multicolumn{2}{c}{Local occupation $\lambda_{\bZ ii \sigma}$} &
    \multicolumn{2}{c}{Average energy $\evo{h}_{\bZ i \sigma}$ (\si{\electronvolt})} &
    \multicolumn{2}{c}{Spatial spread $\evo{r^2}_{\bZ i \sigma}$ (\si{\bohr^2})} \\
    \cmidrule{1-2} \cmidrule{3-4} \cmidrule{5-6}
    $\bk = 6 \times 6 \times 6$ & $\bk = 8 \times 8 \times 8$ &
    $\bk = 6 \times 6 \times 6$ & $\bk = 8 \times 8 \times 8$ &
    $\bk = 6 \times 6 \times 6$ & $\bk = 8 \times 8 \times 8$ \\
    \midrule
    1.0000 & 1.0000 & $-$3.075373   & $-$3.065085   & 1.827005 & 1.840974 \\
    0.9995 & 0.9997 & 4.702881      &  3.041265     & 2.313304 & 3.948855 \\
    0.9995 & 0.9988 & 4.703143      &  5.511713     & 2.313211 & 3.489266 \\
    0.9995 & 0.9988 & 4.703387      &  5.532760     & 2.313190 & 3.553196 \\
    0.0008 & 0.0014 & 14.526515     & 14.583134     & 4.215674 & 4.342439 \\
    0.0004 & 0.0008 & 15.866122     & 15.754140     & 4.308766 & 4.357879 \\
    0.0002 & 0.0004 & 16.851640     & 16.735331     & 3.598073 & 3.649210 \\
    0.0000 & 0.0001 & 18.791317     & 18.544373     & 4.529317 & 4.627405 \\
    \bottomrule
\end{tabular*}
\end{table}

\begin{landscape}
\begin{table}
\centering
\caption{Large organic molecules olLOSC calculation results with $\lambda$ = 0.75 and $\gamma$ = 0.30. (See b and e in Fig. \ref{fig:gl_rel_mol} and Fig. \ref{fig:gl_unscale_mol}) Reference gap (Ref.) is CCSD-T benchmark from Ref\cite{Richard2016}.}
\label{tab:molg3l75}
\begin{small}
\begin{tabular*}{\hsize}{@{\extracolsep{\fill}}rrrrrrrrrrrr}
\toprule
& \multicolumn{2}{c}{Total energy} &
\multicolumn{2}{c}{HOMO (eV)} & \multicolumn{2}{c}{LUMO (eV)} & 
\multicolumn{3}{c}{Gap (eV)} & \multicolumn{2}{c}{olLOSC gap error} \\
\cmidrule{2-3} \cmidrule{4-5} \cmidrule{6-7} \cmidrule{8-10} \cmidrule{11-12} 
System &$E_{PBE}$ (\si{\hartree}) &$\Delta E$ (\si{\milli\hartree}) & PBE & olLOSC & PBE & olLOSC & PBE & olLOSC & Ref. &
eV & \% \\
\midrule
                        Acridine &   -555.067 &   -0.312 &    -5.426 &       -7.966 &    -2.976 &       -0.873 &    2.449 &       7.093 &       7.394 &   -0.301 &           -4.069 \\
                      Anthracene &   -539.021 &    1.462 &    -4.979 &       -7.422 &    -2.728 &       -0.683 &    2.251 &       6.739 &       7.214 &   -0.475 &           -6.579 \\
                         Azulene &   -385.478 &    8.534 &    -4.970 &       -7.499 &    -2.881 &       -0.531 &    2.089 &       6.968 &       7.003 &   -0.035 &           -0.493 \\
                    Benzonitrile &   -324.201 &   -0.087 &    -6.852 &       -9.924 &    -2.521 &        0.069 &    4.331 &       9.992 &      10.168 &   -0.176 &           -1.728 \\
                    Benzoquinone &   -381.160 &    1.487 &    -6.348 &      -10.259 &    -4.566 &       -1.669 &    1.781 &       8.590 &       8.714 &   -0.124 &           -1.428 \\
                        Dichlone &  -1453.559 &   -0.958 &    -6.580 &       -9.899 &    -4.385 &       -2.069 &    2.195 &       7.829 &       8.066 &   -0.237 &           -2.934 \\
             Dinitrobenzonitrile &   -733.009 &   -0.448 &    -7.725 &      -11.315 &    -4.583 &       -2.006 &    3.142 &       9.309 &       9.408 &   -0.099 &           -1.053 \\
                   Fumaronitrile &   -262.854 &   -0.312 &    -7.851 &      -11.310 &    -4.094 &       -1.183 &    3.757 &      10.128 &      10.507 &   -0.379 &           -3.611 \\
                 Maleicanhydride &   -379.039 &   -0.029 &    -7.129 &      -11.342 &    -4.199 &       -1.176 &    2.930 &      10.167 &      10.308 &   -0.141 &           -1.372 \\
                            NDCA &   -686.067 &    3.383 &    -6.468 &       -9.023 &    -3.658 &       -1.394 &    2.810 &       7.629 &       7.918 &   -0.289 &           -3.655 \\
                Naphthalenedione &   -534.682 &    1.114 &    -6.151 &       -9.659 &    -4.171 &       -1.671 &    1.979 &       7.988 &       8.403 &   -0.415 &           -4.935 \\
                    Nitrobenzene &   -436.435 &   -0.581 &    -6.816 &      -10.760 &    -3.485 &       -0.536 &    3.332 &      10.224 &       9.697 &    0.527 &            5.432 \\
               Nitrobenzonitrile &   -528.608 &    0.651 &    -7.301 &      -11.042 &    -4.136 &       -1.451 &    3.165 &       9.590 &       9.348 &    0.242 &            2.591 \\
                       Phenazine &   -571.106 &    0.464 &    -5.735 &       -8.959 &    -3.404 &       -1.180 &    2.331 &       7.779 &       7.387 &    0.392 &            5.306 \\
               Phthalicanhydride &   -532.560 &   -0.336 &    -6.900 &      -10.579 &    -3.601 &       -0.986 &    3.299 &       9.593 &       9.700 &   -0.107 &           -1.107 \\
                     Phthalimide &   -512.703 &   -0.964 &    -6.350 &       -9.972 &    -3.299 &       -0.741 &    3.051 &       9.231 &       9.482 &   -0.251 &           -2.650 \\
                            TCNE &   -447.173 &    3.614 &    -8.597 &      -11.754 &    -5.780 &       -3.244 &    2.817 &       8.510 &       8.954 &   -0.444 &           -4.954 \\
                            TCNQ &   -678.022 &   16.349 &    -7.050 &       -9.377 &    -5.514 &       -3.666 &    1.536 &       5.711 &       6.262 &   -0.551 &           -8.804 \\
 Tetrachloro-isobenzofuranedione &  -2370.300 &    3.504 &    -7.106 &       -9.777 &    -4.062 &       -1.721 &    3.044 &       8.057 &       8.430 &   -0.373 &           -4.427 \\
         Tetrachlorobenzoquinone &  -2218.912 &    3.724 &    -6.983 &       -9.787 &    -5.010 &       -2.604 &    1.973 &       7.184 &       7.762 &   -0.578 &           -7.451 \\
Tetrafluorobenzenedicarbonitrile &   -813.119 &    1.210 &    -7.444 &      -10.070 &    -4.133 &       -1.703 &    3.311 &       8.367 &       9.151 &   -0.784 &           -8.569 \\
         Tetrafluorobenzoquinone &   -777.913 &    2.135 &    -7.374 &      -10.343 &    -5.127 &       -2.261 &    2.248 &       8.082 &       8.861 &   -0.779 &           -8.791 \\
                           mDCBN &   -416.374 &    0.572 &    -7.393 &      -10.240 &    -3.288 &       -0.812 &    4.105 &       9.428 &       9.832 &   -0.404 &           -4.112 \\
\bottomrule
\end{tabular*}
\end{small}
\end{table}
\end{landscape}

\begin{landscape}
\begin{table}
\centering
\caption{Large organic molecules olLOSC calculation results with $\lambda$ = 1.0 and $\gamma$ = 0.47714. (See b and e in Fig. \ref{fig:gl_rel_mol} and Fig. \ref{fig:gl_unscale_mol}) Reference gap (Ref.) is CCSD-T benchmark from Ref\cite{Richard2016}.}
\label{tab:molg48l1}
\begin{small}
\begin{tabular*}{\hsize}{@{\extracolsep{\fill}}rrrrrrrrrrrr}
\toprule
& \multicolumn{2}{c}{Total energy} &
\multicolumn{2}{c}{HOMO (eV)} & \multicolumn{2}{c}{LUMO (eV)} & 
\multicolumn{3}{c}{Gap (eV)} & \multicolumn{2}{c}{olLOSC gap error} \\
\cmidrule{2-3} \cmidrule{4-5} \cmidrule{6-7} \cmidrule{8-10} \cmidrule{11-12} 
System &$E_{PBE}$ (\si{\hartree}) &$\Delta E$ (\si{\milli\hartree}) & PBE & olLOSC & PBE & olLOSC & PBE & olLOSC & Ref. &
eV & \% \\
\midrule
                        Acridine &   -555.067 &   -0.935 &    -5.426 &       -7.809 &    -2.976 &       -0.830 &    2.449 &       6.979 &       7.394 &   -0.415 &           -5.618 \\
                      Anthracene &   -539.021 &    0.266 &    -4.979 &       -7.378 &    -2.728 &       -0.662 &    2.251 &       6.716 &       7.214 &   -0.498 &           -6.907 \\
                         Azulene &   -385.478 &    2.300 &    -4.970 &       -7.582 &    -2.881 &       -0.432 &    2.089 &       7.149 &       7.003 &    0.146 &            2.089 \\
                    Benzonitrile &   -324.201 &   -0.216 &    -6.852 &       -9.848 &    -2.521 &        0.129 &    4.331 &       9.977 &      10.168 &   -0.191 &           -1.874 \\
                    Benzoquinone &   -381.160 &   -0.022 &    -6.348 &      -10.256 &    -4.566 &       -1.674 &    1.781 &       8.582 &       8.714 &   -0.132 &           -1.517 \\
                        Dichlone &  -1453.559 &   -1.344 &    -6.580 &       -9.914 &    -4.385 &       -2.094 &    2.195 &       7.819 &       8.066 &   -0.247 &           -3.056 \\
             Dinitrobenzonitrile &   -733.009 &    0.171 &    -7.725 &      -11.345 &    -4.583 &       -1.924 &    3.142 &       9.421 &       9.408 &    0.013 &            0.139 \\
                   Fumaronitrile &   -262.854 &   -0.331 &    -7.851 &      -11.256 &    -4.094 &       -1.113 &    3.757 &      10.143 &      10.507 &   -0.364 &           -3.469 \\
                 Maleicanhydride &   -379.039 &    0.071 &    -7.129 &      -11.342 &    -4.199 &       -1.083 &    2.930 &      10.259 &      10.308 &   -0.049 &           -0.477 \\
                            NDCA &   -686.067 &    0.346 &    -6.468 &       -9.019 &    -3.658 &       -1.341 &    2.810 &       7.678 &       7.918 &   -0.240 &           -3.036 \\
                Naphthalenedione &   -534.682 &    0.036 &    -6.151 &       -9.668 &    -4.171 &       -1.676 &    1.979 &       7.992 &       8.403 &   -0.411 &           -4.888 \\
                    Nitrobenzene &   -436.435 &   -0.010 &    -6.816 &      -10.809 &    -3.485 &       -0.462 &    3.332 &      10.347 &       9.697 &    0.650 &            6.704 \\
               Nitrobenzonitrile &   -528.608 &   -0.013 &    -7.301 &      -11.082 &    -4.136 &       -1.385 &    3.165 &       9.696 &       9.348 &    0.348 &            3.725 \\
                       Phenazine &   -571.106 &    0.090 &    -5.735 &       -8.998 &    -3.404 &       -1.110 &    2.331 &       7.887 &       7.387 &    0.500 &            6.773 \\
               Phthalicanhydride &   -532.560 &   -0.428 &    -6.900 &      -10.598 &    -3.601 &       -0.945 &    3.299 &       9.653 &       9.700 &   -0.047 &           -0.485 \\
                     Phthalimide &   -512.703 &   -0.751 &    -6.350 &       -9.990 &    -3.299 &       -0.674 &    3.051 &       9.315 &       9.482 &   -0.167 &           -1.757 \\
                            TCNE &   -447.173 &    1.231 &    -8.597 &      -11.816 &    -5.780 &       -3.165 &    2.817 &       8.651 &       8.954 &   -0.303 &           -3.383 \\
                            TCNQ &   -678.022 &   11.383 &    -7.050 &       -9.031 &    -5.514 &       -3.587 &    1.536 &       5.444 &       6.262 &   -0.818 &          -13.070 \\
 Tetrachloro-isobenzofuranedione &  -2370.300 &    1.541 &    -7.106 &       -9.671 &    -4.062 &       -1.763 &    3.044 &       7.908 &       8.430 &   -0.522 &           -6.191 \\
         Tetrachlorobenzoquinone &  -2218.912 &    0.299 &    -6.983 &       -9.490 &    -5.010 &       -2.631 &    1.973 &       6.859 &       7.762 &   -0.903 &          -11.637 \\
Tetrafluorobenzenedicarbonitrile &   -813.119 &    0.201 &    -7.444 &       -9.995 &    -4.133 &       -1.652 &    3.311 &       8.343 &       9.151 &   -0.808 &           -8.829 \\
         Tetrafluorobenzoquinone &   -777.913 &    0.071 &    -7.374 &      -10.361 &    -5.127 &       -2.348 &    2.248 &       8.013 &       8.861 &   -0.848 &           -9.571 \\
                           mDCBN &   -416.374 &   -0.150 &    -7.393 &      -10.172 &    -3.288 &       -0.764 &    4.105 &       9.409 &       9.832 &   -0.423 &           -4.307 \\
\bottomrule
\end{tabular*}
\end{small}
\end{table}
\end{landscape}

\begin{landscape}
\begin{table}
\centering
\caption{Small molecules olLOSC calculation results with $\lambda$ = 0.75 and $\gamma$ = 0.3. (See a and d in Fig. \ref{fig:gl_rel_mol} and Fig. \ref{fig:gl_unscale_mol}) Reference gap (Ref.) is CCSD-T benchmark from Ref\cite{Richard2016}.}
\label{tab:smolg3l75}
\begin{tabular*}{\hsize}{@{\extracolsep{\fill}}rrrrrrrrrrrr}
\toprule
& \multicolumn{2}{c}{Total energy} &
\multicolumn{2}{c}{HOMO (eV)} & \multicolumn{2}{c}{LUMO (eV)} & 
\multicolumn{3}{c}{Gap (eV)} & \multicolumn{2}{c}{olLOSC gap error} \\
\cmidrule{2-3} \cmidrule{4-5} \cmidrule{6-7} \cmidrule{8-10} \cmidrule{11-12} 
System &$E_\text{PBE}$ (\si{\hartree}) &$\Delta E$ (\si{\milli\hartree}) & PBE & olLOSC & PBE & olLOSC & PBE & olLOSC & Ref. &
eV & \% \\
\midrule
 CF2 &    -237.57 &     0.03 &     -7.36 &       -12.22 &     -3.73 &         0.68 &     3.63 &       12.90 &       12.68 &     0.22 &             1.71 \\
CH3O &    -114.96 &     0.05 &     -6.06 &       -11.18 &     -5.34 &        -1.01 &     0.73 &       10.17 &        9.51 &     0.66 &             6.93 \\
  CN &     -92.64 &     1.12 &     -9.38 &       -14.47 &     -7.67 &        -3.64 &     1.71 &       10.83 &       10.29 &     0.54 &             5.28 \\
 Cl2 &    -920.04 &    -0.00 &     -7.42 &       -11.73 &     -4.66 &        -0.92 &     2.76 &       10.81 &       10.61 &     0.20 &             1.89 \\
 H2C &     -39.11 &     0.01 &     -5.73 &       -10.93 &     -3.32 &        -0.46 &     2.41 &       10.47 &        9.57 &     0.90 &             9.40 \\
H2CS &    -437.27 &     0.02 &     -5.54 &        -9.41 &     -3.64 &        -0.11 &     1.90 &        9.29 &        9.16 &     0.13 &             1.44 \\
 H2N &     -55.83 &     0.03 &     -7.27 &       -12.90 &     -4.54 &        -0.28 &     2.73 &       12.62 &       11.37 &     1.25 &            10.96 \\
 H2P &    -342.35 &     0.03 &     -6.05 &       -10.11 &     -4.20 &        -1.20 &     1.85 &        8.91 &        8.56 &     0.35 &             4.09 \\
H2Si &    -290.46 &     0.07 &     -5.86 &        -9.43 &     -4.06 &        -1.24 &     1.80 &        8.19 &        8.48 &    -0.29 &            -3.42 \\
 H3C &     -39.79 &     0.00 &     -5.42 &       -10.44 &     -2.95 &         0.24 &     2.48 &       10.68 &        9.84 &     0.84 &             8.52 \\
H3Si &    -291.08 &     0.13 &     -5.36 &        -9.12 &     -3.55 &        -1.12 &     1.82 &        7.99 &        7.93 &     0.06 &             0.79 \\
 HCO &    -113.78 &     0.14 &     -5.00 &        -9.66 &     -3.30 &         0.49 &     1.70 &       10.15 &        9.44 &     0.71 &             7.53 \\
  HN &     -55.17 &     0.03 &     -7.91 &       -14.23 &     -4.32 &         0.02 &     3.60 &       14.26 &       13.15 &     1.11 &             8.40 \\
  HO &     -75.68 &     0.01 &     -7.36 &       -14.01 &     -6.44 &        -1.14 &     0.92 &       12.87 &       11.24 &     1.63 &            14.51 \\
  HP &    -341.73 &     0.03 &     -6.19 &       -10.30 &     -3.97 &        -1.05 &     2.21 &        9.25 &        9.16 &     0.09 &             0.96 \\
  HS &    -398.59 &     0.02 &     -6.22 &       -10.53 &     -5.79 &        -2.24 &     0.43 &        8.29 &        8.05 &     0.24 &             3.01 \\
  S2 &    -796.07 &    -0.01 &     -5.91 &        -9.56 &     -4.61 &        -1.30 &     1.30 &        8.26 &        7.92 &     0.34 &             4.29 \\
\bottomrule
\end{tabular*}
\end{table}
\end{landscape}

\begin{landscape}
\begin{table}
\centering
\caption{Small molecules olLOSC calculation results with $\lambda$ = 1.0 and $\gamma$ = 0.47714. (See a and d in Fig. \ref{fig:gl_rel_mol} and Fig. \ref{fig:gl_unscale_mol}) Reference gap (Ref.) is CCSD-T benchmark from Ref\cite{Richard2016}.}
\label{tab:smolg48l1}
\begin{tabular*}{\hsize}{@{\extracolsep{\fill}}rrrrrrrrrrrr}
\toprule
& \multicolumn{2}{c}{Total energy} &
\multicolumn{2}{c}{HOMO (eV)} & \multicolumn{2}{c}{LUMO (eV)} & 
\multicolumn{3}{c}{Gap (eV)} & \multicolumn{2}{c}{olLOSC gap error} \\
\cmidrule{2-3} \cmidrule{4-5} \cmidrule{6-7} \cmidrule{8-10} \cmidrule{11-12} 
System &$E_\text{PBE}$ (\si{\hartree}) &$\Delta E$ (\si{\milli\hartree}) & PBE & olLOSC & PBE & olLOSC & PBE & olLOSC & Ref. &
eV & \% \\
\midrule
 CF2 &    -237.57 &     0.00 &     -7.36 &       -12.30 &     -3.73 &         0.80 &     3.63 &       13.11 &       12.68 &     0.43 &             3.37 \\
CH3O &    -114.96 &     0.01 &     -6.06 &       -11.26 &     -5.34 &        -0.90 &     0.73 &       10.35 &        9.51 &     0.84 &             8.84 \\
  CN &     -92.64 &     0.12 &     -9.38 &       -14.61 &     -7.67 &        -3.49 &     1.71 &       11.12 &       10.29 &     0.83 &             8.05 \\
 Cl2 &    -920.04 &    -0.00 &     -7.42 &       -11.41 &     -4.66 &        -0.83 &     2.76 &       10.59 &       10.61 &    -0.02 &            -0.20 \\
 H2C &     -39.11 &     0.00 &     -5.73 &       -11.06 &     -3.32 &        -0.32 &     2.41 &       10.74 &        9.57 &     1.17 &            12.22 \\
H2CS &    -437.27 &     0.00 &     -5.54 &        -9.49 &     -3.64 &        -0.02 &     1.90 &        9.46 &        9.16 &     0.30 &             3.32 \\
 H2N &     -55.83 &     0.01 &     -7.27 &       -13.05 &     -4.54 &        -0.13 &     2.73 &       12.92 &       11.37 &     1.55 &            13.64 \\
 H2P &    -342.35 &     0.01 &     -6.05 &       -10.21 &     -4.20 &        -1.10 &     1.85 &        9.10 &        8.56 &     0.54 &             6.36 \\
H2Si &    -290.46 &     0.02 &     -5.86 &        -9.51 &     -4.06 &        -1.16 &     1.80 &        8.34 &        8.48 &    -0.14 &            -1.63 \\
 H3C &     -39.79 &     0.00 &     -5.42 &       -10.57 &     -2.95 &         0.36 &     2.48 &       10.93 &        9.84 &     1.09 &            11.05 \\
H3Si &    -291.08 &     0.03 &     -5.36 &        -9.19 &     -3.55 &        -1.01 &     1.82 &        8.18 &        7.93 &     0.25 &             3.19 \\
 HCO &    -113.78 &     0.04 &     -5.00 &        -9.78 &     -3.30 &         0.64 &     1.70 &       10.42 &        9.44 &     0.98 &            10.36 \\
  HN &     -55.17 &     0.01 &     -7.91 &       -14.41 &     -4.32 &         0.17 &     3.60 &       14.59 &       13.15 &     1.44 &            10.91 \\
  HO &     -75.68 &     0.00 &     -7.36 &       -14.21 &     -6.44 &        -0.95 &     0.92 &       13.26 &       11.24 &     2.02 &            17.97 \\
  HP &    -341.73 &     0.01 &     -6.19 &       -10.40 &     -3.97 &        -0.95 &     2.21 &        9.45 &        9.16 &     0.29 &             3.16 \\
  HS &    -398.59 &     0.00 &     -6.22 &       -10.65 &     -5.79 &        -2.13 &     0.43 &        8.53 &        8.05 &     0.48 &             5.94 \\
  S2 &    -796.07 &    -0.00 &     -5.91 &        -9.64 &     -4.61 &        -1.22 &     1.30 &        8.43 &        7.92 &     0.51 &             6.39 \\
\bottomrule
\end{tabular*}
\end{table}
\end{landscape}

\begin{table}
\centering
\caption{All trans-polyacetylene (\ce{H(C2H2)_nH} olLOSC HOMO energies with \{$\lambda = 0.75$ and $\gamma = 0.30$\} and \{$\lambda = 1.00$ and $\gamma = 0.47714$\} (See a, b, e, f in Fig. \ref{fig:gl_polymer}) Reference energies is RASPT2 calculated negative IP energies from Ref\cite{Shahi2009}.}
\label{tab:molg3l75_poly}
\begin{tabular*}{\hsize}{@{\extracolsep{\fill}}rrrrrrrrrrrr}
\toprule
& & &\multicolumn{2}{c}{Total energy} &
\multicolumn{3}{c}{HOMO (eV)} &  \multicolumn{2}{c}{olLOSC error} \\
\cmidrule{4-5} \cmidrule{6-8} \cmidrule{9-10} 
\ce{H(C2H2)_nH} & LOSC $\gamma$ & vW $\lambda$ & $E_\text{PBE}$ (\si{\hartree}) &$\Delta E$ (\si{\milli\hartree}) & PBE & olLOSC & Ref. &
eV & \% \\
\midrule
   1 &  0.300 &   0.75 &    -78.501 &    0.000 &    -6.722 &      -10.870 &       -10.481 &       -0.389 &           3.712 \\
   1 &  0.477 &   1.00 &    -78.501 &    0.000 &    -6.722 &      -10.975 &       -10.481 &       -0.494 &           4.712 \\
   2 &  0.300 &   0.75 &   -155.831 &    0.165 &    -5.819 &       -9.306 &        -9.176 &       -0.130 &           1.415 \\
   2 &  0.477 &   1.00 &   -155.831 &    0.038 &    -5.819 &       -9.385 &        -9.176 &       -0.209 &           2.282 \\
   3 &  0.300 &   0.75 &   -233.162 &    1.026 &    -5.366 &       -8.390 &        -8.177 &       -0.213 &           2.599 \\
   3 &  0.477 &   1.00 &   -233.162 &    0.498 &    -5.366 &       -8.415 &        -8.177 &       -0.238 &           2.915 \\
   4 &  0.300 &   0.75 &   -310.494 &    2.130 &    -5.094 &       -7.832 &        -7.691 &       -0.141 &           1.835 \\
   4 &  0.477 &   1.00 &   -310.494 &    1.173 &    -5.094 &       -7.877 &        -7.691 &       -0.186 &           2.414 \\
   5 &  0.300 &   0.75 &   -387.827 &    4.028 &    -4.913 &       -7.372 &        -7.329 &       -0.043 &           0.581 \\
   5 &  0.477 &   1.00 &   -387.827 &    2.149 &    -4.913 &       -7.314 &        -7.329 &        0.015 &          -0.202 \\
   6 &  0.300 &   0.75 &   -465.160 &    4.596 &    -4.783 &       -7.076 &        -7.044 &       -0.032 &           0.451 \\
   6 &  0.477 &   1.00 &   -465.160 &    2.587 &    -4.783 &       -7.028 &        -7.044 &        0.016 &          -0.232 \\
   7 &  0.300 &   0.75 &   -542.493 &    6.656 &    -4.685 &       -6.795 &        -6.852 &        0.057 &          -0.836 \\
   7 &  0.477 &   1.00 &   -542.493 &    3.667 &    -4.685 &       -6.764 &        -6.852 &        0.088 &          -1.285 \\
   8 &  0.300 &   0.75 &   -619.826 &    7.781 &    -4.609 &       -6.631 &        -6.662 &        0.031 &          -0.471 \\
   8 &  0.477 &   1.00 &   -619.826 &    4.656 &    -4.609 &       -6.585 &        -6.662 &        0.077 &          -1.158 \\
   9 &  0.300 &   0.75 &   -697.159 &    9.879 &    -4.548 &       -6.440 &        -6.555 &        0.115 &          -1.751 \\
   9 &  0.477 &   1.00 &   -697.159 &    5.750 &    -4.548 &       -6.400 &        -6.555 &        0.155 &          -2.362 \\
  10 &  0.300 &   0.75 &   -774.493 &   10.759 &    -4.499 &       -6.313 &        -6.410 &        0.097 &          -1.515 \\
  10 &  0.477 &   1.00 &   -774.493 &    6.569 &    -4.499 &       -6.263 &        -6.410 &        0.147 &          -2.294 \\

\bottomrule
\end{tabular*}
\end{table}

\begin{table}[t!]
\centering
\caption{Comparison of the default olLOSC routine (within random phase approximation) with the olLOSC routine beyond random phase approximation across various categories of molecular systems. Also see Fig. \ref{fig:gl_rel_mol} - \ref{fig:gl_ofxc_unscale_mol} for paramter set other than \{$\gamma = 0.3, \lambda = 0.75$\} and \{$\gamma = 0.477114, \lambda = 1.0$\}. Raw data and visualization scripts are in the SI dataset.}
\label{tab:rpa}
\begin{tabular*}{\hsize}{@{\extracolsep{\fill}}rrrrrrr}
\toprule
$\gamma$ & $\lambda$ & isRPA & MAE (eV) & MSE (eV) & MARE & MSRE \\
\midrule
\multicolumn{7}{c}{Small Molecules (17 molecules)} \\
\midrule
0.30 & 0.75 & TRUE & 0.562 & 0.528 & 5.479\% & 5.077\% \\
0.30 & 0.75 & FALSE & 0.376 & 0.277 & 3.673\% & 2.526\% \\
0.47714 & 1.0 & TRUE & 0.757 & 0.738 & 7.446\% & 7.231\% \\
0.47714 & 1.0 & FALSE & 0.588 & 0.541 & 5.737\% & 5.219\% \\
\midrule
\multicolumn{7}{c}{Large Molecules (23 molecules)} \\
\midrule
0.30 & 0.75 & TRUE & 0.352 & -0.251 & 4.176\% & -3.017\% \\
0.30 & 0.75 & FALSE & 0.491 & -0.457 & 5.779\% & -5.386\% \\
0.47714 & 1.0 & TRUE & 0.393 & -0.249 & 4.761\% & -3.071\% \\
0.47714 & 1.0 & FALSE & 0.486 & -0.413 & 5.791\% & -4.946\% \\
\midrule
\multicolumn{7}{c}{All Molecules (40 molecules)} \\
\midrule
0.30 & 0.75 & TRUE & 0.442 & 0.080 & 4.730\% & 0.423\% \\
0.30 & 0.75 & FALSE & 0.443 & -0.145 & 4.884\% & -2.024\% \\
0.47714 & 1.0 & TRUE & 0.548 & 0.171 & 5.902\% & 1.307\% \\
0.47714 & 1.0 & FALSE & 0.529 & -0.008 & 5.768\% & -0.626\%\\
\bottomrule
\end{tabular*}
\end{table}

\begin{table}[t!]
\centering
\caption{Mean signed error (MSE) and mean absolute error (MAE) for reaction
barrier heights (in kcal/mol). NHTBH38 is a 38-reaction dataset of non-hydrogen
transfer reaction barrier heights, while HTBH38 contains 38 hydrogen transfer
barrier heights. We take REF1 (which excludes relativistic effect) from
\cite{ZhengRB2009, Peverati2014} as a reference. The first row gives the
reaction barrier calculated by the parent DFT method (PBE); the subsequent
rows detail the olLOSC reaction barrier calculations using 75\% von Weizs\"acker
kinetic energy ($\lambda = 0.75$) at different localization parameters $\gamma$.}
\label{tab:rb}
\begin{tabular}{@{}rrrrrrrr@{}}
\toprule
 &  & \multicolumn{2}{l}{NHTBH38} & \multicolumn{2}{l}{HTBH38} & \multicolumn{2}{l}{All reactions} \\
 \cmidrule{3-4} \cmidrule{5-6} \cmidrule{7-8} 
$\gamma$ & $\lambda$ &  MAE & MSE & MAE & MSE & MAE & MSE \\
\midrule
PBE & N/A & 9.915 & -8.141 & 9.424 & -9.424 & 9.669 & -8.782 \\
0.2 & 0.75 & 7.888 & -2.721 & 9.873 & -9.873 & 8.881 & -6.297 \\
0.3 & 0.75 & 7.012 & -4.937 & 10.116 & -10.116 & 8.564 & -7.527 \\
0.47714 & 0.75 & 9.249 & -7.438 & 9.788 & -9.788 & 9.518 & -8.613 \\
\bottomrule
\end{tabular}
\end{table}

\begin{table}[h!]
    \centering
    \begin{tabular}{rrrr}
    \toprule
        Molecule & Electrons & Basis functions & olLOSC wall time ($\si{\second}$)\\
        \midrule
        $\rm{H(HC=CH)_1H}$ & 16 & 130 & 15.15 \\ 
        $\rm{H(HC=CH)_2H}$ & 30 & 230 & 34.17 \\ 
        $\rm{H(HC=CH)_3H}$ & 44 & 330 & \textbf{65.85} \\ 
        $\rm{H(HC=CH)_4H}$ & 58 & 430 & \textbf{116.92} \\ 
        $\rm{H(HC=CH)_5H}$ & 72 & 530 & \textbf{188.40}\\ 
        $\rm{H(HC=CH)_6H}$ & 86 & 630 & \textbf{291.36} \\ 
        $\rm{H(HC=CH)_7H}$ & 100 & 730 & \textbf{436.64} \\ 
        $\rm{H(HC=CH)_8H}$ & 114 & 830 & \textbf{638.80} \\ 
        $\rm{H(HC=CH)_9H}$ & 128 & 930 & \textbf{868.04}\\ 
        $\rm{H(HC=CH)_{10}H}$ & 142 & 1030 & \textbf{1209.56} \\ 
    \bottomrule
    \end{tabular}
    \caption{Wall times in seconds for the olLOSC $\kappa$ evaluation for polyacetylene using the cc-pVTZ basis set and cc-pVTZ-RIFIT as the auxiliary basis. The olLOSC method is implemented in QM$^4$D and executed on a 32-core Intel(R) Xeon(R) Gold 6226R CPU @ 2.90 GHz, with a NVIDIA A100 GPU employed to accelerate the numerical integration. From the bolded wall times, numerical fitting indicates that the computational cost scales as $\mathcal{O}(N_\text{electrons}^{2.49})$ with $R^2 = 0.9952$ or, alternatively, as $\mathcal{O}(M_\text{CGTOs}^{2.56})$ with $R^2 = 0.9958$.}
\end{table}

\clearpage
\printbibliography

%% file: content.tex

\section*{Introduction}
Density functional theory (DFT) is the ``workhorse of quantum
chemistry and materials science" \cite{teale2022} and probably the method used
most often to predict quantum properties of electronic systems. Among its
greatest appeals is its reasonable accuracy at calculating ground-state
properties, especially the total energy \cite{hohenberg1964}, of molecules and
materials.

Kohn--Sham (KS) DFT \cite{kohn1965} additionally provides single-particle
orbital energies, whose meaning has long been a subject of study.
Janak \cite{janak1978} showed that orbital energies computed by a density
functional approximation (DFA) are equal to the derivative of the ground-state
energy with respect to the orbital occupations, as long as the
exchange-correlation functional $E_{\xc}$ is continuous in the density
$\rho(\br)$. However, this does not connect directly to physical observables.
Two relationships connecting observables and orbital energies have been
established rigorously. First, the ionization potential theorem
\cite{perdew1982, perdew1997} demonstrates that, for the exact (local)
potential, the highest occupied KS orbital (HOMO) energy is the negative of the
first ionization energy (valence band edge) $I$. Second, the ground state
chemical potential theorem \cite{cohen2008a, yang2012b} shows that, in a
ground state KS calculation with $E_{\xc}$ continuous in $\rho$ or a
generalized KS Kohn--Sham (GKS) calculation with $E_{\xc}$ continuous in the
KS density matrix $\gamma(\br, \br')$, the energy of the highest
occupied (lowest unoccupied) molecular orbital is the chemical potential $\mu$
of electron removal (addition). The PPLB linearity condition \cite{perdew1982}
proves that $\mu$ computed by the exact functional are derivatives of the total
energy with respect to the number of electrons; thus, for an $N_0$-electron
system, $\left.\partial E/\partial N\right\rvert_{N_0^-} = -I$ and
$\left.\partial E/\partial N\right\rvert_{N_0^+} = -A$. The frontier orbital
energies thus correspond---in both KS and GKS calculations---to experimentally
observable ionization potentials and electron affinities (in materials, band
edge energies and band gaps). The ground state chemical potential established
for the first time the physical meaning of the lowest unoccupied molecular
orbital (LUMO) and conduction band minimum (CBM) orbital energy in ground-state
(G)KS calculations \cite{cohen2008a}. DFT calculations thus provide estimates
of key properties like the fundamental gap, defined as the difference between
the ionization potential $I$ and the electron affinity $A$,
\begin{equation} \label{eq:gap}
    E_g = I - A
        \approx
        \begin{cases}
            \eps_{\DFT}^{\LUMO} - \eps_{\DFT}^{\HOMO} & \text{in molecules,} 
            \\[0.25\baselineskip]
            \eps_{\DFT}^{\CBM} - \eps_{\DFT}^{\VBM} & \text{in materials,}
        \end{cases}
\end{equation}
as long as $E_{\xc}$ is continuous in $\rho(\br)$ or $\gamma(\br, \br')$.
In molecules, the DFT-computed gap is the difference between the LUMO and
HOMO energies; in materials, the difference is between the analogous
conduction band minimum (CBM) and valence band maximum (VBM) energies.
If the exact density functional were known and used, this result would be
exact \cite{janak1978, perdew1982, cohen2008a} Since it is not,
\eqref{eq:gap} is only approximate, and traditional approaches to its
approximation are quite poor. Common density functional approximations
(DFAs) like LDA \cite{kohn1965, perdew1992, perdew2018} and PBE
\cite{perdew1996} systematically underestimate the fundamental gap by as
much as 40\% \cite{perdew1985}. This problem, as well the underestimation
of chemical reaction energy barriers \cite{johnson2008, kaplan2023a},
unphysically delocalized charge densities \cite{cohen2008}, and DFAs'
qualitative failure to describe molecular dissociation
\cite{zhang1998, vandevondele2005}, are now known to stem from
delocalization error \cite{mori-sanchez2008, cohen2008, bryenton2022}.

The manifestation of delocalization error depends on the size of the
system \cite{cohen2012, mei2021}. In small molecules near their equilibrium
bond lengths, it arises from the failure of the energy $E(N)$ of a DFA, as a
function of the number of electrons, to obey the PPLB linearity
condition \cite{perdew1982, yang2000, ayers2008}
\begin{equation} \label{eq:pplb}
    E(N + \delta) = (1 - \delta) E(N) + \delta E(N + 1), \quad 
    0 \leq \delta < 1.
\end{equation}
In the exact functional, $E(N)$ is piecewise linear, with discontinuities in the
derivative at integer $N$. Because the derivatives of $E$ with respect to the
Kohn--Sham orbital occupations $n_{n\sigma}$ are the orbital energies
\cite{janak1978}, the derivative discontinuities give rise to the expression for
the fundamental gap \eqref{eq:gap}. $E(N)$ is convex instead of piecewise linear,
however, in almost all DFAs used in practice. This convexity softens the
derivative discontinuity and, ultimately, causes the litany of problems due to
delocalization error. (Note that Hartree--Fock theory suffers instead from 
localization error because its $E(N)$ curve is concave \cite{mori-sanchez2008}.
The success of hybrid functionals, which mix a fraction of Hartree--Fock
exchange with a DFA counterpart, is partially due to this cancellation of
errors.)

Small molecules exhibit a pronounced curvature in $E(N)$, but its prominence
decreases for larger systems. In the bulk limit, translational symmetry forces
$E(N)$ to be piecewise linear, but the derivative discontinuities (hence band
gaps) are still underestimated. In fact, the total energy $E(N \pm 1)$ becomes
inaccurate in materials \cite{mori-sanchez2008}. Because of these contrasting
behaviors, it is challenging to correct delocalization error in both molecules
and materials with the same approximation.

Kohn--Sham DFT's underestimation of the (band) gap was perhaps the earliest
hallmark of delocalization error to be recognized \cite{perdew1985, cohen2008},
so there are several fairly mature theories that improve gaps and band structure.
However, a method that corrects delocalization error in all scenarios must not
only correct the band structure; it must also be able to change the total
energy of the system in order to describe the dissociation limit correctly and
to correct the charge density to describe interfacial systems
\cite{johnson_etal_2013, li2018}.

\subsection*{Methods to correct delocalization error}
Range-separated hybrid functionals can mitigate delocalization error in
molecules or materials, but they do so in completely different ways. In
molecules, Coulomb-attenuated functionals \cite{yanai2004} use Hartree--Fock
(exact) exchange at long range; they cancel delocalization error imperfectly,
but improve molecular fundamental gap calculations significantly
\cite{mori-sanchez2006, cohen2007a}. In bulk systems, however, the $1/r$
asymptotic decay of Hartree--Fock exchange is qualitatively wrong. Long-range
correlations decay exponentially in materials with a band gap
\cite{kohn1995, kohn1996, prodan2005}, and exchange is further scaled by the
inverse macroscopic dielectric constant $1/\epsilon_\infty$
\cite{shimazaki2008}. (More generally, it is scaled by the inverse of the
microscopic dielectric function, $\epsilon^{-1}(\br, \br')$; this approach is
used in the $GW$ method \cite{hedin1965, hybertsen1986}.) Range-separated
hybrids for materials therefore use Hartree--Fock exchange only for
short-range interactions, with a fixed screening to describe long-range
interactions \cite{heyd2003, heyd2006, heyd2005}. The correct
mixture of DFT and Hartree--Fock exchange also depends on the particular
system. The optimally tuned range-separated hybrid functional
\cite{refaely-abramson2011, refaely-abramson2013} incorporates both
system-dependent exchange and $\epsilon_\infty$ to correct delocalization error.
With its recent extension to materials \cite{wing2021}, it is effective at
restoring the linearity condition, but the optimal tuning depends not only
on the system but on its geometry; it is therefore difficult to obtain
consistent energetics of (for instance) the reactants and products a chemical
reaction, or of an interface and its separate components. Optimally tuned
range-separated hybrid functionals are used primarily to correct band gaps,
not total energies.

Koopmans-compliant functionals handily correct quasiparticle spectra in
molecules \cite{dabo2010, borghi2014}. So does its periodic extension, the
Koopmans spectral, or Koopmans-compliant Wannier (KCW) functional 
\cite{colonna2018, colonna2022}. KCW is currently implemented
non-self-consistently \cite{linscott2023}, but because it corrects the
Hamiltonian it could be applied self-consistently to provide a density
correction, following a similar prescription as \cite{mei2020a}. The
Wannier--Koopmans method, which shares a similar philosophy, is similarly
effective at improving band gap prediction
\cite{ma2016, ma2016a, weng2017, weng2018, weng2020}.
However, Koopmans-compliant corrections do not correct the total energy of
insulators or typical molecules.

The $GW$ method \cite{hedin1965, hybertsen1986} improves gaps and band
structures significantly, especially when computed with some degree of
self-consistency \cite{rostgaard2010, huser2013a, vansetten2015, liu2019}.
Furthermore, it offers an implicit correction to the density. However, its
description of the total energy is suspect \cite{holm1999} unless computed
fully self-consistently, which is generally prohibitive in computational cost.
This scales na\"ively with the fourth power of the number $N_e$ of electrons;
even after approximations to reduce the scaling \cite{wilhelm2021}, the
prefactor is characteristically large, so fully self-consistent $GW$
calculations are rare. Keeping delocalization error corrections at DFT cost,
$\Ord(N_e^3)$, withreasonable practical time cost, is thus desirable.

Both the Perdew--Zunger self-interaction correction (PZ-SIC) \cite{perdew1981}
and its modern descendant, the Fermi--L\"owdin orbital self-interaction
correction (FLOSIC) correct delocalization error partially, focusing on its
single-electron manifestation. Unlike its earlier predecessor, FLOSIC's energy
correction is invariant under unitary transformation of the occupied orbitals
and size-consistent \cite{pederson2014}. In addition to energy corrections,
FLOSIC can be applied self-consistently \cite{yang2017} and yields a corrected
density \cite{yamamoto2019}. However, the quality of its correction to the
orbital energies is less clear. FLOSIC provides improvement over DFAs for
molecular ionization potentials \cite{schwalbe2018, adhikari2020} and, when
maximally localized Wannier functions \cite{wannier1937, marzari1997} are used to
construct the Fermi orbitals, modest improvement in band gaps of semiconductors
\cite{shinde2020}. However, the improvements are somewhat inconsistent, and
worsen when the DFA is a gradient-corrected functional.

The localized orbital scaling method (LOSC), introduced by \cite{li2018}, is
a delocalization error correction that offers a correction to the total
energy and electron density as well as to band gaps. It does this through the
construction of localized orbitalets (called dually localized Wannier functions in
materials \cite{mahler_etal_2025}) that encode both spatial and spectral
localization in a balanced way \cite{li2018, su2020}. The original implementation
of LOSC performs well for molecular systems, but breaks down in bulk materials
because it lacks a description of dielectric screening \cite{mahler2022b}. More
recently, we developed linear-response (lr)LOSC, adding dielectric screening from
linear-response theory \cite{mei2021} to LOSC. lrLOSC yields semiconductor and
insulator band structures \cite{williams2024}, molecular ionization potentials
and electron affinities \cite{fanEliminatingDelocalizationError2026}, and
core-level binding energies \cite{yu2025} of comparable accuracy to (and in some
cases better than) $GW$ methods. Nevertheless, the linear-response kernel is
expensive to compute, compared with the conventional DFT calculations, preventing
lrLOSC from being broadly applicable to complex bulk systems and interfaces of
interest.

In this work, we present olLOSC, a computationally efficient approximation to
lrLOSC with an orbital-free linear-response curvature. olLOSC is shown as a
unified functional approximation to correct delocalization error in both
molecules and semiconductors. Looking ahead, we seek an approximation that makes
DFT reliably accurate for computing the electronic structure of interfaces,
which combine challenging features of both molecules and materials. olLOSC,
applicable to both finite and bulk systems---each a limiting case of the more
general interface---is an important step toward this goal.

\section*{Theory}
All methods in the LOSC family correct delocalization error by applying a
quadratic-to-linear correction functional to restore \eqref{eq:pplb}
approximately \cite{zheng2011, li2015, li2018, mei2021}. The most advanced
form is the LOSC correction
\begin{equation} \label{eq:losc-de}
    \Delta E_{\LOSC} = 
        \sum_\sigma \sum_{ij} \frac12 \conj{\lambda}_{ij \sigma} 
            \left( \delta_{ij} - \lambda_{ij \sigma} \right)
            \kappa_{ij \sigma},
\end{equation}
where the matrix of local occupations $\lambda_{ij \sigma}$ is the
one-electron density matrix of the noninteracting reference system in
a basis of localized orbitalets $\ket{\phi_{i\sigma}}$ \cite{li2018},
\begin{equation}
    \lambda_{ij\sigma} = \mel{\phi_{i\sigma}}{\rho}{\phi_{j\sigma}}.
\end{equation}
The curvature $\kappa_{ij\sigma}$ measures the magnitude of the
correction for each pair of orbitalets. $\delta_{ij}$ is the Kronecker delta
function, equal to 1 when $i = j$ and to 0 otherwise.
$\Delta E_\text{LOSC}$ is quadratic in $\lambda_{ij\sigma}$, which addresses the
major contribution to delocalization error \cite{hait2018}.\footnote{
Note on notation:
We write $\rho = \textstyle\sum_\sigma \rho^\sigma$ for the total electron density of a
system composed of collinear spin densities $\rho^\sigma$, and
$\ket{\rho_{n \sigma}}$ for the density of a single spin orbital, with
$\rho_{n \sigma}(\br) = \abs{\psi_{n \sigma}(\br)}^2$.
$f_F^{\sigma\tau}(\br, \br') = 
\delta^2 F / \delta\rho^\sigma(\br) \delta\rho^\tau(\br')$ denotes the kernel of
a functional $F[\rho]$. $\conj{z}$ is the complex conjugate of $z$. 

LOSC for materials is implemented in periodic boundary conditions and supports 
nontrivial sampling of the Brillouin zone; for simplicity, we omit the
associated index $\bk$ in this text. We additionally require translational
symmetry of orbitalets in materials, making them (generalized)
Wannier functions \cite{wannier1937, marzari1997, mahler_etal_2025}.
}

The form of $\Delta E_{\LOSC}$ shares similarities to those of other well-known
post-DFT methods, such as the rotationally invariant formulation
\cite{cococcioni2005} of DFT+$U$ \cite{anisimov1997} and the
Koopmans-compliant functionals \cite{dabo2010, nguyen2018}. The curvature
$\kappa_{ij \sigma}$ is the second derivative of the total energy with respect to
the orbital occupation numbers \cite{yang2012a, mei2021}, expressed in the basis of
localized orbitalets. It bears strong similarities to linear-response DFT+$U$
\cite{timrov2018} and to the Koompans-compliant Wannier functional
\cite{colonna2018, colonna2022, linscott2023}. Neither of the latter energy
corrections, however, can be applied to molecules and materials in all
situations because their respective localized orbitals have fixed occupations,
either 1 or 0. Studies on homolytic dissociation demonstrate that the total energy
from DFAs is typically accurate near equilibrium, but yields a qualitatively
incorrect dissociation limit \cite{cohen2008}. DFT+$U$, based on local (often
atomic \ce{d}) orbital occupations, may provide a large correction to the energy
even at molecular equilibrium geometries. On the other hand, the
Koopmans-compliant energy correction is based on the occupations of Kohn--Sham
orbitals or maximally localized Wannier functions, so it can offer no total
energy correction in insulating systems, even stretched molecules. In particular,
they fail to predict the correct dissociation of molecular ions such as \ce{H2+}.
A unified correction to delocalization error therefore requires dynamic
localization, which is LOSC's first key feature.

\subsection*{Localization and orbitalets}
LOSC's localized orbitals mix the occupied and virtual manifolds to balance
localization in the spatial and energy domains. The resulting orbitals, known as
orbitalets for finite systems \cite{li2018, su2020} and dually localized Wannier
functions (DLWFs) for periodic  systems \cite{mahler_etal_2025} are obtained from
the canonical (Kohn--Sham) orbitals $\ket{\psi_{n\sigma}}$ by unitary
transformation,
\begin{equation} \label{eq:loc-unitary}
    \ket{\phi_{i\sigma}} =  
        \sum_n U_{ni}^\sigma \ket{\psi_{n\sigma}},
\end{equation}
with $U^\sigma$ chosen to minimize a cost function $F$. Initially defined as
a penalty based on the Kohn--Sham eigenvalues \cite{li2018}, $F$ was
reformulated in \cite{su2020} to take the form
\begin{equation} \label{eq:localfun}
    F^\sigma = 
    \sum_i \left[ (1 - \gamma) \Delta r_{i\sigma}^2 + 
           C\, \gamma \Delta h_{i\sigma}^2 \right]; \quad 0 \leq \gamma \leq 1.
\end{equation}
$F^\sigma$ is thus a convex sum of the spatial variance
$\Delta r_{i\sigma}^2 = \ev{\br}{\phi_{i\sigma}}^2 - \ev{r^2}{\phi_{i\sigma}}$
of the orbitalets $\ket{\phi_{i\sigma}}$ and their variance in energy
$\Delta h_{i\sigma}^2 = \ev{h}{\phi_{i\sigma}}^2 - \ev{h^2}{\phi_{i\sigma}}$
\cite{gygi2003, su2020}. The constant $C$ ensures that $F^\sigma$ is
dimensionally consistent.\footnote{In our implementation, $\Delta r_{i\sigma}^2$
is in \si{\bohr^2} and $\Delta h_{i\sigma}^2$ is in \si{\electronvolt^2}, and we
set $C = \qty{1}{\bohr^2/\electronvolt^{2}}$.}

The key difference between orbitalets (DLWFs) and all previous localization
approaches is that they are not limited to mixing orbitals with the same
occupation; that is, they allow mixing between occupied (valence) and virtual
(conduction) orbitals. Allowing this mixing means that the diagonal occupations
$\lambda_{\bZ ii\sigma}$ are not constrained to integral values, even in
insulators. It is these noninteger local occupations that allow LOSC to correct
the total energy, necessary for correcting delocalization error in molecular
dissociation.

Molecular orbitalets are dynamic: different molecular environments yield
qualitatively different orbitalets when $F^\sigma$ is minimized. This is
illustrated dramatically by stretched diatomic molecules. Near equilibrium, the
relatively large gap between Kohn--Sham eigenvalues forces the orbitalets to
approximate the canonical orbitals closely; the local and canonical occupations
are essentially the same---integers---and LOSC yields almost no correction to
the total energy. If the molecule is stretched, however, the canonical orbitals
become increasingly delocalized, and the energy gap between them narrows. The
coupled change in spatial and spectral characteristics drives a transition that
localizes the orbitalets on their respective atoms. Their occupations
$\lambda_{ii\sigma}$ become fractional, and LOSC provides a substantial
correction to the total energy that improves the description of dissociation
greatly \cite{li2018}.

\subsection*{Linear-response curvature}
The curvature $\kappa$ measures the deviation from linearity of the DFA energy
with respect to orbital occupation numbers. Yang and coworkers derived this
quantity to second order with respect to the Kohn--Sham (canonical) occupations
$n_{n\sigma} = \ev{\rho}{\psi_{n\sigma}}$ \cite{yang2012a}, finding that
\begin{multline} \label{eq:curv-canonical}
    \frac{\partial^2 E}{\partial n_{n\sigma} \partial n_{m\tau}} = \\
    \mel{\conj{\psi}_{n\sigma} \psi_{m\sigma}}
        {\Big[ f_{\Hxc}^{\sigma\tau} + 
            \sum_{\mu\nu} f_{\Hxc}^{\sigma\mu} \chi^{\mu\nu} f_{\Hxc}^{\nu\tau}
         \Big]}
        {\conj{\psi}_{m\tau}\psi_{m\tau}}.
\end{multline}
Here,
$f_{\Hxc}^{\sigma\tau} = 
\delta^2 E_{\Hxc} / \delta\rho^\sigma \delta \rho^\tau$ is the
Hartree--exchange-correlation (Hxc) kernel, and
$\chi^{\mu\nu} = \delta \rho^\nu / \delta v^\mu$ is the (static) linear response
function of the density to an external perturbing potential. For details of
multiple derivations of $\partial^2 E / \partial n_{n\sigma} \partial n_{m\tau}$,
see the Supporting Information of \cite{mei2021}.

However, as mentioned above, local orbitals with fractional occupations are
required to correct the total energy, necessary to correct delocalization error
size-consistently. The linear-response (lr)LOSC curvature ansatz is
\eqref{eq:curv-canonical} expressed in the orbitalet basis; for collinear spins,
this yields \cite{yu2025, williams2024}
\begin{equation} \label{eq:lr-kappa}
    \kappa_{ij\sigma} =  
        \mel{\rho_{i\sigma}}
            {\Big[ 
                f_{\Hxc}^{\sigma\sigma} + 
                \sum_{\nu\mu} f_{\Hxc}^{\sigma\nu} \chi^{\nu\mu} 
                               f_{\Hxc}^{\mu\sigma}
             \Big]}
            {\rho_{j\sigma}}.
\end{equation}
The linear response function $\chi$ can be written in terms of the
noninteracting linear response function $\chi_s$ with the Dyson equation
\begin{equation} \label{eq:dyson-chi}
    \chi^{\sigma\tau} = 
    \chi_s^{\sigma\tau} + 
        \chi_s^{\sigma\nu} f_{\Hxc}^{\nu\mu} \chi^{\nu\tau},
\end{equation}
where repeated spins (and their corresponding spatial variables) are
integrated over. In Kohn--Sham calculations, $\chi_s$ takes the well-known
form
\begin{equation} \label{eq:chi-ks}
\begin{split}
    \chi_s^{\sigma\tau}(\br, \br') &= 
    \frac{\delta \rho^\sigma(\br)}{\delta v_{\KS}^\tau(\br')} \\ &=
    \delta_{\sigma\tau} \times
    \sum_{ia} \frac{\conj{\psi}_{i\sigma}(\br) \psi_{i\tau}(\br)
                    \conj{\psi}_{i\tau}(\br') \psi_{i\sigma}(\br')}
                   {\eps_{i\sigma} - \eps_{a\sigma}},
\end{split}
\end{equation}  
where $V_{\KS}^\tau = V_{\Hxc}^\tau + v^\tau$ is the effective potential,
which includes both Hxc and external terms. In the second equality, $i$ indexes
occupied and $a$ virtual Kohn--Sham orbitals.

Eq.~\eqref{eq:chi-ks} suggests a self-consistent method for calculating
$\kappa_{ij\sigma}$, but since $\chi$ depends on two spatial coordinates and 
$\chi_s$ includes a sum over many unoccupied orbitals, such a method promises to
be computationally expensive. It turns out that either self-consistency or the
sum over unoccupied states can be avoided. We can rewrite the Dyson equation for
$\chi$ (omitting spin indices) as
\begin{equation} \label{eq:chi-inv}
    \chi^{-1} = \chi_s^{-1} - f_{\Hxc}.
\end{equation}
If $\chi_s$ can be inverted numerically---using, for example, a discretization
enabled by the resolution-of-the-identity (RI) method \cite{ren2012a}---then
$\chi$ can be computed noniteratively. This is the method used in lrLOSC for
molecules \cite{yu2025, fanEliminatingDelocalizationError2026}. On the other
hand, the Sternheimer equation  \cite{sternheimer1954} can be solved in the
manner of density functional perturbation theory, bypassing the sum over
virtual orbitals; this method is used in lrLOSC for materials
\cite{williams2024}.\footnote{In the Sternheimer method, $\chi$ is not
actually computed at all; instead, the linear response to the orbitalet
densities are obtained iteratively via first-order variation of the
occupied Kohn--Sham orbitals.}

Computing $\kappa_{ij\sigma}$ with linear response yields a very accurate
correction to delocalization error for molecules, semiconductors, and
insulators. lrLOSC predicts band gaps ranging from less than
\qty{0.5}{\electronvolt} to more than \qty{15}{\electronvolt} with accuracy
comparable to or better than self-consistent $GW$ \cite{williams2024}; the
fundamental gaps and core-level binding energies of molecules are of similar
accuracy \cite{yu2025}. However, even with the speedup afforded by the
Sternheimer equation, calculating curvature in materials with linear response
remains computationally costly, limiting the application of lrLOSC to crystals
with relatively small unit cells compared to systems of practical interest for
commonly used DFT approximations.

\subsection*{Orbital-free linear-response curvature}
There is another way to think about $\chi$ and $\chi_s$ that enables a
faster calculation with only a modest loss in accuracy: the orbital-free
approach \cite{york1996}. The total energy $E[\rho]$ obeys the stationarity
condition
\begin{equation}
    \frac{\delta E}{\delta\rho^\sigma(\br)} =
    \frac{\delta F_{\HK}}{\delta \rho^\sigma(\br)} + v^\sigma(\br) = \mu^\sigma,
\end{equation}
where $F_{\HK}[\rho] = T_s[\rho] + J[\rho] + E_{\xc}[\rho]$ is the
Hohenberg--Kohn universal functional, $v^\sigma$ is the external potential, and
$\mu^\sigma$ is the spin-$\sigma$ chemical potential. Taking a second
variation yields
\begin{equation}
    \frac{\delta^2 F_{\HK}}{\delta\rho^\sigma(\br) \delta\rho^\tau(\br')} +
        \frac{\delta v^\sigma(\br)}{\delta\rho^\tau(\br')} = 
    \frac{\delta\mu^\sigma}{\delta\rho^\tau(\br')},
\end{equation}
and rearranging gives the Euler--Lagrange equation
\begin{equation} \label{eq:euler-yorkyang}
    \sum_\tau \int d\br'\, \left[ 
        \frac{\delta^2 F_{\HK}}{\delta\rho^\sigma(\br) \delta\rho^\tau(\br')}
    \right] \delta\rho^\tau(\br') =
    \delta\mu^\sigma - \delta v^\sigma(\br)
\end{equation}
for the perturbing potential $\delta v^\sigma(\br)$ and its linear response
$\ket{\delta\rho^\tau}$ \cite{york1996}. The Lagrange
multiplier $\delta\mu^\sigma$ ensures the conservation of charge (equivalently,
electron number). We quickly obtain an expression for the linear-response
function:
\begin{multline}
    \chi^{\sigma\tau}(\br, \br') =
    \left[\frac{\delta v^\sigma(\br)}{\delta \rho^\tau(\br')}\right]^{-1} \\ =
    \left[  
        \delta\mu^\sigma -
        \frac{\delta^2 F_{\HK}}{\delta \rho^\sigma(\br) \delta\rho^\tau(\br')}
    \right]^{-1},
\end{multline}
Finally, in the Kohn--Sham auxiliary system, there is no
Hartree--exchange-correlation energy, so $F_{\HK}$ is exactly the
noninteracting kinetic energy $T_s$. Thus
\begin{multline} \label{eq:chi-kinetic}
    \chi_s^{\sigma\tau}(\br, \br') = 
    \left[ \delta\mu^\sigma -
        \frac{\delta^2 T_s}{\delta \rho^\sigma(\br) \delta\rho^\tau(\br')}
    \right]^{-1} \\ =
    \left[ \delta\mu^\sigma - f_{T_s}^{\sigma\tau}(\br, \br') \right]^{-1}.
\end{multline}

If we had the exact orbital-free kinetic energy functional $T_s[\rho^\sigma]$,
\eqref{eq:dyson-chi} and \eqref{eq:chi-kinetic} would lead to a
$\chi^{\sigma\tau}$ identical to the coupled-perturbed Kohn--Sham equations.
This relationship suggests a computationally simplification; we approximate
$f_{T_s}$ by the kernel of an orbital-free kinetic energy functional. In this
work, we choose the Thomas--Fermi functional \cite{thomas1927, fermi1927} with
the von Weizs\"acker correction
\cite{weizsaecker1935}, obtaining
\begin{equation}
    f_{\TF\vW}^{\sigma\tau}(\br, \br') =
    f_{\TF}^{\sigma\tau}(\br, \br') + \lambda f_{\vW}^{\sigma\tau}(\br, \br'),
\end{equation}
where
$f_{\TF}^{\sigma\tau} = \delta^2 T_{\TF} / \delta\rho^\sigma \delta\rho^\tau$ is
the Thomas--Fermi kernel, $f_{\vW}$ the von Weizs\"acker kernel, and
$\lambda \geq 0$ controls the amount of von Weizs\"acker correction. (The
analytic form of $f_{\TF\vW}$ can be found in the Supporting Information.) Like
its Kohn--Sham counterpart, $f_{\TF\vW}$ is diagonal in spin, with
$f_{\TF\vW}^{\sigma\tau} = f_{\TF\vW}^\sigma \delta_{\sigma\tau}$. The
Thomas--Fermi kernel is local in space, while $f_{\vW}$ includes derivatives of
the Dirac delta distribution; however, the action of $f_{\TF\vW}^{\sigma}$ on
$\ket{\delta\rho_{i \tau}}$ can still be evaluated locally. We thus obtain an
orbital-free approximation
\begin{equation}
    \label{eqn: chi_tfvw}
    \chi_{\TF\vW}^{\sigma\tau}(\br, \br') = 
    \left[\delta\mu^\sigma - f_{\TF\vW}^{\sigma\tau}(\br, \br') \right]^{-1}
\end{equation}
to $\chi_s$. Replacing $\chi_s$ by $\chi_{\TF\vW}$\footnote{Or, although it is
beyond the scope of this work, by some $\chi_{T_\of}$ obtained from another
orbital-free kinetic energy kernel.} in \eqref{eq:dyson-chi} yields the
orbital-free approximation $\chi_{\ol}$ to the interacting response function;
solving for $\chi_{\ol}$ directly or iteratively, we obtain the orbital-free
linear-response (ol)LOSC curvature $\kappa_{ij\sigma}^{\ol}$.

The lrLOSC and olLOSC curvature expressions contain two components: a bare
response to $f_{\Hxc}^{\sigma\sigma}$, indicated by the first
summand in \eqref{eq:lr-kappa}, and a response screened by $\chi$ or $\chi_{\ol}$
from the second summand. It is this screening, or orbital relaxation, that olLOSC
approximates with an orbital-free kinetic energy kernel. Because screening is
a collective, many-electron effect, using $f_{\TF\vW}$ in olLOSC should capture
the effect well; choosing a better $T_s[\rho]$ would likely improve accuracy.

\subsection*{The partial random phase approximation}
We also make a partial random phase approximation (RPA) in olLOSC, replacing
the Hxc kernel $f_{\Hxc}^{\mu\nu}$ by the Hartree kernel
$f_{\Har}^{\mu\nu}(\br, \br') = 1/\abs{\br - \br'}$ in \eqref{eq:dyson-chi}.
There are two reasons for doing so. First, including the exchange-correlation
kernel in $\chi$ induces major numerical instabilities when computing
$\kappa^{\ol}$ in materials, while $\chi_{\ol\RPA}$ can be obtained by
inverting a positive semidefinite matrix. Second, numerical tests on molecules
actually demonstrate better fundamental gaps from $\chi_{\ol\RPA}$ than from
$\chi_{\ol}$. See the Supporting Information for details. This approximation is
akin to the  screening described in the $GW$ approximation
\cite{hedin1965, hybertsen1986}: the operator $W$ accounts only for Coulombic
screening, while exchange-correlation effects are contained in the vertex
function (and therefore usually neglected). Note that exchange-correlation
interactions are excluded only from $\chi$, not entirely removed from the
olLOSC curvature; $f_{\xc}$ is still present in both terms of
\eqref{eq:lr-kappa}.

\subsection*{Hamiltonian and spectrum}
We derive the LOSC correction to the DFA Hamiltonian from \eqref{eq:losc-de}.
At each spin component \cite{li2018, mei2020a},
\begin{multline} \label{eq:losc-dh}
    \Delta h^{\sigma} = 
    \frac{\delta \Delta E}{\delta \rho^\sigma} =
    \sum_{ij} 
        \left[ \frac{\partial \Delta E}{\partial \lambda_{ij\sigma}}
               \frac{\delta \lambda_{ij\sigma}}{\delta \rho^\sigma} +
               \frac{\partial \Delta E}{\partial \conj{\lambda}_{ij \sigma}}
               \frac{\delta \conj{\lambda}_{ij \sigma}}{\delta \rho^\sigma}
        \right] \\ =
    \sum_{ij} \kappa_{ij\sigma}
       \left( \frac12 \delta_{ij} - \Real \lambda_{ij\sigma} \right)
       \ketbra{\phi_{i\sigma}}{\phi_{j\sigma}}.
\end{multline}
Diagonalizing $(h^\sigma + \Delta h^\sigma)$ yields corrected canonical orbitals
$\ket{\wt{\psi}_{n \sigma}}$ and orbital energies $\wt{\eps}_{n \sigma}$. These
eigenvalues are interpreted as quasiparticle energies that produce the
LOSC-corrected band structure, including the fundamental gap
\cite{cohen2008a, yang_ayers_2024, yang_fan_2024a, yang_fan_2024}.

\section*{Results and Discussion}
There are two parameters in olLOSC: the balance between spatial and energy
localization $\gamma$ and the fraction $\lambda$ of von Weizs\"acker kinetic
energy. We found that $\gamma = 0.30$ and $\lambda = 0.75$ gave the best
results when comparing between molecules and materials. Increasing the fraction
$\lambda$ of von Weizs\"acker kinetic energy tends to increase the fundamental
gap slightly (decreases the screened response $\braket{\delta\rho}{V}$), while
increasing the spatial delocalization $\gamma$ has the opposite effect. We
note that most previous implementations of the LOSC method use
$\gamma = 0.47714$, but in some cases this decreased spatial localization leads
the orbitalets of materials to become trapped in local minima (see the Supporting
Information for details). All results in this section use
$(\gamma, \lambda) = (0.30, 0.75)$. Underlying density functional calculations
are performed with the PBE functional \cite{perdew1996}.

\begin{figure*}[ht]
    \begin{subfigure}[t]{0.45\textwidth} 
        \centering
        \includegraphics[width=\linewidth]{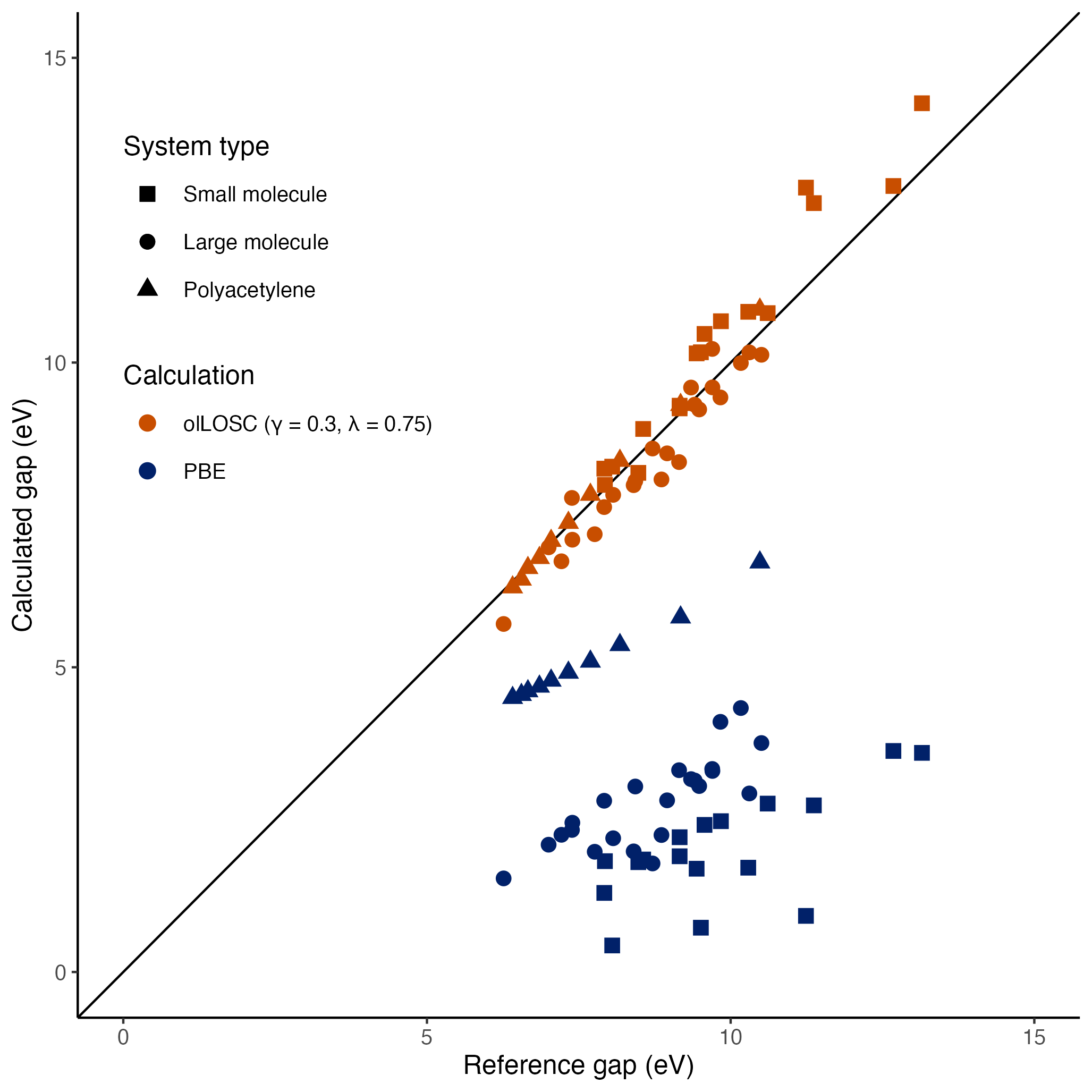}
        \subcaption{olLOSC in molecules.}
        \label{fig:cve-mol}
    \end{subfigure}
    \hfill
    \begin{subfigure}[t]{0.45\textwidth}
        \centering
        \includegraphics[width=\linewidth]{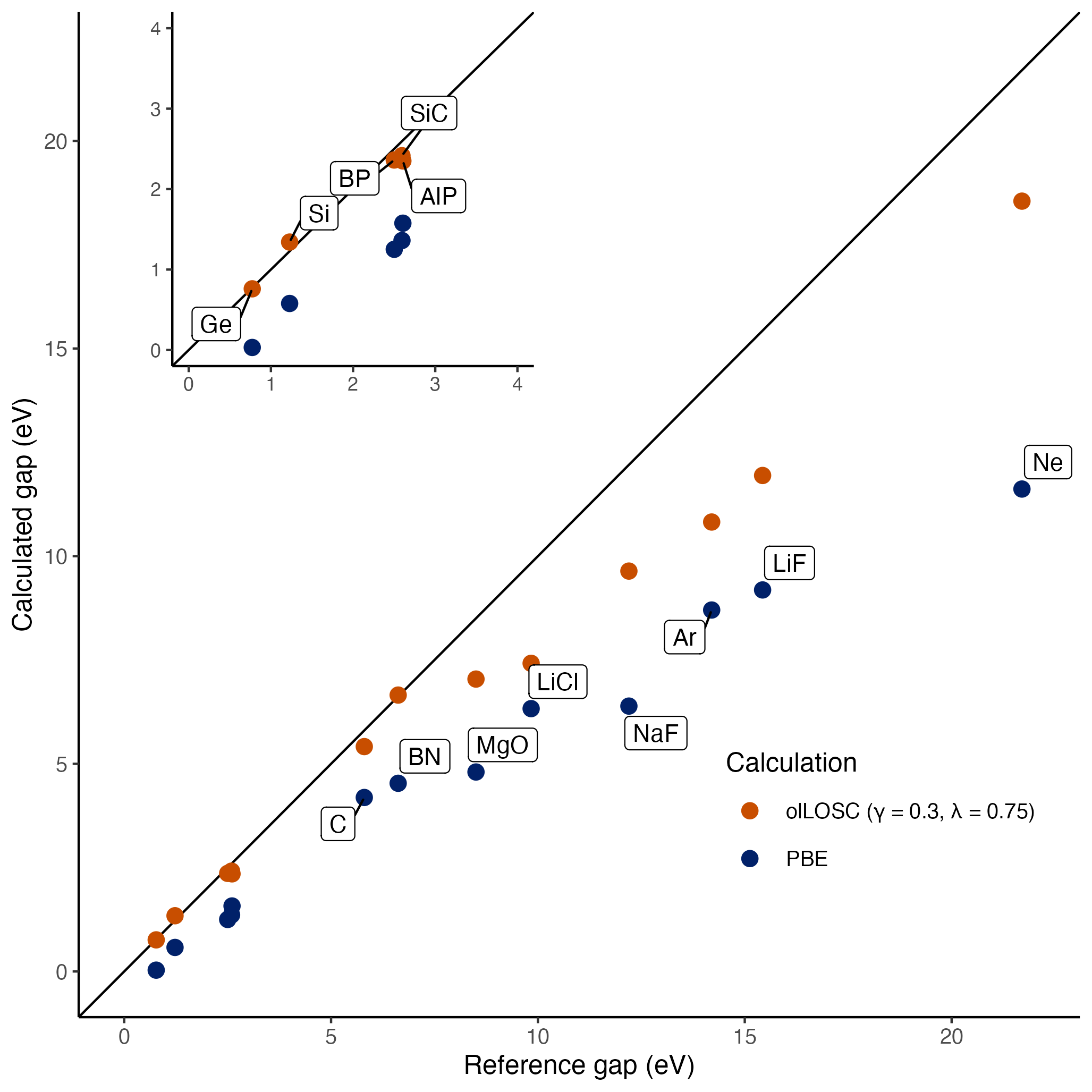}
        \subcaption{olLOSC in materials. Inset: gaps
                    $\leq \SI{4}{\electronvolt}$.}
        \label{fig:cve-bulk}
    \end{subfigure}
    \caption{DFA (blue) and olLOSC (orange) fundamental gaps (ionization
             potentials for polyacetylenes), in (a) molecules and polymers,
             (b) materials. Reference values are from (a) CCSD(T) and RASPT2,
             (b) experiments, corrected for zero-point renormalization. The
             orbitalet space/energy mixing parameter $\gamma = 0.30$, and the
             kinetic energy functional is
             $\TF + 75\%\,\vW$ ($\lambda = 0.75$).}
    \label{fig:cve}
\end{figure*}

We tested the molecular implementation of olLOSC on the fundamental gaps of 
small and large molecules and on the ionization potentials of a set of polymers.
The small molecules are a 17-member subset of the G2/97 test set
\cite{Curtiss1997} such that
$\eps_{\text{PBE}}^{\LUMO} \leq \SI{-2}{\electronvolt}$. The reference values
for small molecules are fundamental gaps calculated by CCSD(T) in the
Supporting Information of \cite{su2020}. Large molecules are organic acceptors
with reference CCSD(T) gaps from \cite{Richard2016}. The polymers we study are
polyacetylene, \ce{H(C2H2)_nH}, where $1 \leq n \leq 10$; reference ionization
potentials are computed with RASPT2 in \cite{Shahi2009}.

The bulk implementation was tested on a set of thirteen semiconductors and
large-gap insulators, with lattice constants given by the experimental
values found in \cite{heyd2005}. The reference values are electronic
gaps---experimental gaps adjusted for lattice-coupling effects via
zero-point renormalization \cite{miglio2020, shang2021, engel2022};\footnote{
Ne and Ar are not adjusted for zero-point renormalization.} they
range from less than 1 to more than 21 \si{\electronvolt}.

\begin{table}[t!] \label{tab:mae}
\centering
\caption{Relative errors from olLOSC ($\gamma = 0.30$, $\lambda = 0.75$)
of gaps (molecules, materials) and ionization potentials (polymers). 
Materials' experimental gaps are renormalized for zero-point energy. Mean
signed errors (MSE) and mean absolute errors (MAE) are in \si{\electronvolt};
mean absolute relative errors (MARE) are percentages.}
\begin{tabular*}{\hsize}{@{\extracolsep{\fill}}rrrr}
\toprule
 & MSE (\si{\electronvolt}) & MAE (\si{\electronvolt}) & MARE ($\%$) \\
\midrule
& \multicolumn{3}{c}{Molecules} \\
\cmidrule{2-4}
Small                               & $0.528$  & $0.562$ & $5.48$ \\
Large                               & $-0.251$ & $0.352$ & $4.18$ \\
All                                 & $0.080$  & $0.441$ & $4.73$ \\
\midrule
\ce{H(C2H2)_nH} & \multicolumn{3}{c}{Polymers (IP only)} \\
\cmidrule{2-4} 
($1 \leq n \leq 10$) & $ 0.064$ & $0.125$ & $ 1.52$ \\ 
\midrule
& \multicolumn{3}{c}{Materials} \\
\cmidrule{2-4}
Gap $\leq \SI{8}{\electronvolt}$    & $-0.118$ & $0.160 $ & $5.77$ \\
Gap $>    \SI{8}{\electronvolt}$    & $-2.742$ & $2.742$ & $20.6$ \\
All                                 & $-1.329$ & $1.352$ & $12.6$ \\
\bottomrule
\end{tabular*}
\end{table}

As is clear from Fig.~\ref{fig:cve-mol}, olLOSC provides substantial
improvement in molecular ionization potentials and fundamental gaps compared
with PBE. With the chosen parameters, we see little systematic error and a
greatly improved absolute error. For example, the ionization potential of
polyacetylene oligomers (triangles) are both more accurate and better follow
the true trendline as their length is increased. Both small and large
molecules have far better band gaps. We note a slight overestimation in the
gap of molecules with especially large band gaps, in contrast to the
situation for bulk insulators (see below).

The band gaps of semiconductors and moderately gapped insulators, up to about
\SI{8}{\electronvolt}, are similarly improved (Fig.~\ref{fig:cve-bulk}). For
large-gapped insulators (in this dataset, LiCl, NaF, LiF, Ar, and Ne), band
gaps are systematically underestimated, although the olLOSC predictions are
still better than those of PBE.

The behavior of olLOSC in large-gapped systems differs between molecules and
materials. In molecules, we see a small overestimation of the gap; in
materials, the opposite effect is observed. Thus, we cannot tune the olLOSC
parameters to eliminate the inaccuracy; at best, we can achieve a compromise.
We attribute the difference to the limited accuracy of the relatively simple
Thomas--Fermi--von Weizs\"acker kinetic energy functional. Investigating
more sophisticated functionals, such as those discussed in \cite{mi2023}, is
a promising avenue for future work. However, the TFvW functional is sufficiently
accurate in the regime of interest for interfacial systems: inorganic and
organic molecules, especially those with modest gaps, and semiconductors.
We also expect that olLOSC with $f_{\TF\vW}$ will perform adequately when
extended to metals, which must account for perturbations to the Fermi energy
\cite[section II.C.4]{baroni2001}.

\section*{Conclusion}
olLOSC corrects delocalization error consistently across system sizes, from
small molecules to bulk semiconductors and insulators (with small and moderate
band gaps). Besides this, the LOSC family has three crucial ingredients, each
necessary for accurately calculating interfaces with DFT: correcting the band
structure or quasiparticle energies \cite{mahler2022b, yu2025, williams2024};
modifying the total energy size-consistently \cite{li2018, su2020}; and
correcting the charge density \cite{mei2020a}. 
It is equally important to offer corrections to the total energy
to describe chemical reactions occurring at interfaces. More fundamentally,
approximate density functional calculations with delocalization error are not
size-consistent \cite{li2018}, an error that could be exacerbated in interfaces.
Finally, changes to the charge density are necessary to capture the charge
transfer that can occur across an interface
\cite{zhu2009, bai2014, zhu2015, zhu2015a, otero2017}. To the best of our
knowledge, LOSC is the only method that can satisfy all three requirements in a
unified approximation. In future work, we will explore the impact of other
kinetic energy functionals \cite{mi2023} and add the
long-wavelength correction to the Fermi energy in metals. Finally, we will
implement the procedure of \cite{mei2020a} in olLOSC to obtain a
self-consistently optimized density. We expect that olLOSC will develop into a
robust and efficient DFT approach for modeling interfaces between molecules and
surfaces.


\section*{Materials and methods}

\subsection*{Curvature implementation: Molecules}
In molecular systems, we discretize $\chi_{\TF\vW}$ and $\chi_{\ol}$ in a
low-rank function space inspired by the resolution-of-the-identity (RI)
approximation \cite{whitten_1973,dunlap_etal_1979,vahtras_etal_1993,ren2012a};
in the resulting auxiliary basis, we can invert them directly. We call our
auxiliary, atom-centered basis functions $P_\sigma(\br)$. In the $P_\sigma$
basis, the matrix elements of the orbital-free kinetic energy kernel are
\begin{equation}
    f_{\TF\vW}^{P\sigma,Q\tau} =
    \left[\chi_{\TF\vW}^{P\sigma,Q\tau}\right]^{-1} =
    \mel{P_\sigma}{f_{\TF\vW}^{\sigma\tau}}{Q_\tau}.
\end{equation}
The Hartree kernel is similarly
\begin{equation}
    f_{\Har}^{P\sigma,Q\tau} = \mel{P_\sigma}{f_{\Har}^{\sigma\tau}}{Q_\tau} =
    \iint d\br\, d\br'\, \frac{P_\sigma(\br) Q_\tau(\br')}{\abs{\br-\br'}};
\end{equation}
these matrix elements are computed readily with the RI-V machinery of
\cite{ren2012a}. Applying this discretization and the partial RPA to solve
\eqref{eq:chi-inv}, we obtain the orbital-free linear-response function
\begin{equation}
    \chi_{\ol}^{P\sigma,Q\tau} = 
    (f_{\Har}^{P\sigma,Q\tau} + f_{\TF\vW}^{P\sigma,Q\tau})^{-1}.
\end{equation}

As mentioned before, we must impose charge conservation in $\chi$. However,
given $\chi_{\ol}^{P\sigma,Q\tau}$, we can conserve charge without computing
the chemical potential perturbation $\delta\mu^\sigma$ directly. The total charge
of $P_\sigma(\br)$ is
\begin{equation}
    d_{P\sigma} = \int d\br\, P_\sigma(\br);
\end{equation}
then, following \cite{york1996}, the charge-conserving linear-response function
$\bar{\chi}$ has matrix elements
\begin{equation}
    \mel{P_\sigma}{\bar{\chi}}{Q_\tau} =
    \bar{\chi}_\text{ol}^{P\sigma,Q\tau} = 
    \frac{\sum_{RS}
            \chi_\text{ol}^{P\sigma,R\tau} 
            d_{R\tau} d_{S\sigma} 
            \chi^{S\sigma,Q\tau}_\text{ol}
         }
         {\sum_{RS} d_{R\sigma} \chi_\text{ol}^{R\sigma,S\tau} d_{S\tau}} 
        - \chi_\text{ol}^{P\sigma,Q\tau}.
\end{equation}
The olLOSC curvature is thus computed as
\begin{equation} \label{eq:ri-kappa}
    \kappa_{ij\sigma}^{\ol} =  
    \mel{\rho_{i\sigma}}{f_{\Hxc}^{\sigma\sigma}}{\rho_{j\sigma}} + 
        \sum_{P\mu,Q\nu} 
            f_\Hxc^{i\sigma,P\mu} 
            \bar{\chi}_{\ol}^{P\mu,Q\nu}
            f_\Hxc^{j\sigma,Q\nu},
\end{equation}
where
$f_\Hxc^{i\sigma,P\mu} = \mel{\rho_{i\sigma}}{f_{\Hxc}^{\sigma\mu}}{P_\mu}$
is a Hxc kernel matrix element between the orbitalet
and auxiliary bases.

\subsection*{Curvature implementation: Materials}
Following \cite{timrov2018, colonna2022}, we decompose
\begin{equation} \label{eq:kap-mat}
\begin{split}
    \kappa_{ij\sigma}^{\ol} &= 
    \mel{\rho_{i\sigma}}{f_{\Hxc}}{\rho_{j\sigma}} +
        \sum_{\mu\nu} 
            \mel{\rho_{i\sigma}}
                {f_{\Hxc}^{\sigma\mu} \chi_{\ol}^{\mu\nu} f_{\Hxc}^{\nu\sigma}}
                {\rho_{j\sigma}} \\ &=
    \braket{\rho_{i\sigma}}{V_{j\sigma}} + 
        \sum_\nu \braket{\delta\rho_{i\nu}}{V_{j\nu}},
\end{split}
\end{equation}
where the potential $\ket{V_{j\nu}}$ and screened density response
$\ket{\delta\rho_{i\nu}}$ are given by
\begin{equation}
\begin{split}
    V_{j\nu}(\br) &= 
    \int d\br'\, f_{\Hxc}^{\nu\sigma}(\br, \br') \rho_{j\sigma}(\br'), \\
    \delta\rho_{i\nu}(\br) &=
    \sum_\mu \iint d\br'\, d\br''\,
        \chi_{\ol}^{\mu\nu}(\br, \br') f_{\Hxc}^{\mu\sigma}(\br', \br'')
            \rho_{i\sigma}(\br'').
\end{split}
\end{equation}
We compute $\ket{\rho_{i\sigma}}$ and $\ket{V_{j\nu}}$ directly. But where 
$\ket{\delta\rho_{i\nu}}$ is computed in KCW \cite{colonna2022} and lrLOSC via
perturbations to the occupied Kohn--Sham orbitals, we obtain it more directly
in olLOSC. We identify the perturbation $\delta\rho^\nu$ of
\eqref{eq:euler-yorkyang} with $\ket{\delta\rho_{i\nu}}$, yielding the linear
system
\begin{equation} \label{eq:drho-bulk}
    \begin{pmatrix}
        f_{\TF\vW}^{\alpha\alpha} + f_{\Har}^{\alpha\alpha} &
        f_{\Har}^{\alpha\beta} \\
        f_{\Har}^{\beta\alpha} &
        f_{\TF\vW}^{\beta\beta} + f_{\Har}^{\beta\beta}
    \end{pmatrix}
    \begin{pmatrix}
        \ket{\delta\rho_{i\alpha}} \\ \ket{\delta\rho_{i\beta}}
    \end{pmatrix} =
    \begin{pmatrix}
        \delta\mu^\alpha - \ket{\delta v^\alpha} \\
        \delta\mu^\beta - \ket{\delta v^\beta}
    \end{pmatrix}.
\end{equation}
(Without the partial RPA, we would replace $f_{\Har}$ by $f_{\Hxc}$.) Choosing
$\delta\rho^\nu = \ket{\delta\rho_{i\nu}}$ also fixes the perturbation to the
external potential, so that $\delta v^\tau = \ket{V_{i\tau}}$.

The matrix on the left-hand side of \eqref{eq:drho-bulk} is nonlocal in either
real or reciprocal space, so it is too large even to store directly. We can,
however, compute its action on a vector and use an iterative solver to obtain
$\ket{\delta\rho_{i\alpha}}$ and $\ket{\delta\rho_{i\beta}}$ simultaneously.
Under the partial RPA, the matrix is positive semidefinite, so we solve it with
the modified conjugate gradient algorithm of \cite{gould2001}. Instead of
directly computing $\delta\mu^\nu$, which changes each iteration, we impose
charge conservation as a constraint via a projected preconditioner
\cite{gould2001}. Without the partial RPA, a linear solver for indefinite
matrices such as MINRES-QLP \cite{choi2006, choi2011, choi2014} would be
required.

\subsection*{Computational details}
olLOSC for molecules is implemented in the in-house code \texttt{QM$^4$D},
which uses Gaussian-type orbitals. Small and large molecules are computed with
correlation-consistent triple-zeta Dunning basis sets augmented with diffuse
functions (aug-cc-pVTZ) \cite{dunning1989a, kendall1992a}; the polymers use
the same basis sets without augmentation (cc-pVTZ). We use (aug)-cc-pVTZ-RIFIT
as the auxiliary basis for $\chi_{\ol}$.

olLOSC for materials is implemented as a module in a locally maintained fork
\cite{williams_yang_2025} of the open-source \texttt{Quantum ESPRESSO} package
\cite{giannozzi2009, giannozzi2017}, version 7.2. The bulk orbitalets are
computed with a fork \cite{mahler2024a} of \texttt{wannier90}
\cite{mostofi2008, mostofi2014, pizzi2020} version 3.1.0.
\texttt{Quantum ESPRESSO}'s \texttt{PWscf} code is used for the underlying DFT
calculations, which use the PBE functional \cite{perdew1996} and optimized
Vanderbilt norm-conserving pseudopotentials with scalar relativistic
corrections \cite{hamann2013} downloaded from the ABINIT PseudoDojo
(\url{http://www.pseudo-dojo.org/}). We use a kinetic energy cutoff of
\SI{75}{\rydberg}; unless otherwise specified, we sample the Brillouin zone with
a $6 \times 6 \times 6$ Monkhorst--Pack (uniform) grid \cite{monkhorst1976}
centered at the origin $\Gamma$ of reciprocal space. The macroscopic dielectric
constant $\epsilon_\infty$, needed for the Gygi--Baldereschi correction
\cite{gygi1986} to the Coulomb singularity, is computed by density functional
perturbation theory \cite{baroni2001} in the \texttt{PHonon} module of
\texttt{Quantum ESPRESSO}, with the same parameters. The eigenvalues of the
olLOSC Hamiltonian, used for corrected band structures, are computed along the
paths specified by \cite{setyawan2010} for the systems' respective lattices.

\subsection*{Computational complexity}
Computing the curvature---in particular, the screened response---is the most
computationally demanding part of a LOSC calculation. Thus, we restrict our
complexity analysis to this bottleneck step.

In materials, the monochromatic decomposition means that the density response
$\{\ket{\delta\rho_{i \tau}}\}_\tau$, which encodes the screening, must be
computed $N_w \times N_k$ times, where $N_w$ is the number of DLWFs and
$N_k$ is the number of $\bk$-points sampled in the Brillouin zone. Each 
calculation of $\ket{\delta\rho_{i \tau}}$ requires a matrix inversion, computed
by the projected preconditioned conjugate gradient \cite{gould2001}; each
iteration scales linearly in the number of plane waves $N_G$ (the length of
$\ket{\delta\rho_{i\tau}}$) \cite{choi2014}. Each iteration also requires
several conversions between real and reciprocal space (because the kinetic
energy kernel is semilocal in $\br$ former, while the Coulomb kernel is local
in $\bG$) via the fast Fourier transform, which scales as $N_G \log N_G$.
Thus, olLOSC for materials scales as $\Ord{(N_w N_k N_G \log{N_G})}$. This saves
a factor of $N_k$ relative to lrLOSC \cite{williams2024}; even with the
Sternheimer equation bypassing direct\ computation of $\chi_s$, constructing
$\ket{\delta\rho_{i\tau}}$ in lrLOSC requires coupled pairs of $\bk$-points.

In molecules, the interacting linear response function $\chi$ is computed
directly; the bottleneck step is inverting $\chi$ and $\chi_{\ol}$. Within the
orbital-free-approximation, all involved matrices are evaluated in a
$N_{\aux} \times N_{\aux}$ basis of auxiliary functions, where $N_{\aux}$ is
proportional to the number of orbitalets $N_w$. We invert $\chi$ with an LU
decomposition, so olLOSC for molecules scales as $\Ord(N_w^3)$. In addition to
the evaluation of the linear response function, the numerical integration of
the exchange-correlation kernel can be a potential time-determining step
because of its relatively large pre-factor. In olLOSC, this integral is
constructed as shown in Equation \ref{eq:ri-kappa}; it scales as
$\Ord{N_w^2 N_\text{grid}}$, where $N_\text{grid}$ is the number of grid points.
lrLOSC, by comparison, requires the inversion of a rank-4 tensor instead of a
rank-2 tensor because $\chi_s$ (unlike $\chi_{\TF\vW}$) depends on the
difference between occupied and virtual orbital energies.

%% file: olLOSC.bib
@article{adhikari2020,
  title = {The {{Fermi}}--{{L{\"o}wdin}} Self-Interaction Correction for Ionization Energies of Organic Molecules},
  author = {Adhikari, Santosh and Santra, Biswajit and Ruan, Shiqi and Bhattarai, Puskar and Nepal, Niraj K. and Jackson, Koblar A. and Ruzsinszky, Adrienn},
  year = {2020},
  month = nov,
  journal = {The Journal of Chemical Physics},
  volume = {153},
  number = {18},
  pages = {184303},
  issn = {0021-9606},
  doi = {10.1063/5.0024776},
  urldate = {2024-04-24}
}

@article{anisimov1997,
  title = {First-Principles Calculations of the Electronic Structure and Spectra of Strongly Correlated Systems: The {{LDA}}+ {{U}} Method},
  shorttitle = {First-Principles Calculations of the Electronic Structure and Spectra of Strongly Correlated Systems},
  author = {Anisimov, Vladimir I. and Aryasetiawan, F. and Lichtenstein, A. I.},
  year = {1997},
  month = jan,
  journal = {J. Phys.: Condens. Matter},
  volume = {9},
  number = {4},
  pages = {767},
  issn = {0953-8984},
  doi = {10.1088/0953-8984/9/4/002},
  urldate = {2024-04-12},
  langid = {english}
}

@article{ayers2008,
  title = {The Dependence on and Continuity of the Energy and Other Molecular Properties with Respect to the Number of Electrons},
  author = {Ayers, Paul W.},
  year = {2008},
  month = jan,
  journal = {J Math Chem},
  volume = {43},
  number = {1},
  pages = {285--303},
  issn = {1572-8897},
  doi = {10.1007/s10910-006-9195-5},
  urldate = {2021-02-09},
  langid = {english}
}

@article{bai2014,
  title = {Charge Transfer Kinetics at the Solid--Solid Interface in Porous Electrodes},
  author = {Bai, Peng and Bazant, Martin Z.},
  year = {2014},
  month = apr,
  journal = {Nat Commun},
  volume = {5},
  number = {1},
  pages = {3585},
  publisher = {Nature Publishing Group},
  issn = {2041-1723},
  doi = {10.1038/ncomms4585},
  urldate = {2024-04-23},
  copyright = {2014 Springer Nature Limited},
  langid = {english}
}

@article{baroni2001,
  title = {Phonons and Related Crystal Properties from Density-Functional Perturbation Theory},
  author = {Baroni, Stefano and {de Gironcoli}, Stefano and Dal Corso, Andrea and Giannozzi, Paolo},
  year = {2001},
  month = jul,
  journal = {Rev. Mod. Phys.},
  volume = {73},
  number = {2},
  pages = {515--562},
  publisher = {American Physical Society},
  doi = {10.1103/RevModPhys.73.515},
  urldate = {2022-04-07}
}

@article{bloch1929,
  title = {{{\"U}ber die Quantenmechanik der Elektronen in Kristallgittern}},
  author = {Bloch, Felix},
  year = {1929},
  month = jul,
  journal = {Z. Physik},
  volume = {52},
  number = {7},
  pages = {555--600},
  issn = {0044-3328},
  doi = {10.1007/BF01339455},
  urldate = {2023-09-18},
  langid = {ngerman}
}

@article{borghi2014,
  title = {Koopmans-Compliant Functionals and Their Performance against Reference Molecular Data},
  author = {Borghi, Giovanni and Ferretti, Andrea and Nguyen, Ngoc Linh and Dabo, Ismaila and Marzari, Nicola},
  year = {2014},
  month = aug,
  journal = {Phys. Rev. B},
  volume = {90},
  number = {7},
  pages = {075135},
  publisher = {American Physical Society},
  doi = {10.1103/PhysRevB.90.075135},
  urldate = {2022-03-22}
}

@article{bryenton2022,
  title = {Delocalization Error: {{The}} Greatest Outstanding Challenge in Density-Functional Theory},
  shorttitle = {Delocalization Error},
  author = {Bryenton, Kyle R. and Adeleke, Adebayo A. and Dale, Stephen G. and Johnson, Erin R.},
  year = {2022},
  journal = {WIREs Computational Molecular Science},
  volume = {n/a},
  number = {n/a},
  pages = {e1631},
  issn = {1759-0884},
  doi = {10.1002/wcms.1631},
  urldate = {2022-07-19},
  langid = {english}
}

@article{chattaraj2007,
  title = {Local Hardness: A Critical Account},
  shorttitle = {Local Hardness},
  author = {Chattaraj, Pratim K. and Roy, Debesh R. and Geerlings, Paul and {Torrent-Sucarrat}, Miquel},
  year = {2007},
  month = dec,
  journal = {Theor Chem Account},
  volume = {118},
  number = {5},
  pages = {923--930},
  issn = {1432-2234},
  doi = {10.1007/s00214-007-0373-8},
  urldate = {2022-12-16},
  langid = {english}
}

@phdthesis{choi2006,
  title = {Iterative {{Methods}} for {{Singular Linear Equations}} and {{Least-Squares Problems}}},
  author = {Choi, Sou-Cheng},
  year = {2006},
  month = dec,
  address = {Stanford, CA},
  langid = {english},
  school = {Stanford University}
}

@article{choi2011,
  title = {{{MINRES-QLP}}: {{A Krylov Subspace Method}} for {{Indefinite}} or {{Singular Symmetric Systems}}},
  shorttitle = {{{MINRES-QLP}}},
  author = {Choi, Sou-Cheng T. and Paige, Christopher C. and Saunders, Michael A.},
  year = {2011},
  month = jan,
  journal = {SIAM J. Sci. Comput.},
  volume = {33},
  number = {4},
  pages = {1810--1836},
  publisher = {{Society for Industrial and Applied Mathematics}},
  issn = {1064-8275},
  doi = {10.1137/100787921},
  urldate = {2024-01-19}
}

@article{choi2014,
  title = {Algorithm 937: {{MINRES-QLP}} for Symmetric and {{Hermitian}} Linear Equations and Least-Squares Problems},
  shorttitle = {Algorithm 937},
  author = {Choi, Sou-Cheng T. and Saunders, Michael A.},
  year = {2014},
  month = mar,
  journal = {ACM Trans. Math. Softw.},
  volume = {40},
  number = {2},
  pages = {16:1--16:12},
  issn = {0098-3500},
  doi = {10.1145/2527267},
  urldate = {2024-01-19}
}

@article{cococcioni2005,
  title = {Linear Response Approach to the Calculation of the Effective Interaction Parameters in the {$\mathrm{LDA}+U$} Method},
  author = {Cococcioni, Matteo and {de Gironcoli}, Stefano},
  year = {2005},
  month = jan,
  journal = {Phys. Rev. B},
  volume = {71},
  number = {3},
  pages = {035105},
  publisher = {American Physical Society},
  doi = {10.1103/PhysRevB.71.035105},
  urldate = {2021-02-23}
}

@article{cohen2007a,
  title = {Development of Exchange-Correlation Functionals with Minimal Many-Electron Self-Interaction Error},
  author = {Cohen, Aron J. and {Mori-S{\'a}nchez}, Paula and Yang, Weitao},
  year = {2007},
  month = may,
  journal = {The Journal of Chemical Physics},
  volume = {126},
  number = {19},
  pages = {191109},
  issn = {0021-9606},
  doi = {10.1063/1.2741248},
  urldate = {2023-11-29}
}

@article{cohen2008,
  title = {Insights into {{Current Limitations}} of {{Density Functional Theory}}},
  author = {Cohen, A. J. and {Mori-Sanchez}, P. and Yang, W.},
  year = {2008},
  month = aug,
  journal = {Science},
  volume = {321},
  number = {5890},
  pages = {792--794},
  issn = {0036-8075, 1095-9203},
  doi = {10.1126/science.1158722},
  urldate = {2020-06-16},
  langid = {english}
}

@article{cohen2008a,
  title = {Fractional Charge Perspective on the Band Gap in Density-Functional Theory},
  author = {Cohen, Aron J. and {Mori-S{\'a}nchez}, Paula and Yang, Weitao},
  year = {2008},
  month = mar,
  journal = {Phys. Rev. B},
  volume = {77},
  number = {11},
  pages = {115123},
  publisher = {American Physical Society},
  doi = {10.1103/PhysRevB.77.115123},
  urldate = {2021-02-12}
}

@article{cohen2012,
  title = {Challenges for {{Density Functional Theory}}},
  author = {Cohen, Aron J. and {Mori-S{\'a}nchez}, Paula and Yang, Weitao},
  year = {2012},
  month = jan,
  journal = {Chem. Rev.},
  volume = {112},
  number = {1},
  pages = {289--320},
  issn = {0009-2665, 1520-6890},
  doi = {10.1021/cr200107z},
  urldate = {2020-02-21},
  langid = {english}
}

@article{colonna2018,
  title = {Screening in {{Orbital-Density-Dependent Functionals}}},
  author = {Colonna, Nicola and Nguyen, Ngoc Linh and Ferretti, Andrea and Marzari, Nicola},
  year = {2018},
  month = may,
  journal = {J. Chem. Theory Comput.},
  volume = {14},
  number = {5},
  pages = {2549--2557},
  publisher = {American Chemical Society},
  issn = {1549-9618},
  doi = {10.1021/acs.jctc.7b01116},
  urldate = {2022-02-22}
}

@article{colonna2022,
  title = {Koopmans {{Spectral Functionals}} in {{Periodic Boundary Conditions}}},
  author = {Colonna, Nicola and De Gennaro, Riccardo and Linscott, Edward and Marzari, Nicola},
  year = {2022},
  month = sep,
  journal = {J. Chem. Theory Comput.},
  volume = {18},
  number = {9},
  pages = {5435--5448},
  publisher = {American Chemical Society},
  issn = {1549-9618},
  doi = {10.1021/acs.jctc.2c00161},
  urldate = {2022-10-17}
}

@article{dabo2010,
  title = {Koopmans' Condition for Density-Functional Theory},
  author = {Dabo, Ismaila and Ferretti, Andrea and Poilvert, Nicolas and Li, Yanli and Marzari, Nicola and Cococcioni, Matteo},
  year = {2010},
  month = sep,
  journal = {Phys. Rev. B},
  volume = {82},
  number = {11},
  pages = {115121},
  publisher = {American Physical Society},
  doi = {10.1103/PhysRevB.82.115121},
  urldate = {2021-12-16}
}

@article{damle2015,
  title = {Compressed {{Representation}} of {{Kohn}}--{{Sham Orbitals}} via {{Selected Columns}} of the {{Density Matrix}}},
  author = {Damle, Anil and Lin, Lin and Ying, Lexing},
  year = {2015},
  month = apr,
  journal = {J. Chem. Theory Comput.},
  volume = {11},
  number = {4},
  pages = {1463--1469},
  issn = {1549-9618, 1549-9626},
  doi = {10.1021/ct500985f},
  urldate = {2020-07-13},
  langid = {english}
}

@article{damle2017a,
  title = {{{SCDM-k}}: {{Localized}} Orbitals for Solids via Selected Columns of the Density Matrix},
  shorttitle = {{{SCDM-k}}},
  author = {Damle, Anil and Lin, Lin and Ying, Lexing},
  year = {2017},
  month = apr,
  journal = {Journal of Computational Physics},
  volume = {334},
  pages = {1--15},
  issn = {00219991},
  doi = {10.1016/j.jcp.2016.12.053},
  urldate = {2020-07-13},
  langid = {english}
}

@article{dellasala2022,
  title = {Orbital-Free Methods for Plasmonics: {{Linear}} Response},
  shorttitle = {Orbital-Free Methods for Plasmonics},
  author = {Della Sala, Fabio},
  year = {2022},
  month = sep,
  journal = {J. Chem. Phys.},
  volume = {157},
  number = {10},
  pages = {104101},
  publisher = {American Institute of Physics},
  issn = {0021-9606},
  doi = {10.1063/5.0100797},
  urldate = {2023-02-02}
}

@article{engel2022,
  title = {Zero-Point Renormalization of the Band Gap of Semiconductors and Insulators Using the Projector Augmented Wave Method},
  author = {Engel, Manuel and Miranda, Henrique and Chaput, Laurent and Togo, Atsushi and Verdi, Carla and Marsman, Martijn and Kresse, Georg},
  year = {2022},
  month = sep,
  journal = {Phys. Rev. B},
  volume = {106},
  number = {9},
  pages = {094316},
  publisher = {American Physical Society},
  doi = {10.1103/PhysRevB.106.094316},
  urldate = {2023-01-19}
}

@article{fermi1927,
  title = {Un {{Metodo Statistico}} per La {{Determinazione}} Di Alcune {{Priopriet{\`a}}} Dell'{{Atomo}}},
  author = {Fermi, Enrico},
  year = {1927},
  journal = {Rend. Accad. Naz. Lincei.},
  volume = {6},
  pages = {602--607}
}

@article{foster1960,
  title = {Canonical {{Configurational Interaction Procedure}}},
  author = {Foster, J. M. and Boys, S. F.},
  year = {1960},
  month = apr,
  journal = {Rev. Mod. Phys.},
  volume = {32},
  number = {2},
  pages = {300--302},
  publisher = {American Physical Society},
  doi = {10.1103/RevModPhys.32.300},
  urldate = {2021-02-16}
}

@article{giannozzi2009,
  title = {{{QUANTUM ESPRESSO}}: A Modular and Open-Source Software Project for Quantum Simulations of Materials},
  shorttitle = {{{QUANTUM ESPRESSO}}},
  author = {Giannozzi, Paolo and Baroni, Stefano and Bonini, Nicola and Calandra, Matteo and Car, Roberto and Cavazzoni, Carlo and Ceresoli, Davide and Chiarotti, Guido L and Cococcioni, Matteo and Dabo, Ismaila and Dal Corso, Andrea and {de Gironcoli}, Stefano and Fabris, Stefano and Fratesi, Guido and Gebauer, Ralph and Gerstmann, Uwe and Gougoussis, Christos and Kokalj, Anton and Lazzeri, Michele and {Martin-Samos}, Layla and Marzari, Nicola and Mauri, Francesco and Mazzarello, Riccardo and Paolini, Stefano and Pasquarello, Alfredo and Paulatto, Lorenzo and Sbraccia, Carlo and Scandolo, Sandro and Sclauzero, Gabriele and Seitsonen, Ari P and Smogunov, Alexander and Umari, Paolo and Wentzcovitch, Renata M},
  year = {2009},
  month = sep,
  journal = {J. Phys.: Condens. Matter},
  volume = {21},
  number = {39},
  pages = {395502},
  issn = {0953-8984, 1361-648X},
  doi = {10.1088/0953-8984/21/39/395502},
  urldate = {2020-04-25},
  langid = {english}
}

@article{giannozzi2017,
  title = {Advanced Capabilities for Materials Modelling with {{Quantum ESPRESSO}}},
  author = {Giannozzi, P and Andreussi, O and Brumme, T and Bunau, O and Buongiorno Nardelli, M and Calandra, M and Car, R and Cavazzoni, C and Ceresoli, D and Cococcioni, M and Colonna, N and Carnimeo, I and Dal Corso, A and {de Gironcoli}, S and Delugas, P and DiStasio, R A and Ferretti, A and Floris, A and Fratesi, G and Fugallo, G and Gebauer, R and Gerstmann, U and Giustino, F and Gorni, T and Jia, J and Kawamura, M and Ko, H-Y and Kokalj, A and K{\"u}{\c c}{\"u}kbenli, E and Lazzeri, M and Marsili, M and Marzari, N and Mauri, F and Nguyen, N L and Nguyen, H-V and {Otero-de-la-Roza}, A and Paulatto, L and Ponc{\'e}, S and Rocca, D and Sabatini, R and Santra, B and Schlipf, M and Seitsonen, A P and Smogunov, A and Timrov, I and Thonhauser, T and Umari, P and Vast, N and Wu, X and Baroni, S},
  year = {2017},
  month = nov,
  journal = {J. Phys.: Condens. Matter},
  volume = {29},
  number = {46},
  pages = {465901},
  issn = {0953-8984, 1361-648X},
  doi = {10.1088/1361-648X/aa8f79},
  urldate = {2020-04-25},
  langid = {english}
}

@article{gould2001,
  title = {On the {{Solution}} of {{Equality Constrained Quadratic Programming Problems Arising}} in {{Optimization}}},
  author = {Gould, Nicholas I. M. and Hribar, Mary E. and Nocedal, Jorge},
  year = {2001},
  month = jan,
  journal = {SIAM J. Sci. Comput.},
  volume = {23},
  number = {4},
  pages = {1376--1395},
  publisher = {{Society for Industrial and Applied Mathematics}},
  issn = {1064-8275},
  doi = {10.1137/S1064827598345667},
  urldate = {2023-10-31}
}

@article{gygi1986,
  title = {Self-Consistent {{Hartree-Fock}} and Screened-Exchange Calculations in Solids: {{Application}} to Silicon},
  shorttitle = {Self-Consistent {{Hartree-Fock}} and Screened-Exchange Calculations in Solids},
  author = {Gygi, F. and Baldereschi, A.},
  year = {1986},
  month = sep,
  journal = {Phys. Rev. B},
  volume = {34},
  number = {6},
  pages = {4405--4408},
  issn = {0163-1829},
  doi = {10.1103/PhysRevB.34.4405},
  urldate = {2020-01-15},
  langid = {english}
}

@article{gygi2003,
  title = {Computation of {{Maximally Localized Wannier Functions}} Using a Simultaneous Diagonalization Algorithm},
  author = {Gygi, Fran{\c c}ois and Fattebert, Jean-Luc and Schwegler, Eric},
  year = {2003},
  month = sep,
  journal = {Computer Physics Communications},
  volume = {155},
  number = {1},
  pages = {1--6},
  issn = {00104655},
  doi = {10.1016/S0010-4655(03)00315-1},
  urldate = {2020-02-16},
  langid = {english}
}

@article{hait2018,
  title = {Delocalization {{Errors}} in {{Density Functional Theory Are Essentially Quadratic}} in {{Fractional Occupation Number}}},
  author = {Hait, Diptarka and {Head-Gordon}, Martin},
  year = {2018},
  month = nov,
  journal = {J. Phys. Chem. Lett.},
  volume = {9},
  number = {21},
  pages = {6280--6288},
  issn = {1948-7185},
  doi = {10.1021/acs.jpclett.8b02417},
  urldate = {2020-06-25},
  langid = {english}
}

@article{hamann2013,
  title = {Optimized Norm-Conserving {{Vanderbilt}} Pseudopotentials},
  author = {Hamann, D. R.},
  year = {2013},
  month = aug,
  journal = {Phys. Rev. B},
  volume = {88},
  number = {8},
  pages = {085117},
  publisher = {American Physical Society},
  doi = {10.1103/PhysRevB.88.085117},
  urldate = {2021-02-17}
}

@article{hedin1965,
  title = {New {{Method}} for {{Calculating}} the {{One-Particle Green}}'s {{Function}} with {{Application}} to the {{Electron-Gas Problem}}},
  author = {Hedin, Lars},
  year = {1965},
  month = aug,
  journal = {Phys. Rev.},
  volume = {139},
  number = {3A},
  pages = {A796-A823},
  publisher = {American Physical Society},
  doi = {10.1103/PhysRev.139.A796},
  urldate = {2021-02-05}
}

@article{heyd2003,
  title = {Hybrid Functionals Based on a Screened {{Coulomb}} Potential},
  author = {Heyd, Jochen and Scuseria, Gustavo E. and Ernzerhof, Matthias},
  year = {2003},
  month = apr,
  journal = {J. Chem. Phys.},
  volume = {118},
  number = {18},
  pages = {8207--8215},
  publisher = {American Institute of Physics},
  issn = {0021-9606},
  doi = {10.1063/1.1564060},
  urldate = {2021-03-12}
}

@article{heyd2005,
  title = {Energy Band Gaps and Lattice Parameters Evaluated with the {{Heyd-Scuseria-Ernzerhof}} Screened Hybrid Functional},
  author = {Heyd, Jochen and Peralta, Juan E. and Scuseria, Gustavo E. and Martin, Richard L.},
  year = {2005},
  month = oct,
  journal = {J. Chem. Phys.},
  volume = {123},
  number = {17},
  pages = {174101},
  publisher = {American Institute of Physics},
  issn = {0021-9606},
  doi = {10.1063/1.2085170},
  urldate = {2021-05-05}
}

@article{heyd2006,
  title = {Erratum: ``{{Hybrid}} Functionals Based on a Screened {{Coulomb}} Potential'' [{{J}}. {{Chem}}. {{Phys}}. 118, 8207 (2003)]},
  shorttitle = {Erratum},
  author = {Heyd, Jochen and Scuseria, Gustavo E. and Ernzerhof, Matthias},
  year = {2006},
  month = jun,
  journal = {The Journal of Chemical Physics},
  volume = {124},
  number = {21},
  pages = {219906},
  issn = {0021-9606},
  doi = {10.1063/1.2204597},
  urldate = {2024-01-16}
}

@article{hirata1999,
  title = {Time-Dependent Density Functional Theory for Radicals {{An}} Improved Description of Excited States with Substantial Double Excitation Character},
  author = {Hirata, So and {Head-Gordon}, Martin},
  year = {1999},
  journal = {Chem. Phys. Lett.},
  volume = {302},
  pages = {375--382},
  doi = {10.1016/S0009-2614(99)00137-2},
  langid = {english}
}

@article{hohenberg1964,
  title = {Inhomogeneous {{Electron Gas}}},
  author = {Hohenberg, P. and Kohn, W.},
  year = {1964},
  month = nov,
  journal = {Phys. Rev.},
  volume = {136},
  number = {3B},
  pages = {B864-B871},
  issn = {0031-899X},
  doi = {10.1103/PhysRev.136.B864},
  urldate = {2020-08-03},
  langid = {english}
}

@article{holm1999,
  title = {Total {{Energies}} from \${\textbackslash}mathit\{\vphantom\}{{GW}}\vphantom\{\}\$ {{Calculations}}},
  author = {Holm, Bengt},
  year = {1999},
  month = jul,
  journal = {Phys. Rev. Lett.},
  volume = {83},
  number = {4},
  pages = {788--791},
  publisher = {American Physical Society},
  doi = {10.1103/PhysRevLett.83.788},
  urldate = {2024-04-04}
}

@article{huser2013a,
  title = {Quasiparticle {{GW}} Calculations for Solids, Molecules, and Two-Dimensional Materials},
  author = {H{\"u}ser, Falco and Olsen, Thomas and Thygesen, Kristian S.},
  year = {2013},
  month = jun,
  journal = {Phys. Rev. B},
  volume = {87},
  number = {23},
  pages = {235132},
  publisher = {American Physical Society},
  doi = {10.1103/PhysRevB.87.235132},
  urldate = {2024-04-23}
}

@article{hybertsen1986,
  title = {Electron Correlation in Semiconductors and Insulators: {{Band}} Gaps and Quasiparticle Energies},
  shorttitle = {Electron Correlation in Semiconductors and Insulators},
  author = {Hybertsen, Mark S. and Louie, Steven G.},
  year = {1986},
  month = oct,
  journal = {Phys. Rev. B},
  volume = {34},
  number = {8},
  pages = {5390--5413},
  publisher = {American Physical Society},
  doi = {10.1103/PhysRevB.34.5390},
  urldate = {2021-02-05}
}

@article{janak1978,
  title = {Proof that $\partial E/\partial n_i = \epsilon_i$ in density-functional theory},
  author = {Janak, J. F.},
  year = {1978},
  month = dec,
  journal = {Phys. Rev. B},
  volume = {18},
  number = {12},
  pages = {7165--7168},
  publisher = {American Physical Society},
  doi = {10.1103/PhysRevB.18.7165},
  urldate = {2020-12-08}
}

@article{johnson2008,
  title = {Delocalization Errors in Density Functionals and Implications for Main-Group Thermochemistry},
  author = {Johnson, Erin R. and {Mori-S{\'a}nchez}, Paula and Cohen, Aron J. and Yang, Weitao},
  year = {2008},
  month = nov,
  journal = {The Journal of Chemical Physics},
  volume = {129},
  number = {20},
  pages = {204112},
  issn = {0021-9606},
  doi = {10.1063/1.3021474},
  urldate = {2023-10-11}
}

@article{kaplan2023a,
  title = {Understanding {{Density-Driven Errors}} for {{Reaction Barrier Heights}}},
  author = {Kaplan, Aaron D. and Shahi, Chandra and Bhetwal, Pradeep and Sah, Raj K. and Perdew, John P.},
  year = {2023},
  month = jan,
  journal = {J. Chem. Theory Comput.},
  volume = {19},
  number = {2},
  pages = {532--543},
  publisher = {American Chemical Society},
  issn = {1549-9618},
  doi = {10.1021/acs.jctc.2c00953},
  urldate = {2024-04-29}
}

@article{kohn1965,
  title = {Self-{{Consistent Equations Including Exchange}} and {{Correlation Effects}}},
  author = {Kohn, W. and Sham, L. J.},
  year = {1965},
  month = nov,
  journal = {Phys. Rev.},
  volume = {140},
  number = {4A},
  pages = {A1133-A1138},
  issn = {0031-899X},
  doi = {10.1103/PhysRev.140.A1133},
  urldate = {2020-07-31},
  langid = {english}
}

@article{kohn1995,
  title = {Density Functional Theory for Systems of Very Many Atoms},
  author = {Kohn, W.},
  year = {1995},
  journal = {International Journal of Quantum Chemistry},
  volume = {56},
  number = {4},
  pages = {229--232},
  issn = {1097-461X},
  doi = {10.1002/qua.560560407},
  urldate = {2024-04-11},
  copyright = {Copyright {\copyright} 1995 John Wiley \& Sons, Inc.},
  langid = {english}
}

@article{kohn1996,
  title = {Density {{Functional}} and {{Density Matrix Method Scaling Linearly}} with the {{Number}} of {{Atoms}}},
  author = {Kohn, W.},
  year = {1996},
  month = apr,
  journal = {Phys. Rev. Lett.},
  volume = {76},
  number = {17},
  pages = {3168--3171},
  publisher = {American Physical Society},
  doi = {10.1103/PhysRevLett.76.3168},
  urldate = {2024-04-11}
}

@article{li2018,
  title = {Localized Orbital Scaling Correction for Systematic Elimination of Delocalization Error in Density Functional Approximations},
  author = {Li, Chen and Zheng, Xiao and Su, Neil Qiang and Yang, Weitao},
  year = {2018},
  month = mar,
  journal = {Nat. Sci. Rev.},
  volume = {5},
  number = {2},
  pages = {203--215},
  issn = {2095-5138, 2053-714X},
  doi = {10.1093/nsr/nwx111},
  urldate = {2020-01-20},
  langid = {english}
}

@article{linscott2023,
  title = {Koopmans: {{An Open-Source Package}} for {{Accurately}} and {{Efficiently Predicting Spectral Properties}} with {{Koopmans Functionals}}},
  shorttitle = {Koopmans},
  author = {Linscott, Edward B. and Colonna, Nicola and De Gennaro, Riccardo and Nguyen, Ngoc Linh and Borghi, Giovanni and Ferretti, Andrea and Dabo, Ismaila and Marzari, Nicola},
  year = {2023},
  month = aug,
  journal = {J. Chem. Theory Comput.},
  volume = {19},
  number = {20},
  pages = {7097--7111},
  publisher = {American Chemical Society},
  issn = {1549-9618},
  doi = {10.1021/acs.jctc.3c00652},
  urldate = {2023-08-24}
}

@article{ma2016,
  title = {Using {{Wannier}} Functions to Improve Solid Band Gap Predictions in Density Functional Theory},
  author = {Ma, Jie and Wang, Lin-Wang},
  year = {2016},
  month = apr,
  journal = {Sci. Rep.},
  volume = {6},
  number = {1},
  pages = {24924},
  publisher = {Nature Publishing Group},
  issn = {2045-2322},
  doi = {10.1038/srep24924},
  urldate = {2021-02-07},
  copyright = {2016 The Author(s)},
  langid = {english}
}

@article{ma2016a,
  title = {The Energy Level Alignment at Metal--Molecule Interfaces Using {{Wannier}}--{{Koopmans}} Method},
  author = {Ma, Jie and Liu, Zhen-Fei and Neaton, Jeffrey B. and Wang, Lin-Wang},
  year = {2016},
  month = jun,
  journal = {Applied Physics Letters},
  volume = {108},
  number = {26},
  pages = {262104},
  issn = {0003-6951},
  doi = {10.1063/1.4955128},
  urldate = {2024-04-04}
}

@article{mahler2022b,
  title = {Localized Orbital Scaling Correction for Periodic Systems},
  author = {Mahler, Aaron and Williams, Jacob and Su, Neil Qiang and Yang, Weitao},
  year = {2022},
  month = jul,
  journal = {Phys. Rev. B},
  volume = {106},
  number = {3},
  pages = {035147},
  publisher = {American Physical Society},
  doi = {10.1103/PhysRevB.106.035147},
  urldate = {2022-07-28}
}

@article{marzari1997,
  title = {Maximally Localized Generalized {{Wannier}} Functions for Composite Energy Bands},
  author = {Marzari, Nicola and Vanderbilt, David},
  year = {1997},
  month = nov,
  journal = {Phys. Rev. B},
  volume = {56},
  number = {20},
  pages = {12847--12865},
  issn = {0163-1829, 1095-3795},
  doi = {10.1103/PhysRevB.56.12847},
  urldate = {2020-03-28},
  langid = {english}
}

@article{mei2020a,
  title = {Self-{{Consistent Calculation}} of the {{Localized Orbital Scaling Correction}} for {{Correct Electron Densities}} and {{Energy-Level Alignments}} in {{Density Functional Theory}}},
  author = {Mei, Yuncai and Chen, Zehua and Yang, Weitao},
  year = {2020},
  month = nov,
  journal = {J. Phys. Chem. Lett.},
  volume = {11},
  pages = {10269--10277},
  issn = {1948-7185, 1948-7185},
  doi = {10.1021/acs.jpclett.0c03133},
  urldate = {2020-11-22},
  langid = {english}
}

@article{mei2021,
  title = {Exact {{Second-Order Corrections}} and {{Accurate Quasiparticle Energy Calculations}} in {{Density Functional Theory}}},
  author = {Mei, Yuncai and Chen, Zehua and Yang, Weitao},
  year = {2021},
  month = aug,
  journal = {J. Phys. Chem. Lett.},
  volume = {12},
  number = {30},
  pages = {7236--7244},
  publisher = {American Chemical Society},
  doi = {10.1021/acs.jpclett.1c01962},
  urldate = {2021-08-17}
}

@article{mi2023,
  title = {Orbital-{{Free Density Functional Theory}}: {{An Attractive Electronic Structure Method}} for {{Large-Scale First-Principles Simulations}}},
  shorttitle = {Orbital-{{Free Density Functional Theory}}},
  author = {Mi, Wenhui and Luo, Kai and Trickey, S. B. and Pavanello, Michele},
  year = {2023},
  month = nov,
  journal = {Chem. Rev.},
  volume = {123},
  number = {21},
  pages = {12039--12104},
  publisher = {American Chemical Society},
  issn = {0009-2665},
  doi = {10.1021/acs.chemrev.2c00758},
  urldate = {2024-04-17}
}

@article{miglio2020,
  title = {Predominance of Non-Adiabatic Effects in Zero-Point Renormalization of the Electronic Band Gap},
  author = {Miglio, Anna and {Brousseau-Couture}, V{\'e}ronique and Godbout, Emile and Antonius, Gabriel and Chan, Yang-Hao and Louie, Steven G. and C{\^o}t{\'e}, Michel and Giantomassi, Matteo and Gonze, Xavier},
  year = {2020},
  month = nov,
  journal = {npj Comput Mater},
  volume = {6},
  number = {1},
  pages = {1--8},
  publisher = {Nature Publishing Group},
  issn = {2057-3960},
  doi = {10.1038/s41524-020-00434-z},
  urldate = {2023-01-19},
  copyright = {2020 The Author(s)},
  langid = {english}
}

@article{monaco2018,
  title = {Optimal {{Decay}} of {{Wannier}} Functions in {{Chern}} and {{Quantum Hall Insulators}}},
  author = {Monaco, Domenico and Panati, Gianluca and Pisante, Adriano and Teufel, Stefan},
  year = {2018},
  month = apr,
  journal = {Commun. Math. Phys.},
  volume = {359},
  number = {1},
  pages = {61--100},
  issn = {1432-0916},
  doi = {10.1007/s00220-017-3067-7},
  urldate = {2020-12-10},
  langid = {english}
}

@article{monkhorst1976,
  title = {Special Points for {{Brillouin-zone}} Integrations},
  author = {Monkhorst, Hendrik J. and Pack, James D.},
  year = {1976},
  month = jun,
  journal = {Phys. Rev. B},
  volume = {13},
  number = {12},
  pages = {5188--5192},
  issn = {0556-2805},
  doi = {10.1103/PhysRevB.13.5188},
  urldate = {2020-06-17},
  langid = {english}
}

@article{mori-sanchez2006,
  title = {Many-Electron Self-Interaction Error in Approximate Density Functionals},
  author = {{Mori-S{\'a}nchez}, Paula and Cohen, Aron J. and Yang, Weitao},
  year = {2006},
  month = nov,
  journal = {J. Chem. Phys.},
  volume = {125},
  number = {20},
  pages = {201102},
  publisher = {American Institute of Physics},
  issn = {0021-9606},
  doi = {10.1063/1.2403848},
  urldate = {2021-02-12}
}

@article{mori-sanchez2008,
  title = {Localization and {{Delocalization Errors}} in {{Density Functional Theory}} and {{Implications}} for {{Band-Gap Prediction}}},
  author = {{Mori-S{\'a}nchez}, Paula and Cohen, Aron J. and Yang, Weitao},
  year = {2008},
  month = apr,
  journal = {Phys. Rev. Lett.},
  volume = {100},
  number = {14},
  pages = {146401},
  publisher = {American Physical Society},
  doi = {10.1103/PhysRevLett.100.146401},
  urldate = {2023-04-28}
}

@article{mostofi2008,
  title = {Wannier90: {{A}} Tool for Obtaining Maximally-Localised {{Wannier}} Functions},
  shorttitle = {Wannier90},
  author = {Mostofi, Arash A. and Yates, Jonathan R. and Lee, Young-Su and Souza, Ivo and Vanderbilt, David and Marzari, Nicola},
  year = {2008},
  month = may,
  journal = {Computer Physics Communications},
  volume = {178},
  number = {9},
  pages = {685--699},
  issn = {0010-4655},
  doi = {10.1016/j.cpc.2007.11.016},
  urldate = {2020-03-06},
  langid = {english}
}

@article{mostofi2014,
  title = {An Updated Version of Wannier90: {{A}} Tool for Obtaining Maximally-Localised {{Wannier}} Functions},
  shorttitle = {An Updated Version of Wannier90},
  author = {Mostofi, Arash A. and Yates, Jonathan R. and Pizzi, Giovanni and Lee, Young-Su and Souza, Ivo and Vanderbilt, David and Marzari, Nicola},
  year = {2014},
  month = aug,
  journal = {Computer Physics Communications},
  volume = {185},
  number = {8},
  pages = {2309--2310},
  issn = {0010-4655},
  doi = {10.1016/j.cpc.2014.05.003},
  urldate = {2021-02-14},
  langid = {english}
}

@article{nguyen2018,
  title = {Koopmans-{{Compliant Spectral Functionals}} for {{Extended Systems}}},
  author = {Nguyen, Ngoc Linh and Colonna, Nicola and Ferretti, Andrea and Marzari, Nicola},
  year = {2018},
  month = may,
  journal = {Phys. Rev. X},
  volume = {8},
  number = {2},
  pages = {021051},
  publisher = {American Physical Society},
  doi = {10.1103/PhysRevX.8.021051},
  urldate = {2022-02-22}
}

@article{oliver1979,
  title = {Spin-Density Gradient Expansion for the Kinetic Energy},
  author = {Oliver, G. L. and Perdew, J. P.},
  year = {1979},
  month = aug,
  journal = {Phys. Rev. A},
  volume = {20},
  number = {2},
  pages = {397--403},
  publisher = {American Physical Society},
  doi = {10.1103/PhysRevA.20.397},
  urldate = {2023-07-13}
}

@article{otero2017,
  title = {Electronic, Structural and Chemical Effects of Charge-Transfer at Organic/Inorganic Interfaces},
  author = {Otero, R. and {V{\'a}zquez de Parga}, A. L. and Gallego, J. M.},
  year = {2017},
  month = jul,
  journal = {Surface Science Reports},
  volume = {72},
  number = {3},
  pages = {105--145},
  issn = {0167-5729},
  doi = {10.1016/j.surfrep.2017.03.001},
  urldate = {2024-04-23}
}

@article{pederson2014,
  title = {Communication: {{Self-interaction}} Correction with Unitary Invariance in Density Functional Theory},
  shorttitle = {Communication},
  author = {Pederson, Mark R. and Ruzsinszky, Adrienn and Perdew, John P.},
  year = {2014},
  month = mar,
  journal = {The Journal of Chemical Physics},
  volume = {140},
  number = {12},
  pages = {121103},
  issn = {0021-9606},
  doi = {10.1063/1.4869581},
  urldate = {2024-01-18}
}

@article{perdew1981,
  title = {Self-Interaction Correction to Density-Functional Approximations for Many-Electron Systems},
  author = {Perdew, J. P. and Zunger, Alex},
  year = {1981},
  month = may,
  journal = {Phys. Rev. B},
  volume = {23},
  number = {10},
  pages = {5048--5079},
  publisher = {American Physical Society},
  doi = {10.1103/PhysRevB.23.5048},
  urldate = {2021-02-23}
}

@article{perdew1982,
  title = {Density-{{Functional Theory}} for {{Fractional Particle Number}}: {{Derivative Discontinuities}} of the {{Energy}}},
  shorttitle = {Density-{{Functional Theory}} for {{Fractional Particle Number}}},
  author = {Perdew, John P. and Parr, Robert G. and Levy, Mel and Balduz, Jose L.},
  year = {1982},
  month = dec,
  journal = {Phys. Rev. Lett.},
  volume = {49},
  number = {23},
  pages = {1691--1694},
  issn = {0031-9007},
  doi = {10.1103/PhysRevLett.49.1691},
  urldate = {2020-01-20},
  langid = {english}
}

@article{perdew1985,
  title = {Density Functional Theory and the Band Gap Problem},
  author = {Perdew, John P.},
  year = {1985},
  journal = {Int. J. Quantum Chem.},
  volume = {28},
  number = {S19},
  pages = {497--523},
  issn = {1097-461X},
  doi = {10.1002/qua.560280846},
  urldate = {2021-02-09},
  copyright = {Copyright {\copyright} 1985 John Wiley \& Sons, Inc.},
  langid = {english}
}

@article{perdew1992,
  title = {Accurate and Simple Analytic Representation of the Electron-Gas Correlation Energy},
  author = {Perdew, John P. and Wang, Yue},
  year = {1992},
  month = jun,
  journal = {Phys. Rev. B},
  volume = {45},
  number = {23},
  pages = {13244--13249},
  publisher = {American Physical Society},
  doi = {10.1103/PhysRevB.45.13244},
  urldate = {2021-02-23}
}

@article{perdew1996,
  title = {Generalized {{Gradient Approximation Made Simple}}},
  author = {Perdew, John P. and Burke, Kieron and Ernzerhof, Matthias},
  year = {1996},
  month = oct,
  journal = {Phys. Rev. Lett.},
  volume = {77},
  number = {18},
  pages = {3865--3868},
  publisher = {American Physical Society},
  doi = {10.1103/PhysRevLett.77.3865},
  urldate = {2020-10-07}
}

@article{perdew2018,
  title = {Erratum: {{Accurate}} and Simple Analytic Representation of the Electron-Gas Correlation Energy [{{Phys}}. {{Rev}}. {{B}} 45, 13244 (1992)]},
  shorttitle = {Erratum},
  author = {Perdew, John P. and Wang, Yue},
  year = {2018},
  month = aug,
  journal = {Phys. Rev. B},
  volume = {98},
  number = {7},
  pages = {079904},
  publisher = {American Physical Society},
  doi = {10.1103/PhysRevB.98.079904},
  urldate = {2023-08-18}
}

@article{pizzi2020,
  title = {Wannier90 as a Community Code: New Features and Applications},
  shorttitle = {Wannier90 as a Community Code},
  author = {Pizzi, Giovanni and Vitale, Valerio and Arita, Ryotaro and Bl{\"u}gel, Stefan and Freimuth, Frank and G{\'e}ranton, Guillaume and Gibertini, Marco and Gresch, Dominik and Johnson, Charles and Koretsune, Takashi and {Iba{\~n}ez-Azpiroz}, Julen and Lee, Hyungjun and Lihm, Jae-Mo and Marchand, Daniel and Marrazzo, Antimo and Mokrousov, Yuriy and Mustafa, Jamal I and Nohara, Yoshiro and Nomura, Yusuke and Paulatto, Lorenzo and Ponc{\'e}, Samuel and Ponweiser, Thomas and Qiao, Junfeng and Th{\"o}le, Florian and Tsirkin, Stepan S and Wierzbowska, Ma{\l}gorzata and Marzari, Nicola and Vanderbilt, David and Souza, Ivo and Mostofi, Arash A and Yates, Jonathan R},
  year = {2020},
  month = apr,
  journal = {J. Phys.: Condens. Matter},
  volume = {32},
  number = {16},
  pages = {165902},
  issn = {0953-8984, 1361-648X},
  doi = {10.1088/1361-648X/ab51ff},
  urldate = {2020-03-06},
  langid = {english}
}

@article{poole1975,
  title = {Electronic Band Structure of the Alkali Halides. {{I}}. {{Experimental}} Parameters},
  author = {Poole, R. T. and Jenkin, J. G. and Liesegang, J. and Leckey, R. C. G.},
  year = {1975},
  month = jun,
  journal = {Phys. Rev. B},
  volume = {11},
  number = {12},
  pages = {5179--5189},
  publisher = {American Physical Society},
  doi = {10.1103/PhysRevB.11.5179},
  urldate = {2023-04-12}
}

@article{prodan2005,
  title = {Nearsightedness of Electronic Matter},
  author = {Prodan, E and Kohn, W},
  year = {2005},
  journal = {PNAS},
  volume = {102},
  number = {33},
  pages = {11635--11638},
  doi = {10.1073/pnas.0505436102},
  langid = {english}
}

@article{refaely-abramson2011,
  title = {Fundamental and Excitation Gaps in Molecules of Relevance for Organic Photovoltaics from an Optimally Tuned Range-Separated Hybrid Functional},
  author = {{Refaely-Abramson}, Sivan and Baer, Roi and Kronik, Leeor},
  year = {2011},
  month = aug,
  journal = {Phys. Rev. B},
  volume = {84},
  number = {7},
  pages = {075144},
  publisher = {American Physical Society},
  doi = {10.1103/PhysRevB.84.075144},
  urldate = {2024-01-16}
}

@article{refaely-abramson2013,
  title = {Gap Renormalization of Molecular Crystals from Density-Functional Theory},
  author = {{Refaely-Abramson}, Sivan and Sharifzadeh, Sahar and Jain, Manish and Baer, Roi and Neaton, Jeffrey B. and Kronik, Leeor},
  year = {2013},
  month = aug,
  journal = {Phys. Rev. B},
  volume = {88},
  number = {8},
  pages = {081204},
  publisher = {American Physical Society},
  doi = {10.1103/PhysRevB.88.081204},
  urldate = {2024-04-04}
}

@article{ren2012a,
  title = {Resolution-of-Identity Approach to {Hartree--Fock}, Hybrid Density Functionals, {RPA}, {MP2} and $GW$ with Numeric Atom-Centered Orbital Basis Functions},
  author = {Ren, Xinguo and Rinke, Patrick and Blum, Volker and Wieferink, J{\"u}rgen and Tkatchenko, Alexandre and Sanfilippo, Andrea and Reuter, Karsten and Scheffler, Matthias},
  year = {2012},
  month = may,
  journal = {New J. Phys.},
  volume = {14},
  number = {5},
  pages = {053020},
  publisher = {IOP Publishing},
  issn = {1367-2630},
  doi = {10.1088/1367-2630/14/5/053020},
  urldate = {2021-01-09},
  langid = {english}
}

@article{rostgaard2010,
  title = {Fully Self-Consistent {{GW}} Calculations for Molecules},
  author = {Rostgaard, C. and Jacobsen, K. W. and Thygesen, K. S.},
  year = {2010},
  month = feb,
  journal = {Phys. Rev. B},
  volume = {81},
  number = {8},
  pages = {085103},
  publisher = {American Physical Society},
  doi = {10.1103/PhysRevB.81.085103},
  urldate = {2024-04-23}
}

@article{schwalbe2018,
  title = {Fermi-{{L{\"o}wdin}} Orbital Self-Interaction Corrected Density Functional Theory: {{Ionization}} Potentials and Enthalpies of Formation},
  shorttitle = {Fermi-{{L{\"o}wdin}} Orbital Self-Interaction Corrected Density Functional Theory},
  author = {Schwalbe, Sebastian and Hahn, Torsten and Liebing, Simon and Trepte, Kai and Kortus, Jens},
  year = {2018},
  journal = {Journal of Computational Chemistry},
  volume = {39},
  number = {29},
  pages = {2463--2471},
  issn = {1096-987X},
  doi = {10.1002/jcc.25586},
  urldate = {2024-04-24},
  copyright = {{\copyright} 2018 Wiley Periodicals, Inc.},
  langid = {english}
}

@article{setyawan2010,
  title = {High-Throughput Electronic Band Structure Calculations: {{Challenges}} and Tools},
  shorttitle = {High-Throughput Electronic Band Structure Calculations},
  author = {Setyawan, Wahyu and Curtarolo, Stefano},
  year = {2010},
  month = aug,
  journal = {Computational Materials Science},
  volume = {49},
  number = {2},
  pages = {299--312},
  issn = {0927-0256},
  doi = {10.1016/j.commatsci.2010.05.010},
  urldate = {2022-09-08},
  langid = {english}
}

@article{shang2021,
  title = {Assessment of the {{Mass Factor}} for the {{Electron}}--{{Phonon Coupling}} in {{Solids}}},
  author = {Shang, Honghui and Zhao, Jin and Yang, Jinlong},
  year = {2021},
  month = mar,
  journal = {J. Phys. Chem. C},
  volume = {125},
  number = {11},
  pages = {6479--6485},
  publisher = {American Chemical Society},
  issn = {1932-7447},
  doi = {10.1021/acs.jpcc.1c00861},
  urldate = {2023-02-14}
}

@article{shimazaki2008,
  title = {Band Structure Calculations Based on Screened {{Fock}} Exchange Method},
  author = {Shimazaki, Tomomi and Asai, Yoshihiro},
  year = {2008},
  month = nov,
  journal = {Chemical Physics Letters},
  volume = {466},
  number = {1},
  pages = {91--94},
  issn = {0009-2614},
  doi = {10.1016/j.cplett.2008.10.012},
  urldate = {2024-04-16}
}

@article{shinde2020,
  title = {Improved Band Gaps and Structural Properties from {{Wannier}}--{{Fermi}}--{{L{\"o}wdin}} Self-Interaction Corrections for Periodic Systems},
  author = {Shinde, Ravindra and Yamijala, Sharma S. R. K. C. and Wong, Bryan M.},
  year = {2020},
  month = dec,
  journal = {J. Phys.: Condens. Matter},
  volume = {33},
  number = {11},
  pages = {115501},
  publisher = {IOP Publishing},
  issn = {0953-8984},
  doi = {10.1088/1361-648X/abc407},
  urldate = {2024-04-24},
  langid = {english}
}

@article{souza2001,
  title = {Maximally Localized {{Wannier}} Functions for Entangled Energy Bands},
  author = {Souza, Ivo and Marzari, Nicola and Vanderbilt, David},
  year = {2001},
  month = dec,
  journal = {Phys. Rev. B},
  volume = {65},
  number = {3},
  pages = {035109},
  issn = {0163-1829, 1095-3795},
  doi = {10.1103/PhysRevB.65.035109},
  urldate = {2020-07-23},
  langid = {english}
}

@article{su2020,
  title = {Preserving {{Symmetry}} and {{Degeneracy}} in the {{Localized Orbital Scaling Correction Approach}}},
  author = {Su, Neil Qiang and Mahler, Aaron and Yang, Weitao},
  year = {2020},
  month = feb,
  journal = {J. Phys. Chem. Lett.},
  volume = {11},
  number = {4},
  pages = {1528--1535},
  issn = {1948-7185, 1948-7185},
  doi = {10.1021/acs.jpclett.9b03888},
  urldate = {2020-02-21},
  langid = {english}
}

@article{teale2022,
  title = {{{DFT}} Exchange: Sharing Perspectives on the Workhorse of Quantum Chemistry and Materials Science},
  shorttitle = {{{DFT}} Exchange},
  author = {Teale, Andrew M. and Helgaker, Trygve and Savin, Andreas and Adamo, Carlo and Aradi, B{\'a}lint and Arbuznikov, Alexei V. and Ayers, Paul W. and Baerends, Evert Jan and Barone, Vincenzo and Calaminici, Patrizia and Canc{\`e}s, Eric and Carter, Emily A. and Chattaraj, Pratim Kumar and Chermette, Henry and Ciofini, Ilaria and Crawford, T. Daniel and Proft, Frank De and Dobson, John F. and Draxl, Claudia and Frauenheim, Thomas and Fromager, Emmanuel and Fuentealba, Patricio and Gagliardi, Laura and Galli, Giulia and Gao, Jiali and Geerlings, Paul and Gidopoulos, Nikitas and Gill, Peter M. W. and {Gori-Giorgi}, Paola and G{\"o}rling, Andreas and Gould, Tim and Grimme, Stefan and Gritsenko, Oleg and Jensen, Hans J{\o}rgen Aagaard and Johnson, Erin R. and Jones, Robert O. and Kaupp, Martin and K{\"o}ster, Andreas M. and Kronik, Leeor and Krylov, Anna I. and Kvaal, Simen and Laestadius, Andre and Levy, Mel and Lewin, Mathieu and Liu, Shubin and Loos, Pierre-Fran{\c c}ois and Maitra, Neepa T. and Neese, Frank and Perdew, John P. and Pernal, Katarzyna and Pernot, Pascal and Piecuch, Piotr and Rebolini, Elisa and Reining, Lucia and Romaniello, Pina and Ruzsinszky, Adrienn and Salahub, Dennis R. and Scheffler, Matthias and Schwerdtfeger, Peter and Staroverov, Viktor N. and Sun, Jianwei and Tellgren, Erik and Tozer, David J. and Trickey, Samuel B. and Ullrich, Carsten A. and Vela, Alberto and Vignale, Giovanni and Wesolowski, Tomasz A. and Xu, Xin and Yang, Weitao},
  year = {2022},
  month = dec,
  journal = {Phys. Chem. Chem. Phys.},
  volume = {24},
  number = {47},
  pages = {28700--28781},
  publisher = {The Royal Society of Chemistry},
  issn = {1463-9084},
  doi = {10.1039/D2CP02827A},
  urldate = {2024-04-09},
  langid = {english}
}

@article{thomas1927,
  title = {The Calculation of Atomic Fields},
  author = {Thomas, L. H.},
  year = {1927},
  month = jan,
  journal = {Mathematical Proceedings of the Cambridge Philosophical Society},
  volume = {23},
  number = {5},
  pages = {542--548},
  publisher = {Cambridge University Press},
  issn = {1469-8064, 0305-0041},
  doi = {10.1017/S0305004100011683},
  urldate = {2023-04-21},
  langid = {english}
}

@article{timrov2018,
  title = {Hubbard Parameters from Density-Functional Perturbation Theory},
  author = {Timrov, Iurii and Marzari, Nicola and Cococcioni, Matteo},
  year = {2018},
  month = aug,
  journal = {Phys. Rev. B},
  volume = {98},
  number = {8},
  pages = {085127},
  publisher = {American Physical Society},
  doi = {10.1103/PhysRevB.98.085127},
  urldate = {2022-05-11}
}

@article{tran2009,
  title = {Accurate {{Band Gaps}} of {{Semiconductors}} and {{Insulators}} with a {{Semilocal Exchange-Correlation Potential}}},
  author = {Tran, Fabien and Blaha, Peter},
  year = {2009},
  month = jun,
  journal = {Phys. Rev. Lett.},
  volume = {102},
  number = {22},
  pages = {226401},
  issn = {0031-9007, 1079-7114},
  doi = {10.1103/PhysRevLett.102.226401},
  urldate = {2020-01-28},
  langid = {english}
}

@article{vandevondele2005,
  title = {A Molecular Dynamics Study of the Hydroxyl Radical in Solution Applying Self-Interaction-Corrected Density Functional Methods},
  author = {VandeVondele, Joost and Sprik, Michiel},
  year = {2005},
  journal = {Physical Chemistry Chemical Physics},
  volume = {7},
  number = {7},
  pages = {1363--1367},
  publisher = {Royal Society of Chemistry},
  doi = {10.1039/B501603G},
  urldate = {2024-04-29},
  langid = {english}
}

@article{vansetten2015,
  title = {{{GW100}}: {{Benchmarking G0W0}} for {{Molecular Systems}}},
  shorttitle = {{{GW100}}},
  author = {{van Setten}, Michiel J. and Caruso, Fabio and Sharifzadeh, Sahar and Ren, Xinguo and Scheffler, Matthias and Liu, Fang and Lischner, Johannes and Lin, Lin and Deslippe, Jack R. and Louie, Steven G. and Yang, Chao and Weigend, Florian and Neaton, Jeffrey B. and Evers, Ferdinand and Rinke, Patrick},
  year = {2015},
  month = dec,
  journal = {J. Chem. Theory Comput.},
  volume = {11},
  number = {12},
  pages = {5665--5687},
  publisher = {American Chemical Society},
  issn = {1549-9618},
  doi = {10.1021/acs.jctc.5b00453},
  urldate = {2024-04-23}
}

@article{vitale2020,
  title = {Automated High-Throughput {{Wannierisation}}},
  author = {Vitale, Valerio and Pizzi, Giovanni and Marrazzo, Antimo and Yates, Jonathan R. and Marzari, Nicola and Mostofi, Arash A.},
  year = {2020},
  month = jun,
  journal = {npj Comput Mater},
  volume = {6},
  number = {1},
  pages = {1--18},
  publisher = {Nature Publishing Group},
  issn = {2057-3960},
  doi = {10.1038/s41524-020-0312-y},
  urldate = {2024-09-30},
  copyright = {2020 The Author(s)},
  langid = {english}
}

@article{wannier1937,
  title = {The {{Structure}} of {{Electronic Excitation Levels}} in {{Insulating Crystals}}},
  author = {Wannier, Gregory H.},
  year = {1937},
  month = aug,
  journal = {Phys. Rev.},
  volume = {52},
  number = {3},
  pages = {191--197},
  publisher = {American Physical Society},
  doi = {10.1103/PhysRev.52.191},
  urldate = {2023-02-23}
}

@article{weizsaecker1935,
  title = {{Zur Theorie der Kernmassen}},
  author = {v. Weizs{\"a}cker, C. F.},
  year = {1935},
  month = jul,
  journal = {Z. Physik},
  volume = {96},
  number = {7},
  pages = {431--458},
  issn = {0044-3328},
  doi = {10.1007/BF01337700},
  urldate = {2023-04-21},
  langid = {ngerman}
}

@article{weng2017,
  title = {Wannier {{Koopman}} Method Calculations of the Band Gaps of Alkali Halides},
  author = {Weng, Mouyi and Li, Sibai and Ma, Jie and Zheng, Jiaxin and Pan, Feng and Wang, Lin-Wang},
  year = {2017},
  month = jul,
  journal = {Applied Physics Letters},
  volume = {111},
  number = {5},
  pages = {054101},
  issn = {0003-6951},
  doi = {10.1063/1.4996743},
  urldate = {2023-06-21}
}

@article{weng2018,
  title = {Wannier {{Koopmans Method Calculations}} of {{2D Material Band Gaps}}},
  author = {Weng, Mouyi and Li, Sibai and Zheng, Jiaxin and Pan, Feng and Wang, Lin-Wang},
  year = {2018},
  month = jan,
  journal = {J. Phys. Chem. Lett.},
  volume = {9},
  number = {2},
  pages = {281--285},
  publisher = {American Chemical Society},
  doi = {10.1021/acs.jpclett.7b03041},
  urldate = {2023-06-21}
}

@article{weng2020,
  title = {Wannier--{{Koopmans}} Method Calculations for Transition Metal Oxide Band Gaps},
  author = {Weng, Mouyi and Pan, Feng and Wang, Lin-Wang},
  year = {2020},
  month = apr,
  journal = {npj Comput Mater},
  volume = {6},
  number = {1},
  pages = {1--8},
  publisher = {Nature Publishing Group},
  issn = {2057-3960},
  doi = {10.1038/s41524-020-0302-0},
  urldate = {2023-06-21},
  copyright = {2020 The Author(s)},
  langid = {english}
}

@article{wilhelm2021,
  title = {Low-{{Scaling GW}} with {{Benchmark Accuracy}} and {{Application}} to {{Phosphorene Nanosheets}}},
  author = {Wilhelm, Jan and Seewald, Patrick and Golze, Dorothea},
  year = {2021},
  month = mar,
  journal = {J. Chem. Theory Comput.},
  volume = {17},
  number = {3},
  pages = {1662--1677},
  publisher = {American Chemical Society},
  issn = {1549-9618},
  doi = {10.1021/acs.jctc.0c01282},
  urldate = {2024-04-24}
}

@misc{williams2024,
  title = {Correcting {{Delocalization Error}} in {{Materials}} with {{Localized Orbitals}} and {{Linear-Response Screening}}},
  author = {Williams, Jacob Z. and Yang, Weitao},
  year = {2024},
  month = jul,
  number = {arXiv:2406.07351},
  eprint = {2406.07351},
  primaryclass = {cond-mat, physics:physics},
  publisher = {arXiv},
  doi = {10.48550/arXiv.2406.07351},
  urldate = {2024-07-03},
  archiveprefix = {arXiv}
}

@article{wing2021,
  title = {Band Gaps of Crystalline Solids from {{Wannier-localization}}--Based Optimal Tuning of a Screened Range-Separated Hybrid Functional},
  author = {Wing, Dahvyd and Ohad, Guy and Haber, Jonah B. and Filip, Marina R. and Gant, Stephen E. and Neaton, Jeffrey B. and Kronik, Leeor},
  year = {2021},
  month = aug,
  journal = {Proc Natl Acad Sci USA},
  volume = {118},
  number = {34},
  pages = {e2104556118},
  issn = {0027-8424, 1091-6490},
  doi = {10.1073/pnas.2104556118},
  urldate = {2022-03-04},
  langid = {english}
}

@article{wirtinger1927,
  title = {{Zur formalen Theorie der Funktionen von mehr komplexen Ver{\"a}nderlichen}},
  author = {Wirtinger, W.},
  year = {1927},
  month = dec,
  journal = {Math. Ann.},
  volume = {97},
  number = {1},
  pages = {357--375},
  issn = {1432-1807},
  doi = {10.1007/BF01447872},
  urldate = {2022-01-19},
  langid = {ngerman}
}

@book{wyckoff1973,
  title = {Crystal {{Structures}}},
  author = {Wyckoff, {\relax Ralph}. W. G.},
  year = {1973},
  edition = {2nd},
  publisher = {Wiley},
  address = {New York},
  langid = {english}
}

@article{yamamoto2019,
  title = {Fermi-{{L{\"o}wdin}} Orbital Self-Interaction Correction Using the Strongly Constrained and Appropriately Normed Meta-{{GGA}} Functional},
  author = {Yamamoto, Yoh and Diaz, Carlos M. and Basurto, Luis and Jackson, Koblar A. and Baruah, Tunna and Zope, Rajendra R.},
  year = {2019},
  month = oct,
  journal = {The Journal of Chemical Physics},
  volume = {151},
  number = {15},
  pages = {154105},
  issn = {0021-9606},
  doi = {10.1063/1.5120532},
  urldate = {2024-04-24}
}

@article{yanai2004,
  title = {A New Hybrid Exchange--Correlation Functional Using the {{Coulomb-attenuating}} Method ({{CAM-B3LYP}})},
  author = {Yanai, Takeshi and Tew, David P and Handy, Nicholas C},
  year = {2004},
  month = jul,
  journal = {Chemical Physics Letters},
  volume = {393},
  number = {1},
  pages = {51--57},
  issn = {0009-2614},
  doi = {10.1016/j.cplett.2004.06.011},
  urldate = {2024-04-12}
}

@article{yang2000,
  title = {Degenerate {{Ground States}} and a {{Fractional Number}} of {{Electrons}} in {{Density}} and {{Reduced Density Matrix Functional Theory}}},
  author = {Yang, Weitao and Zhang, Yingkai and Ayers, Paul W.},
  year = {2000},
  month = may,
  journal = {Phys. Rev. Lett.},
  volume = {84},
  number = {22},
  pages = {5172--5175},
  issn = {0031-9007, 1079-7114},
  doi = {10.1103/PhysRevLett.84.5172},
  urldate = {2020-01-20},
  langid = {english}
}

@article{yang2012a,
  title = {Analytical Evaluation of {{Fukui}} Functions and Real-Space Linear Response Function},
  author = {Yang, Weitao and Cohen, Aron J. and De Proft, Frank and Geerlings, Paul},
  year = {2012},
  month = apr,
  journal = {J. Chem. Phys.},
  volume = {136},
  number = {14},
  pages = {144110},
  publisher = {American Institute of Physics},
  issn = {0021-9606},
  doi = {10.1063/1.3701562},
  urldate = {2021-08-05}
}

@article{yang2017,
  title = {Full Self-Consistency in the {{Fermi-orbital}} Self-Interaction Correction},
  author = {Yang, Zeng-hui and Pederson, Mark R. and Perdew, John P.},
  year = {2017},
  month = may,
  journal = {Phys. Rev. A},
  volume = {95},
  number = {5},
  pages = {052505},
  publisher = {American Physical Society},
  doi = {10.1103/PhysRevA.95.052505},
  urldate = {2024-04-12}
}

@article{york1996,
  title = {A Chemical Potential Equalization Method for Molecular Simulations},
  author = {York, Darrin M. and Yang, Weitao},
  year = {1996},
  month = jan,
  journal = {J. Chem. Phys.},
  volume = {104},
  number = {1},
  pages = {159--172},
  publisher = {American Institute of Physics},
  issn = {0021-9606},
  doi = {10.1063/1.470886},
  urldate = {2021-07-27}
}

@article{zhang1998,
  title = {A Challenge for Density Functionals: {Self-interaction} Error Increases for Systems with a Noninteger Number of Electrons},
  shorttitle = {A Challenge for Density Functionals},
  author = {Zhang, Yingkai and Yang, Weitao},
  year = {1998},
  month = aug,
  journal = {J. Chem. Phys.},
  volume = {109},
  number = {7},
  pages = {2604--2608},
  publisher = {American Institute of Physics},
  issn = {0021-9606},
  doi = {10.1063/1.476859},
  urldate = {2021-02-12}
}

@article{zhu2009,
  title = {Charge-{{Transfer Excitons}} at {{Organic Semiconductor Surfaces}} and {{Interfaces}}},
  author = {Zhu, X.-Y. and Yang, Q. and Muntwiler, M.},
  year = {2009},
  month = nov,
  journal = {Acc. Chem. Res.},
  volume = {42},
  number = {11},
  pages = {1779--1787},
  publisher = {American Chemical Society},
  issn = {0001-4842},
  doi = {10.1021/ar800269u},
  urldate = {2024-04-23}
}

@article{zhu2015,
  title = {Charge {{Transfer Excitons}} at van Der {{Waals Interfaces}}},
  author = {Zhu, Xiaoyang and Monahan, Nicholas R. and Gong, Zizhou and Zhu, Haiming and Williams, Kristopher W. and Nelson, Cory A.},
  year = {2015},
  month = jul,
  journal = {J. Am. Chem. Soc.},
  volume = {137},
  number = {26},
  pages = {8313--8320},
  publisher = {American Chemical Society},
  issn = {0002-7863},
  doi = {10.1021/jacs.5b03141},
  urldate = {2024-04-23}
}

@article{zhu2015a,
  title = {Correction to ``{{Charge Transfer Excitons}} at van Der {{Waals Interfaces}}''},
  author = {Zhu, Xiaoyang and Monahan, Nicholas R. and Gong, Zizhou and Zhu, Haiming and Williams, Kristopher W. and Nelson, Cory A.},
  year = {2015},
  month = nov,
  journal = {J. Am. Chem. Soc.},
  volume = {137},
  number = {44},
  pages = {14230--14230},
  publisher = {American Chemical Society},
  issn = {0002-7863},
  doi = {10.1021/jacs.5b09894},
  urldate = {2024-04-23}
}

@article{sternheimer1954,
  title = {Electronic {{Polarizabilities}} of {{Ions}} from the {{Hartree-Fock Wave Functions}}},
  author = {Sternheimer, R. M.},
  year = {1954},
  journal = {Phys. Rev.},
  volume = {96},
  number = {4},
  pages = {951--968},
  publisher = {American Physical Society},
  doi = {10.1103/PhysRev.96.951},
  url = {https://link.aps.org/doi/10.1103/PhysRev.96.951},
  urldate = {2026-02-03}
}

@article{yu2025,
  title = {Accurate {{Prediction}} of {{Core-Level Binding Energies}} from {{Ground-State Density Functional Calculations}}: {{The Importance}} of {{Localization}} and {{Screening}}},
  shorttitle = {Accurate {{Prediction}} of {{Core-Level Binding Energies}} from {{Ground-State Density Functional Calculations}}},
  author = {Yu, Jincheng and Mei, Yuncai and Chen, Zehua and Fan, Yichen and Yang, Weitao},
  year = 2025,
  month = mar,
  journal = {The Journal of Physical Chemistry Letters},
  volume = {16},
  number = {10},
  pages = {2492--2500},
  publisher = {American Chemical Society},
  doi = {10.1021/acs.jpclett.5c00120},
  urldate = {2025-07-14}
}

@article{johnson_etal_2013,
  title = {Extreme Density-Driven Delocalization Error for a Model Solvated-Electron System},
  author = {Johnson, Erin R. and {Otero-de-la-Roza}, A. and Dale, Stephen G.},
  year = 2013,
  month = nov,
  journal = {J. Chem. Phys.},
  volume = {139},
  number = {18},
  pages = {184116},
  issn = {0021-9606},
  doi = {10.1063/1.4829642},
  urldate = {2026-02-06}
}

@article{Shahi2009,
   author = {Abdul Rehaman Moughal Shahi and Christopher J. Cramer and Laura Gagliardi},
   doi = {10.1039/b912607d},
   issn = {14639076},
   issue = {46},
   journal = {Physical Chemistry Chemical Physics},
   pages = {10964-10972},
   title = {Second-order perturbation theory with complete and restricted active space reference functions applied to oligomeric unsaturated hydrocarbons},
   volume = {11},
   year = {2009},
}

@article{Richard2016,
   author = {Ryan M. Richard and Michael S. Marshall and O. Dolgounitcheva and J. V. Ortiz and Jean Luc Brédas and Noa Marom and C. David Sherrill},
   doi = {10.1021/ACS.JCTC.5B00875/SUPPL_FILE/CT5B00875_SI_002.PDF},
   issn = {15499626},
   issue = {2},
   journal = {Journal of Chemical Theory and Computation},
   month = {2},
   pages = {595-604},
   publisher = {American Chemical Society},
   title = {Accurate Ionization Potentials and Electron Affinities of Acceptor Molecules I. Reference Data at the CCSD(T) Complete Basis Set Limit},
   volume = {12},
   url = {https://pubs.acs.org/doi/full/10.1021/acs.jctc.5b00875},
   year = {2016},
}

@article{Curtiss1997,
   author = {Larry A. Curtiss and Krishnan Raghavachari and Paul C. Redfern and John A. Pople},
   doi = {10.1063/1.473182},
   issn = {0021-9606},
   issue = {3},
   journal = {Journal of Chemical Physics},
   month = {1},
   pages = {1063-1079},
   publisher = {AIP Publishing},
   title = {Assessment of Gaussian-2 and density functional theories for the computation of enthalpies of formation},
   volume = {106},
   url = {/aip/jcp/article/106/3/1063/478756/Assessment-of-Gaussian-2-and-density-functional},
   year = {1997},
}

@article{dunning1989a,
    author = {Dunning, Thom H.},
    title = {Gaussian basis sets for use in correlated molecular calculations. I. The atoms boron through neon and hydrogen},
    journal = {J. Chem. Phys.},
    volume = {90},
    pages = {1007-1023},
    year = {1989},
    doi = {10.1063/1.456153}
}

@article{kendall1992a,
    author = {Kendall, Rick A. and Dunning, Thom H. and Harrison, Robert J.},
    title = {Electron affinities of the first-row atoms revisited. Systematic basis sets and wave functions},
    journal = {J. Chem. Phys.},
    volume = {96},
    pages = {6796-6806},
    year = {1992},
    doi = {10.1063/1.462569}
}

@article{ZhengRB2009,
author = {Zheng, Jingjing and Zhao, Yan and Truhlar, Donald G.},
title = {The DBH24/08 Database and Its Use to Assess Electronic Structure Model Chemistries for Chemical Reaction Barrier Heights},
journal = {Journal of Chemical Theory and Computation},
volume = {5},
number = {4},
pages = {808-821},
year = {2009},
doi = {10.1021/ct800568m},}

@article{Peverati2014,
   author = {Roberto Peverati and Donald G. Truhlar},
   doi = {10.1098/RSTA.2012.0476},
   issn = {1364503X},
   issue = {2011},
   journal = {Philosophical Transactions of the Royal Society A: Mathematical, Physical and Engineering Sciences},
   month = {3},
   publisher = {The Royal Society Publishing},
   title = {Quest for a universal density functional: the accuracy of density functionals across a broad spectrum of databases in chemistry and physics},
   volume = {372},
   url = {https://royalsocietypublishing.org/doi/10.1098/rsta.2012.0476},
   year = {2014},
}

@misc{fanEliminatingDelocalizationError2026,
  title = {Eliminating {{Delocalization Error}} through {{Localized Orbital Scaling Correction}} with {{Orbital Relaxation}} from {{Linear Response}}},
  author = {Fan, Yichen and Yu, Jincheng and Du, Jiayi and Yang, Weitao},
  year = 2026,
  month = feb,
  number = {arXiv:2602.11003},
  eprint = {2602.11003},
  primaryclass = {physics},
  publisher = {arXiv},
  doi = {10.48550/arXiv.2602.11003},
  urldate = {2026-02-12},
  note = {arXiv:2602.11003},
  archiveprefix = {arXiv}
}

@book{madelung2004,
  title = {Semiconductors: {{Data Handbook}}},
  shorttitle = {Semiconductors},
  author = {Madelung, Otfried},
  year = {2004},
  publisher = {Springer Berlin Heidelberg},
  location = {Berlin, Heidelberg},
  doi = {10.1007/978-3-642-18865-7},
  url = {http://link.springer.com/10.1007/978-3-642-18865-7},
  urldate = {2025-12-10},
  isbn = {978-3-642-18865-7},
  langid = {english}
}

@article{perdew1997,
  title = {Comment on ``{{Significance}} of the Highest Occupied {{Kohn-Sham}} Eigenvalue''},
  author = {Perdew, John P. and Levy, Mel},
  year = 1997,
  month = dec,
  journal = {Phys. Rev. B},
  volume = {56},
  number = {24},
  pages = {16021--16028},
  issn = {0163-1829, 1095-3795},
  doi = {10.1103/PhysRevB.56.16021},
  urldate = {2024-06-28},
  copyright = {http://link.aps.org/licenses/aps-default-license}
}

@article{yang2012b,
  title = {Derivative Discontinuity, Bandgap and Lowest Unoccupied Molecular Orbital in Density Functional Theory},
  author = {Yang, Weitao and Cohen, Aron J and {Mori-S{\'a}nchez}, Paula},
  year = 2012,
  journal = {J. Chem. Phys},
  volume = {136},
  number = {20},
  issn = {00219606},
  doi = {10.1063/1.3702391},
  pmid = {22667544}
}

@article{mahler_etal_2025,
  title = {Wannier Functions Dually Localized in Space and Energy},
  author = {Mahler, Aaron and Williams, Jacob Z. and Su, Neil Qiang and Yang, Weitao},
  date = {2025-08-18},
  journal = {Phys. Rev. B},
  volume = {112},
  number = {7},
  pages = {075137},
  publisher = {American Physical Society},
  doi = {10.1103/16r3-hjc7},
  url = {https://link.aps.org/doi/10.1103/16r3-hjc7},
  urldate = {2025-12-11},
}

@misc{mahler2024a,
  title = {\texttt{wannier90:costSE}},
  author = {Mahler, Aaron},
  year = {2024},
  urldate = {2024-07-31},
  howpublished = {\url{https://github.com/mtesseracted/wannier90}}
}

@misc{williams_yang_2025,
  title = {\texttt{ollosc}},
  author = {Williams, Jacob and Yang, Weitao},
  year = {2026},
  howpublished = {\url{https://gitlab.oit.duke.edu/jzw5/ollosc}},
  urldate = {2026-03-09}
}

@article{liu2019,
  title = {Accelerating {{{\emph{GW}}}} -{{Based Energy Level Alignment Calculations}} for {{Molecule}}--{{Metal Interfaces Using}} a {{Substrate Screening Approach}}},
  author = {Liu, Zhen-Fei and Da Jornada, Felipe H. and Louie, Steven G. and Neaton, Jeffrey B.},
  year = 2019,
  month = jul,
  journal = {J. Chem. Theor. Comput.},
  volume = {15},
  number = {7},
  pages = {4218--4227},
  issn = {1549-9618, 1549-9626},
  doi = {10.1021/acs.jctc.9b00326},
  urldate = {2026-03-09},
  copyright = {https://doi.org/10.15223/policy-029}
}

@article{whitten_1973,
  title = {Coulombic Potential Energy Integrals and Approximations},
  author = {Whitten, J. L.},
  date = {1973-05-15},
  journal = {J. Chem. Phys.},
  volume = {58},
  number = {10},
  pages = {4496--4501},
  issn = {0021-9606},
  doi = {10.1063/1.1679012},
  url = {https://doi.org/10.1063/1.1679012},
  urldate = {2026-03-09},
}

@article{dunlap_etal_1979,
  title = {On Some Approximations in Applications of $X \alpha$ Theory},
  author = {Dunlap, B. I. and Connolly, J. W. D. and Sabin, J. R.},
  date = {1979-10-15},
  journal = {J. Chem. Phys.},
  volume = {71},
  number = {8},
  pages = {3396--3402},
  issn = {0021-9606},
  doi = {10.1063/1.438728},
  url = {https://doi.org/10.1063/1.438728},
  urldate = {2026-03-09},
}

@article{vahtras_etal_1993,
  title = {Integral Approximations for {{LCAO-SCF}} Calculations},
  author = {Vahtras, O. and Almlöf, J. and Feyereisen, M. W.},
  date = {1993-10-15},
  journal = {Chem. Phys. Lett.},
  volume = {213},
  number = {5},
  pages = {514--518},
  issn = {0009-2614},
  doi = {10.1016/0009-2614(93)89151-7},
  url = {https://www.sciencedirect.com/science/article/pii/0009261493891517},
  urldate = {2026-03-09},
}

@article{zheng2011,
  title = {Improving Band Gap Prediction in Density Functional Theory from Molecules to Solids},
  author = {Zheng, Xiao and Cohen, Aron J. and {Mori-S{\'a}nchez}, Paula and Hu, Xiangqian and Yang, Weitao},
  year = 2011,
  month = jul,
  journal = {Phys. Rev. Lett.},
  volume = {107},
  number = {2},
  issn = {00319007},
  doi = {10.1103/PhysRevLett.107.026403}
}

@article{li2015,
  title = {Local {{Scaling Correction}} for {{Reducing Delocalization Error}} in {{Density Functional Approximations}}},
  author = {Li, Chen and Zheng, Xiao and Cohen, Aron J and {Mori-S{\'a}nchez}, Paula and Yang, Weitao},
  year = 2015,
  month = feb,
  journal = {Phys. Rev. Lett.},
  volume = {114},
  number = {5},
  pages = {53001},
  issn = {0031-9007, 1079-7114},
  doi = {10.1103/PhysRevLett.114.053001},
  langid = {english}
}

@misc{mahler2022,
  title = {Wannier {{Functions Dually Localized}} in {{Space}} and {{Energy}}},
  author = {Mahler, Aaron and Williams, Jacob Z. and Su, Neil Qiang and Yang, Weitao},
  year = {2022},
  month = jan,
  eprint = {2201.07751v1},
  primaryclass = {cond-mat.mtrl-sci},
  publisher = {arXiv},
  urldate = {2022-01-20},
  archiveprefix = {arXiv}
}

@misc{yang_ayers_2024,
  title = {Foundation for the $\Delta$SCF Approach in {{Density Functional Theory}}},
  author = {Yang, Weitao and Ayers, Paul W.},
  date = {2024-03-07},
  eprint = {2403.04604},
  eprinttype = {arXiv},
  eprintclass = {physics},
  doi = {10.48550/arXiv.2403.04604},
  url = {http://arxiv.org/abs/2403.04604},
  urldate = {2026-03-09},
  pubstate = {prepublished},
}

@misc{yang_fan_2024a,
  title = {Fractional {{Charges}}, {{Linear Conditions}} and {{Chemical Potentials}} for {{Excited States}} in $\Delta$SCF {{Theory}}},
  author = {Yang, Weitao and Fan, Yichen},
  date = {2024-08-19},
  eprint = {2408.08443},
  eprinttype = {arXiv},
  eprintclass = {physics},
  doi = {10.48550/arXiv.2408.08443},
  url = {http://arxiv.org/abs/2408.08443},
  urldate = {2026-03-09},
  pubstate = {prepublished},
}

@misc{yang_fan_2024,
  title = {Orbital {{Energies Are Chemical Potentials}} in {{Ground-State Density Functional Theory}} and {{Excited-State}} $\Delta$SCF Theory},
  author = {Yang, Weitao and Fan, Yichen},
  date = {2024-08-19},
  eprint = {2408.10059},
  eprinttype = {arXiv},
  eprintclass = {physics},
  doi = {10.48550/arXiv.2408.10059},
  url = {http://arxiv.org/abs/2408.10059},
  urldate = {2026-03-09},
  pubstate = {prepublished},
}

@misc{fan2026,
  author = {Fan, Yichen and Williams, Jacob Z. and Yang, Weitao},
  title = {Data and scripts from: olLOSC: Unified and efficient density functional approximation to correct
       delocalization error in molecules and periodic materials [Data set]},
  year = {2026},
  doi = {10.7924/r4r489},
  note = "{Duke Research Data Repository}",
}
